\documentclass[usenatbib]{aa}  
\usepackage{threeparttable}
\usepackage{adjustbox}
\usepackage{rotating}
\usepackage{amsmath}
\usepackage{color,soulutf8}
\usepackage{colortbl}
\usepackage{subfig}
\soulregister\ref7
\soulregister\cite7
\soulregister\citet7
\soulregister\citep7
\soulregister\citealp7

\usepackage{arydshln}
\usepackage{graphicx}
\usepackage{txfonts}
\usepackage[breaklinks=true,colorlinks=true,allcolors=blue,pagebackref=true,draft=false]{hyperref}
\usepackage{orcidlink}

\usepackage[normalem]{ulem}

\newcommand{\mypm}{\mathbin{\smash{%
\raisebox{0.65ex}{%
            $\underset{\raisebox{0.05ex}{$\smash -$}}{\smash+}$%
            }%
        }%
    }%
}

\newcommand{\goodchi}{\protect\raisebox{2pt}{$\chi$}}

\newcommand*\vi{\color{violet}\bf}
\newcommand{\ts}{\,}

\makeatletter
\renewcommand*\aa@pageof{, page \thepage{} of \pageref*{LastPage}}
\makeatother

\usepackage[draft=false]{microtype}

\begin{document} 
\title{
The accretion burst of the massive young stellar object G323.46$-$0.08\thanks{Based on observations made with the NASA/DLR Stratospheric Observatory for Infrared Astronomy (SOFIA) under Proposal ID 75\_0103 and at the European Organization for Astronomical Research in the Southern Hemisphere under ESO programs ID 077.C-0687(A), 083.C-0582(A), and 290.C-5165(A).}}

   \author{V. Wolf\orcidlink{0000-0003-3637-666X}\inst{1}
          \and
          B. Stecklum\orcidlink{0000-0001-6091-163X}\inst{1}
          \and
          A. Caratti o Garatti\orcidlink{0000-0001-8876-6614}
          \inst{2}
          \and
          P.~A. Boley\inst{1, 9}\orcidlink{0000-0001-7923-2979}
          \and
          Ch. Fischer\inst{3}\orcidlink{0000-0003-2649-3707}
          \and 
          T. Harries\inst{4}\orcidlink{0000-0001-8228-9503}
          \and
          J. Eislöffel\inst{1}\orcidlink{0000-0001-6496-0252}
          \and
          H. Linz\inst{5}\orcidlink{0000-0002-8115-8437}
          \and
          A. Ahmadi\inst{6}\orcidlink{0000-0003-4037-5248}
          \and
          J. Kobus\inst{7}\orcidlink{0000-0002-3741-5950}
          \and
          X. Haubois\inst{8}\orcidlink{0000-0001-7878-7278}
          \and
          A. Matter\inst{9}
          \and
          P. Cruzalebes\inst{9}
          }

   \institute{
            Thüringer Landessternwarte Tautenburg, Sternwarte 5, 07778 Tautenburg, Germany,
             \email{verena@tls-tautenburg.de}
        \and
        INAF - Osservatorio Astronomico di Capodimonte, Salita Moiariello 16, 80131, Napoli, Italy
        \and
            Deutsches SOFIA Institut, University of Stuttgart, 70569 Stuttgart, Germany
        \and
            Department of Physics and Astronomy, University of Exeter, Stocker Road, Exeter EX4 4QL, United Kingdom
        \and
            Max Planck Institute for Astronomy, Königstuhl 17, 69117 Heidelberg, Germany
        \and
            ASTRON, Netherlands Institute for Radio Astronomy, Oude Hoogeveensedijk 4, 7991 PD Dwingeloo, The Netherlands
        \and
            Institute of Theoretical Physics and Astrophysics, University of Kiel, Leibnizstraße 15, 24118 Kiel, Germany
        \and
            European Organisation for Astronomical Research in the Southern Hemisphere, Casilla 19001, Santiago 19, Chile  
        \and
            Université Côte d'Azur, Observatoire de la Côte d'Azur, CNRS, Laboratoire Lagrange UMR 7293, Bâtiment H. Fizeau, F-06108 Nice Cedex 2, France  
        }

   \date{Received March 07, 2024; accepted May 15, 2024}

 
  \abstract
  {
   {
Accretion bursts from low-mass young stellar objects (YSOs) have been known
for
many decades. In recent years, the first accretion bursts of massive YSOs (MYSOs) have been observed.
These phases of intense protostellar growth are of particular importance for studying massive star formation. Bursts of MYSOs are accompanied by flares of Class II methanol masers (hereafter masers), which are caused by an increase in exciting mid-infrared (MIR) emission. 
They can lead to long-lasting thermal afterglows of the dust continuum radiation visible at infrared (IR) and (sub)millimeter (hereafter (sub)mm) wavelengths. Furthermore, they might cause a scattered light echo.
The G323.46$-$0.08 (hereafter G323) event, which shows all these features, extends 
the small sample of known MYSO bursts.
   }
   }
   {
   {
Maser observations of the MYSO G323 
show evidence of a flare, which was presumed to be caused by an accretion burst. This should be verified with IR data.
We used time-dependent radiative transfer (TDRT) to characterize the heating and cooling timescales for eruptive MYSOs and to infer the main burst parameters. 
   }
   }
   {
   {
Burst light curves, as well as the pre-burst spectral energy distribution (SED) were 
established from archival IR data.
The properties of the MYSO, including its circumstellar disk and envelope, were derived by using static radiative transfer modeling of pre-burst data. For the first time, TDRT was used to predict the temporal evolution of the SED. Observations with SOFIA/HAWC+ were performed
to constrain the burst energy from the strength of the thermal afterglow. Image subtraction and ratioing were applied to reveal the light echo. 

   }
   }
   {
   {
The G323 accretion burst is confirmed. It reached its peak in late 2013/early 2014 with a $K_{\rm s}$-band increase of ${\sim}${\ts}2.5{\ts}mag.
Both $K_{\rm s}$-band and integrated maser flux densities follow an exponential decay. 
TDRT indicates that the duration of the thermal afterglow in the far-infrared (FIR) can exceed the burst duration by years.
The latter was proved by SOFIA observations, which indicate a flux increase of $(14.2\pm4.6)$\% at $70\, \rm \mu m$ and $(8.5\pm6.1)$\% at $160{\ts}\mu$m in 2022 (2 years after the burst ended). 
A one-sided light echo emerged that was propagating into the interstellar medium.
}
}
{
{
The burst origin of the G323 maser flare has been verified. 
TDRT simulations 
revealed the strong influence of the burst energetics and the local 
dust distribution on the strength and duration of the afterglow.
The G323 burst is probably the most energetic MYSO burst that has been observed so far. Within $8.4 \rm \, yrs$, an energy of $(0.9\mypm_{0.8}^{2.5}) \times 10^{47}\,\rm erg$ was released. The short timescale points to 
the accretion of a compact body, while the burst energy corresponds to an accumulated mass of at least 
$(7\mypm_{6}^{20})\,M_{Jup}$ 
and possibly even more if the protostar is bloated.
In this case, the accretion event might have triggered protostellar pulsations, which give rise to the observed maser periodicity.
The associated IR light echo is the second observed from a MYSO burst. 
}
}
   \keywords{Accretion, accretion disks -- Stars: formation --
                Stars: protostars --
                Stars: individual objects (\object{G323.46$-$0.08},\hfil\newline \object{IRAS15254-5621}) -- Radiative transfer -- Masers
               }
   \authorrunning{V. Wolf et al.}
   \maketitle
%
\,
\section{Introduction}\label{intro}
Episodic accretion events are phases of strongly enhanced mass accumulation during (proto-)stellar growth. These are not restricted to young stars of low and intermediate mass \citep{hartmannFUOrionisPhenomenon1996, audardEpisodicAccretionYoung2014}, but they also occur during the formation of high-mass stars (\citealp{carattiDiskmediatedAccretionBurst2017};  \citealp{hunterExtraordinaryOutburstMassive2017}; \citealp{Stecklum:2021}). The energy released during such accretion bursts strongly affects the massive young stellar object (MYSO) and its circumstellar environment in various ways. Heating of the dust in both the circumstellar disk and envelope 
leads to a temporary increase in the dust continuum emission. Depending on the strength and duration of the burst, this partially or completely affects the spectral energy distribution (SED). The timescales on which the SED changes are wavelength dependent, where different wavelength regions trace different spatial regions (e.g., \citealp{pena:2020}). 
Therefore, accretion bursts provide a unique opportunity for 
"screening" MYSOs, which are often deeply embedded throughout their formation time. 

A specific aspect of MYSO accretion bursts is that their high luminosity leads to the sublimation of volatile substances that were trapped in the ice mantles of dust grains. Thus, for certain molecules, such as methanol in particular, maser emission can occur once the excitation conditions, for example, the specific column density and strong mid-infrared (MIR) radiation field, are satisfied. For this reason, radiatively pumped $\rm CH_3OH$ masers (Class \rm{II}, \citealp{Menten:1991a, Sobolev:1997, Cragg:2005}) are a very good tracer of MYSOs \citep{Breen:2013}. These masers will flare during accretion bursts, which makes them a reliable burst alert. Particularly useful is the 6.7{\ts}GHz transition \citep{Menten:1991b}, as this is usually the brightest.
During the burst, the maser spots can be relocated, thus providing information on the local structures such as spiral arms in a disk (see, e.g., \citealp{ross:2020}).

Even after the burst is over, the far-infrared (FIR) fluxes remain elevated for quite some time, as witnessed for the MYSO G358.93-0.03 burst \citep{Stecklum:2021}.
This thermal afterglow depends on the local dust distribution and is a record of the history of the burst. 
Its detection represents an a posteriori confirmation for MYSO bursts, which were not or could not be detected by other means.
It allows for the energy release to be constrained, which is fundamental for understanding the triggering mechanisms behind the burst.
Until its recent shutdown, the Stratospheric Observatory for Infrared Astronomy (SOFIA, \citealp{1993AdSpR..13..549E, 2012ApJ...749L..17Y}) was the only facility offering the capability to verify the increase in FIR flux caused by the burst, thus allowing for the detection of such an afterglow. 


The object G323, which is also known as IRAS15254-5621, is a massive star-forming region. It is located at 
RA: $\rm 15^h 29^m 19 
.\!^s4{}$, $\rm {DEC\!:~} {-}56\degr 31\arcmin 23\arcsec$, J2000. It is covered by various surveys, since it 
resides in the vicinity of the Galactic center. 
The red MSX survey (RMS; \citealp{lumsden:2013}) and the {\bf A}PEX {\bf T}elescope {\bf L}arge {\bf A}rea {\bf S}urvey of the {\bf Gal}axy (ATLASGAL; \citealp{ATLASGALSchuller2009}) revealed the main properties of the region. A bolometric luminosity of $L{\ts}{\sim}{\ts}(1\,-\,1.3){\ts}{\times}{\ts}10^5{\ts}{\rm L}_\odot$ was derived by integrating its SED \citep{lumsden:2013}. 
G323 is accompanied by the compact ATLASGAL clump AGAL~323.459-00.079 \citep{urquhart:2014}, considered a massive cluster progenitor \citep{csengeri:2017} with a mass of ${\sim}{\ts}600{\ts}{\rm M_\odot}$ that hosts the MYSO. For this reason, the source is included in the ALMAGAL survey, 
a large program on ALMA (2019.1.00195.L, PI: S. Molinari), dedicated to studying the evolution of high-mass protocluster formation in the Galaxy. 
\cite{arayaCH3CNObservationsSouthern2005} found blueshifted and redshifted CS{\ts}(3–2) as well as $^{13}$CO{\ts}(2–1) line wings, indicative of a molecular outflow. A similar signature in the $^{18}$CO{\ts}(2–1) line is observed by the SEDIGISM survey \citep{Yang:2022}. However, \cite{Guerra-Varas;2023} did not detect an outflow in the SiO{\ts}(2-1) line and wings of the HCO$^+${\ts}(2-1) line with the APEX telescope.
Recently, \cite{Yingxiu} identified this clump as a hub for star formation fed by three filaments. For the hub junction, they derived an extent of 2.4\,pc$\ts\times$\ts2.4\,pc and a mass of $(3072\pm1200)\ts M_\odot$.

Associated weak 
radio continuum emission was first reported by \cite{haynesSouthernHemisphereSurvey1978} and later mapped by \cite{Walsh:1998} who did not resolve the source ($<2\arcsec$). \cite{arayaCH3CNObservationsSouthern2005} detected broad radio recombination lines toward this region, indicating the presence of a hypercompact H{\sc ii} region (HCH{\sc ii}). Similar measurements of \cite{murphy:2010}, \cite{Kim:Urquhart:2018} confirmed this finding. 
\cite{murphy:2010} also estimated the extinction toward G323, based on the depth of the silicate absorption feature in a low-resolution IRAS spectrum, which amounts to $A_{\rm V}{\ts}{=}{\ts}(18{\ts}{\pm}{\ts}1)${\ts}mag. The object was observed in the ATOMS survey (ALMA Three-millimeter Observations of Massive Star-forming regions, \citealp{Liu;2020}) in Autumn 2019. The radio recombination line and the continuum data at 3\,mm are used by \cite{Zhang;2023} to characterize the HCH{\sc ii}.

The near-kinematic distance of approximately 4.2{\ts}kpc is generally preferred, rather than the far distance of 9.3{\ts}kpc \citep{lumsden:2013,csengeri:2017}.
Since there is no maser parallax available yet for G323, we re-evaluated its kinematic distance using the radial velocity with respect to the local standard of rest (LSR) of $\varv_{\rm LSR}{\ts}{=}-67.2{\ts}\rm km~s^{-1}$ \citep{proven-adzriDiscoveryPeriodicMethanol2019} together with the kinematic 
model A5 from \cite{reid:2014} which yields
a value of (4.08$\mypm{\ts}^{0.40}_{0.38})${\ts}kpc. This distance is consistent with the largest {\it GAIA}-DR2 stellar distances \citep{bailer-jones:2018} of up to 3.7{\ts}kpc for stars in the foreground of the nebulosity associated with G323. 
It implies an interstellar extinction of  $A_{\rm V}{\ts}{\sim}{\ts}6${\ts}mag according to the model of \cite{Amores:2005}. The higher Av 
of \cite{murphy:2010} is reasonable, since the object is located in a star-forming region. 
{\it GAIA} detected a faint source (ID 5883491191298706432, $G=21.45\pm0.06$) close to the position of G323 with an extremely red color $BP-RP=7.29\pm0.81$ \citep{10.1051/0004-6361/202039657} that points to scattered light from the MYSO. Unfortunately, for such a faint source, the parallax error of {\it GAIA} exceeds 1\,mas \citep{Cantat-Gaudin:2018}, making the distance determination unfeasible.

The massive star-forming region G323 is associated with methanol, water, and hydroxyl masers. The Class \rm{II} methanol 6.7{\ts}GHz transition was first observed by \cite{macleodSearchDetectionMethanol1992}. Monitoring of this transition at Hartebeesthoek Radio Astronomy Observatory (HartRAO) revealed an increase in the total flux density from ${\sim}${\ts}20{\ts}Jy in 2011 \citep{greenExcitedstateHydroxylMaser2015} to ${\sim}${\ts}7\,500{\ts}Jy in 2015 \citep{proven-adzriDiscoveryPeriodicMethanol2019}, 
that is, by a factor of ${\sim}${\ts}430. Interestingly, the decay of the methanol maser, which has been ongoing since then, is characterized by a periodic flux variation with a cycle time of ${\sim}${\ts}93{\ts}d \citep{MacLeod:2021}. The first flare evidence was
obtained from the brightening of the 6.035{\ts}GHz exOH maser \citep{MacLeod:2021}.
The discovery of the maser flare raises the question whether it is due to an accretion burst, similar to S255IR-NIRS3 \citep{carattiDiskmediatedAccretionBurst2017}, NGC6334I-MM1 \citep{hunterExtraordinaryOutburstMassive2017, hunter:2018}, and G358 \citep{brogan:2019,macleod:2019,ross:2020,Stecklum:2021}. 
 The source is classified as "irregular" in the recent NEOWISE-based variability study of 6.7{\ts}GHz maser sources by \cite{song:2023}. Their exclusion of the pre-burst $WISE$ epochs prevents the burst detection.

We study the G323 event by combining archival data with recent SOFIA/HAWC+ measurements and the application of
TDRT
models. 
So far, the question of the heating and cooling times of YSO dust was treated in a simplified fashion by assuming extreme cases of optical thickness together with
energy considerations \citep{johnstone:2013}. 
The use of the time-dependent radiative transfer code TORUS (\citealp{Harries2011,Harries2019}) allows us to estimate the thermal timescales self-consistently. This work represents the first application of TDRT simulations of dust-continuum emission to a real astrophysical object. 

The paper is organized as follows. At first, the 
data obtained by recent observations and from the literature are described in Sect. \ref{obs}.
This is followed by a presentation of the results of the data analysis in Sect. \ref{res}. The theoretical part in Sect. \ref{RTM} contains the radiative transfer (RT) modeling. It starts with the pre-burst static RT modeling. Then, TDRT modeling follows, which predicts the afterglow evolution and main burst parameters for three limiting cases. We draw our conclusions about the burst impact at the end of Section \ref{sec: TDRT res}. 
The Discussion and Conclusion sections evaluate and summarize our findings and put them in the context of current observations of MYSO accretion bursts.


\section{Observations and data reduction}\label{obs}

\subsection{VVV(X) survey imaging}
The {\bf V}ISTA {\bf V}ariables in {\bf V}ia Lactea Survey (VVV, \citealp{minnitiVISTAVariablesLactea2010}) and its e{\bf x}tension VVVX (from 2016 to 2019) are ESO public surveys that were conducted with the 4-m VISTA telescope in the red optical/near-infrared (NIR) domain. They target the Galactic bulge and part of the adjacent plane. Images of the G323 region were retrieved from the VISTA Science Archive. VVV(X) obtained $K_{\rm s}$-band imaging in all of their observing epochs of the region.
Data for the $Z, Y, J$ and $H$ filters are available only for 2010 and one epoch in 2015. Although this does not suffice to produce light curves, a possible color change resulting from the burst can be studied nevertheless. 
NIR images show the bright counterpart of G323 embedded in a scattering nebulosity. 
Photometry was established on VVV(X) images, taking into account the core saturation of its point spread function (PSF) in the $K_{\rm s}$-band. 
The affected pixels were given zero weight when fitting the image profile. This was performed using the MPFit2DPeak function of the IDL Astronomy Users Library \citep{landsmanIDLAstronomyUser1995}, using a tilted Moffat function that is appropriate for PSFs in ground-based images \citep{moffat:1969}. Images with seeing worse than 3\farcs5 were discarded.
The photometry is given in Table{\ts}\ref{tab:Ksmag} of the Appendix.

\subsection{Skymapper survey imaging}\label{ssec: skymap}
The very recent fourth release \citep{Onken:2024} of the Skymapper survey\footnote{\url{https://skymapper.anu.edu.au/}} provides images of the southern sky taken from March 2014 to September 2021 using $uvgriz$ filters. Inspection of the G323 region confirmed the detection of the object during the burst in the $i$ and $z$ bands, although it is not listed in the catalog. After retrieving those images, aperture photometry of G323 was performed and calibrated using the photometric zero points given in the FITS header. The same was done for the three VVV $z$-band images taken at one pre-burst and two burst epochs to supplement the $z$-band photometry. We note, that the filter transmission of 
VISTA and Skymapper are slightly different. The effective wavelengths are 0.877 and $0.912\,\rm \mu m$ respectively, with a band-width of 0.097, and $0.116\,\rm \mu m$. Using the spectral slope in this wavelength range and the difference of the effective wavelengths, the Skymapper magnitudes were tied to the VISTA  $z$-band. The photometry is given in Table{\ts}\ref{tab:Skymp} of the Appendix.

\subsection{ISAAC spectroscopy}
Long-slit spectroscopy with the 
Infrared Spectrometer and Array Camera ISAAC was performed on June 11, 2013,  at the ESO-VLT UT3 telescope, Paranal, Chile.
A slit of 0\farcs3 width was used, producing a spectral resolution of $\sim$8\,900, at a position angle (PA) of 43\fdg8, centered on the source (see Fig. ~\ref{fig:naco}). 
The total integration time was 4 min. Nodding along the slit was performed with a throw of 35\arcsec{} to allow for sky subtraction. The grating was set at a central wavelength of 2.15{\ts}$\mu$m, offering a spectral bandwidth of 0.122{\ts}$\mu$m, covering a wavelength range from 2.086 to 2.210{\ts}$\mu$m. Data reduction was performed using standard IRAF\footnote{IRAF (Image Reduction and Analysis Facility) is distributed by the National
Optical Astronomy Observatories, which are operated by AURA, Inc., in cooperative agreement with the National Science Foundation.} tasks. Each observation was flat-fielded, sky subtracted, and corrected for the distortion caused by long-slit spectroscopy. The atmospheric response was corrected by dividing each spectrum by a telluric standard star (acquired with the same science settings), normalized to the blackbody function at the stellar temperature, and corrected for any absorption feature intrinsic to the star. Originally, we used the telluric standard star for flux calibration. However, probably due to slit flux loss, the resulting calibrated spectrum is $K$ = 6.73, mag, that is, $\sim$ 0.6, mag lower than the VVV photometry around that epoch (see Sect. \ref{sec: IR-col}). Therefore, for calibration, we adopted the closest $K$-band photometric point (16 June 2013). Wavelength calibration was performed using the many OH lines located in this spectral range, resulting in an average uncertainty of 0.2{\ts}\text{\AA} in the extracted spectra. Three spectra, centered on the source and a few arcseconds NE and SW off the source, were extracted from the spectral image, where both continuum and line emission are present. Continuum emission from the final spectral image was
removed with the IRAF task "continuum".
Finally, the radial velocities of the observed lines were calculated using single or multiple Gaussian fits (for multiple components), 
and corrected for the target velocity 
with respect to the LSR.

\subsection{NACO adaptive-optics imaging}
The object G323 was observed on 3 June 2009 using NAOS-CONICA (NACO, \citealp{roussetNAOSFirstAO2003,lenzenNAOSCONICAFirstSky2003}) in the $K_{\rm s}$-band at the ESO-VLT UT4 with a total exposure time of 24{\ts}s at a pixel scale of 0\farcs05. The image was retrieved from the ESO Science Archive Facility. Astrometric calibration was performed using five reference stars from {\it GAIA}-DR2 \citep{gaiacollaborationGaiaDataRelease2018}. For the established world coordinate system (WCS), the mean absolute deviation of the stellar positions from those of {\it GAIA}-DR2 amounts to 0\farcs018.

\subsection{(NEO)WISE photometry}

The $WISE$ mission utilized a cryogenic IR space telescope with a 40-cm aperture \citep{wrightWIDEFIELDINFRAREDSURVEY2010} that carried out an all-sky survey from 2010 to 2011 
in four spectral bands, ranging from 3{\ts}$\mu$m to 25{\ts}$\mu$m. 
After coolant depletion in 2011, the telescope was hibernated until 
reactivation in 2013 to become the NEOWISE mission \citep{ mainzerINITIALPERFORMANCETHENEOWISEREACTIVATION2014}. Because passive cooling is less efficient, since then only the two shortest bands, $W1$ (3.4{\ts}$\mu$m) and $W2$ (4.6{\ts}$\mu$m), can be used.

The (NEO)WISE photometry for G323 covering observations until end of 2022 was retrieved from the NASA/IPAC Infrared Science Archive (IRSA)\footnote{\url{https://irsa.ipac.caltech.edu}} using a search radius of 5\arcsec{} around the RMS position of RA: $\rm 15^h 29^m 19 
.\!^s59{}$, $\rm {DEC\!:~} {-}56\degr 31'{\vi } 21\farcs9$ \citep{mottramRMSSurveyMidinfrared2007}. Since the bright source is saturated in $W1$ and $W2$, a photometric bias correction was applied\footnote{See{\ts}{\ts}\url{http://wise2.ipac.caltech.edu/docs/release/neowise/expsup/sec2\_1civa.html}} to account for the detector warm-up.
Given the duration of the event and the particular time sampling of (NEO)WISE, we use the epoch-averaged magnitudes in the following.
Those for the two $WISE$ epochs preceding the burst amount to $W1{=}4.77{\pm}0.11$ and $W2{=}3.21{\pm}0.12$ which implies a color index $(W1-W2)$ of $1.56{\pm}0.16$. Photometry is given in Table\,\ref{tab:Wmag} of the Appendix.

\subsection{TIMMI2 MIR imaging}

The MIR imaging at 10.4{\ts}$\mu$m (epoch 2006) was performed within the RMS survey \citep{mottramRMSSurveyMidinfrared2007} using the {\bf T}hermal {\bf I}nfrared {\bf M}ulti{\bf M}ode {\bf I}nstrument TIMMI2 \citep{reimannTIMMI2NewMultimode2000} at the ESO 3.6-m telescope on La Silla. The image was retrieved from the corresponding RMS web page\footnote{\url{http://rms.leeds.ac.uk/cgi-bin/public/RMS_SEARCH_RESULTS.cgi?text_field_1=G323.4584-00.0787&radius_field=60&listID=1}}. 
To address its spatial extent, images of the standard star HR~5288, which had been observed before G323, were recovered from the ESO archive and used as PSF reference. 

\subsection{VLTI/MATISSE observations}\label{sec: matisse}

As part of an experiment using the CIAO off-axis mode of the Very Large Telescope Interferometer (VLTI) and the mid-infrared interferometric instrument MATISSE \citep{Lopez2022}, we attempted to observe G323 on the night of May 4, 2023 under technical time (ESO program ID 60.A-9801). While the source itself was overresolved by the interferometer (see below), we summarize the details of this experiment here, as it was the first attempt to use the CIAO subsystem with MATISSE.

Unlike the standard observing mode with MATISSE on the 8-m UT telescopes, which uses the MACAO adaptive optics system for telescope guiding and AO corrections at optical wavelengths, the CIAO system performs off-axis guiding and AO corrections at near-infrared wavelengths (i.e., the $H$, and $K$ band), making observations of embedded targets such as G323 possible. G323 itself does not have a suitably bright optical guide star within the $\sim1$\arcmin{} field accessible to MACAO.

During the observations, UT3 was offline due to technical problems, which means that only the UT1-UT2-UT4 triangle was used. We used the low-resolution standalone mode of MATISSE without chopping, which simultaneously covers the $L$, $M$, and $N$ bands. For the guide star, we used a $K=9.4$~mag star $41.6$\arcsec{} from G323, and we were able to both acquire flux from G323 and track the source without incident on all three telescopes. However, we were unable to acquire fringes for the source after multiple attempts, despite its relative brightness ($\ga$2 Jy in the $L$ band) and successful fringe acquisitions for calibrator stars immediately before and after (HD\,100713 and HD\,135902), implying that G323 was over-resolved at these baselines and wavelengths. For the shortest projected baseline on UT1-UT2 of 52\,m, this implies that the angular size of the compact emission at 3.5\,$\mu$m is roughly larger than ${\sim14}$\,mas (Gaussian full width at half maximum, FWHM). Here we assumed that a source with a correlated flux of less than 0.04\,Jy would be over-resolved by the interferometer (taken from the description of the MATISSE instrument\footnote{\href{https://www.eso.org/sci/facilities/paranal/instruments/matisse/inst.html}{https://www.eso.org/sci/facilities/paranal/instruments/matisse/inst.html}}). We discuss the implications of this constraint further in Sect. \ref{sec: r_inner}.

\subsection{\texorpdfstring{FIR imaging}{}}

As outlined below (cf. Sect.{\ts}\ref{sec:tsed}), key information on the afterglow is provided by the fluxes in the FIR region. Therefore, accurate photometry is required to detect even weak but elevated emission from the afterglow several years after the burst peak. 

G323 is very bright in the wavelength bands of HAWC+. According to the AKARI{\ts}/{\ts}FIS Bright Source Catalogue \citep{2009ASPC..418....3Y}, its pre-burst flux density exceeds 1500{\ts}Jy in this spectral range. 
The Herschel{\ts}/{\ts}PACS point source catalog (HPPSC, \citealp{2017arXiv170505693M}) lists the following fluxes F(70{\ts}$\mu$m){\ts}={\ts}(2459{\ts}$\pm${\ts}23){\ts}Jy and  F(160{\ts}$\mu$m){\ts}={\ts}(1721{\ts}$\pm${\ts}70){\ts}Jy. These will be used for primary comparison, with the HAWC+ fluxes interpolated and transferred to the PACS color system.

To obtain comparison data for the TDRT simulations, Directors' discretionary time (DDT, proposal ID 75\_0103) was granted for HAWC+ observations of G323 with SOFIA during the southern deployment at Christchurch in Cycle 9. HAWC+ \citep{harper:2018} is a FIR camera and imaging polarimeter that allows total and polarized flux imaging in five broad bands with central wavelengths of 53, 62, 89, 154, and 214{\ts}$\mu$m. HAWC+ provides a 64$\times$60 pixel footprint for imaging with pixel sizes ranging from 2\farcs55 to 9\farcs37 from the shortest to the longest wavelengths, producing a field of view (FoV) of 2\farcm8$\times$1\farcm7 to 8\farcm4$\times$6\farcm2.

The observations were carried out on 6 July 2022 (MJD 59766.5852) using all five spectral bands with an integration time of three minutes each.
Total flux imaging observations were made in on-the-fly mapping mode using the Lissajous scan type and employing scan amplitudes ranging from 30\arcsec{} to 90\arcsec{} in each direction. The final map sizes vary from 2\farcm9$\times$4\farcm1 to 10\farcm0$\times$12\farcm9. The data were retrieved from IRSA.
Photometry with variable aperture size was performed on both PACS and HAWC+ images to arrive at a flux comparison with the PACS measurements. Radii that reproduce HPPSC fluxes were derived by varying the aperture size on the PACS images.

\subsection{ALMA observations}
The ALMA Cycle 7 project ALMAGAL aims to observe the 1.3{\ts}mm continuum and lines toward dense molecular clumps in the Galactic Plane at a sensitivity level of $0.1{\ts}\rm mJy$. Measurements are made with the 12-m array and the 7-m Alma Compact Array (ACA). 
Data for G323 were retrieved from the ALMA Science Archive \citep[e.g.,][]{StoehrALMAScienceArchive2014}. For this paper, we used the available 12-m array measurements on ALMAGAL field 767784 (observing date: 30-December-2019), where we worked on the data products obtained by the standard pipeline in CASA (Common Astronomy Software Applications package, \citealp{mcmullin:water:2007}). This data set has 
a typical spatial resolution of 1\farcs2.
In the course of this work, a discrepancy was observed between the flux density at 1.4\,mm obtained from the ALMAGAL data and the 3\,mm measurement by \cite{Zhang;2023}. Therefore, the corresponding data set from \cite{Zhang;2023}, obtained at about the same epoch, was also analyzed using CASA.

\subsection{Archival data and SED}
Apart from the flux densities derived from the observations mentioned above, supplementary data were drawn using the VizieR photometry tool{\footnote{\href{https://vizier.cds.unistra.fr/vizier/sed/}{https://vizier.cds.unistra.fr/vizier/sed/}} } to establish the pre-burst SED of the source. The spectrum was augmented by four flux densities extracted from the IRAS LRS spectrum \citep{IRAS_LRS}.
The values are given in Tab \ref{tab: G323 preSED}.

\section{Data analysis and results}\label{res}

\subsection{IR imaging}\label{sec: IR-ima}

\begin{figure}[htb]
    \includegraphics[width=\hsize]{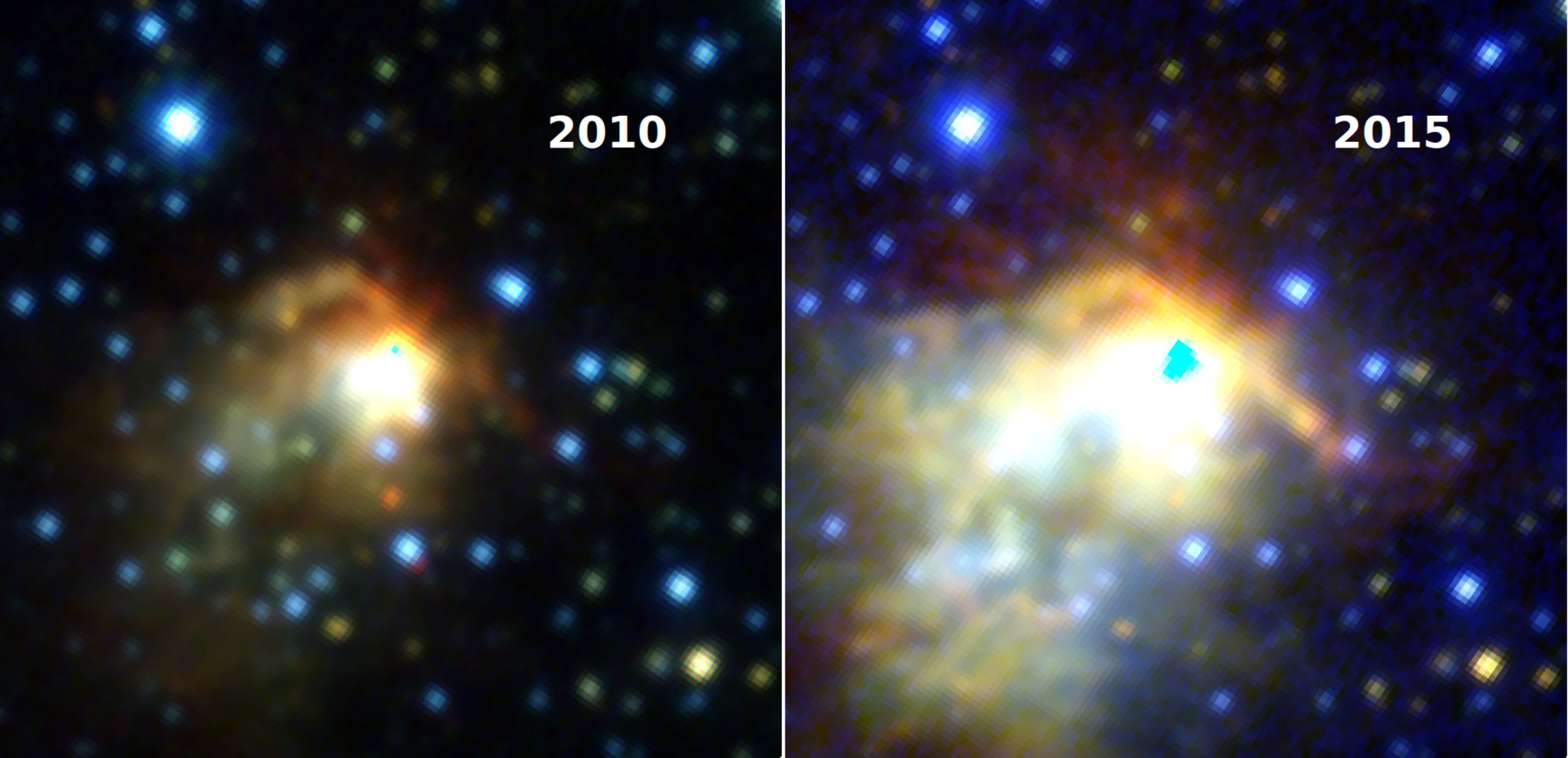}
\caption{VVV $JHK_{\rm s}$ pre-burst (left) and burst (right) color composites (FoV $44\arcsec{\ts}{\times}{\ts}44\arcsec$, north is up and east is to the left). The central cyan-colored areas are due to detector saturation. 
}
\label{fig:burst}
\end{figure}

The VVV $JHK_{\rm s}$ color composite images for the pre-burst and burst epochs are shown in Fig.{\ts}\ref{fig:burst}. The NIR images display the bright counterpart of G323 embedded in a scattering nebulosity. 
The increase in the NIR brightness of the MYSO and its surrounding reflection nebula as a result of the burst is obvious.

\begin{figure}[thb]
    \includegraphics[width=\hsize]{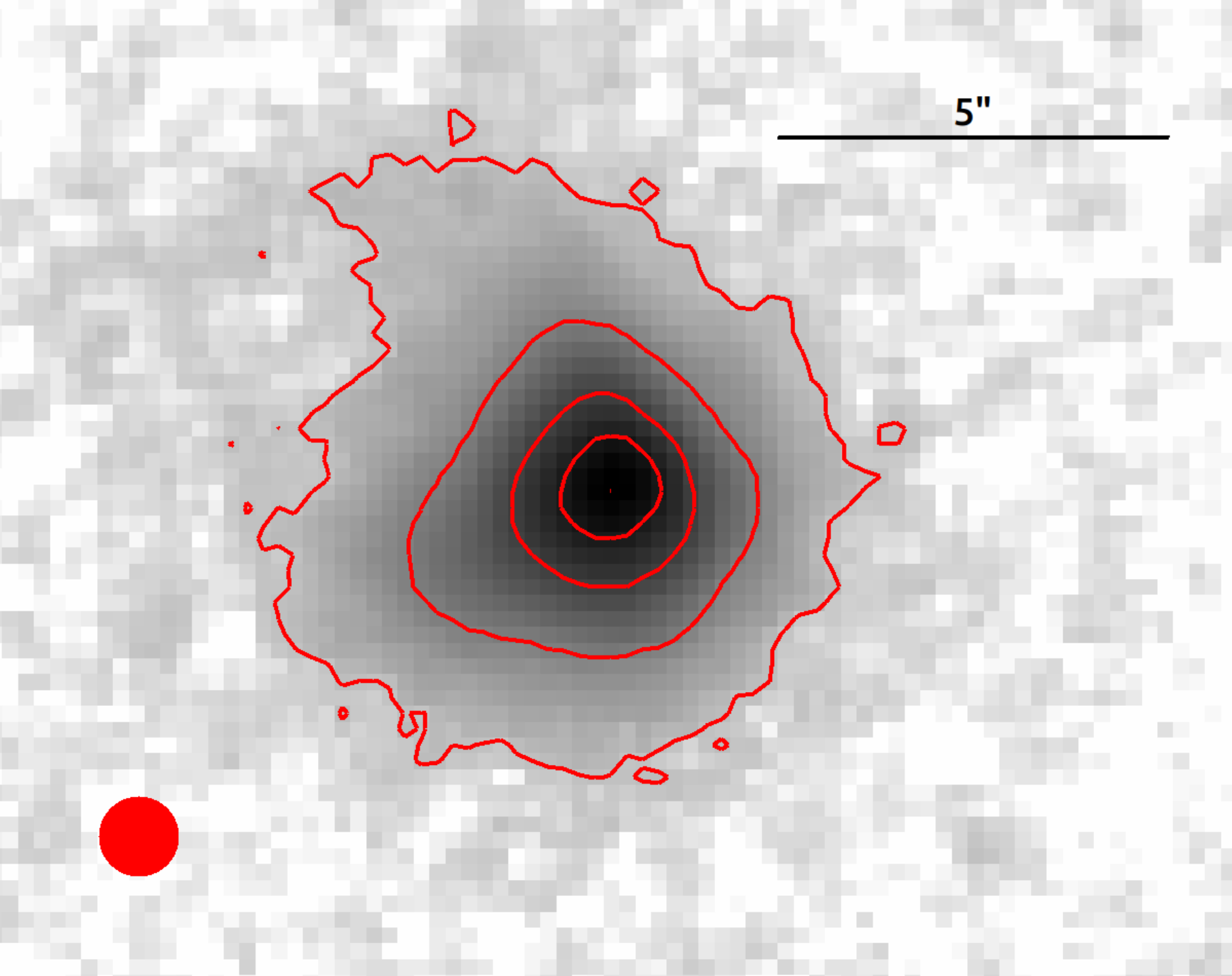}
\caption{TIMMI2 10.4{\ts}$\mu$m image (epoch 2006). The lower left circle shows the FWHM of a standard star measured before the object. The contours delineate $[5, 35, 200, 560]{\ts}{\times}{\ts}{1\sigma}$ levels where the latter corresponds to half of the peak value. No Airy-rings are visible, although the source was classified as point-like. 
}
 \label{fig:t2}
\end{figure}

The TIMMI2 image is shown in Fig.\ts{}\ref{fig:t2}. The contours delineate $[5, 35, 200, 560]{\ts}{\times}{\ts}{1\sigma}$ levels where the latter corresponds to half of the peak value. The lower-left circle shows the FWHM of a standard star measured before the object.
Although the object was classified as point-like, the absence of the first Airy ring, expected for an unresolved source in diffraction-limited imaging, indicates that it is marginally resolved. By subtracting the FWHMs of the standard star in quadrature, a deconvolved size of $(1\farcs20\times1\farcs08)\pm{0\farcs01}$ at a position angle of $102\fdg5\pm{0\farcs1}$ was obtained.

\begin{figure*}[htb]
    \includegraphics[width=\textwidth]{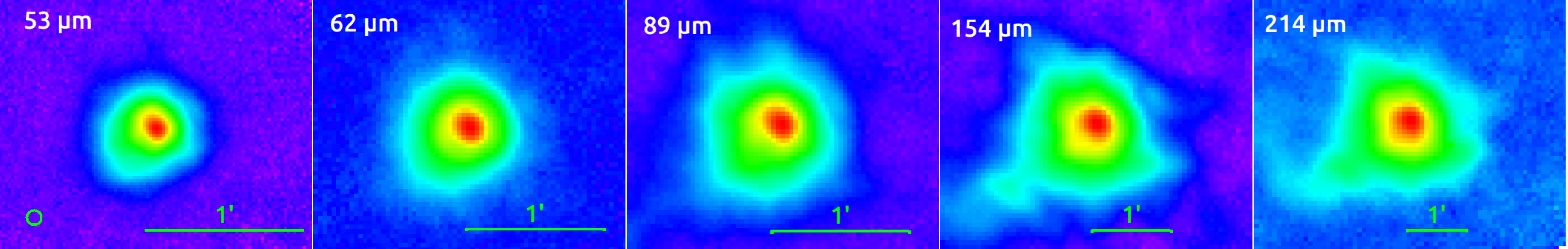}
\caption{HAWC+ log scaled image cutouts, centered on G323 and spatially scaled to the beam FWHM (lower left) for each band. The absence of Airy rings indicates that the source is resolved at all wavelengths. The horizontal line marks an angular size of 1\arcmin.}
 \label{fig:PACS_all}
\end{figure*}

The appearance of the target in the various HAWC+ bands is shown in Fig.{\ts}\ref{fig:PACS_all} where the image size is scaled to apparently `cancel' the wavelength dependence of the PSF. The absence of diffraction rings implies that the source is resolved at all HAWC+ wavelengths.

\subsection{IR light curves and color change}\label{sec: IR-col}

\begin{figure*}
    \includegraphics[width=\textwidth]{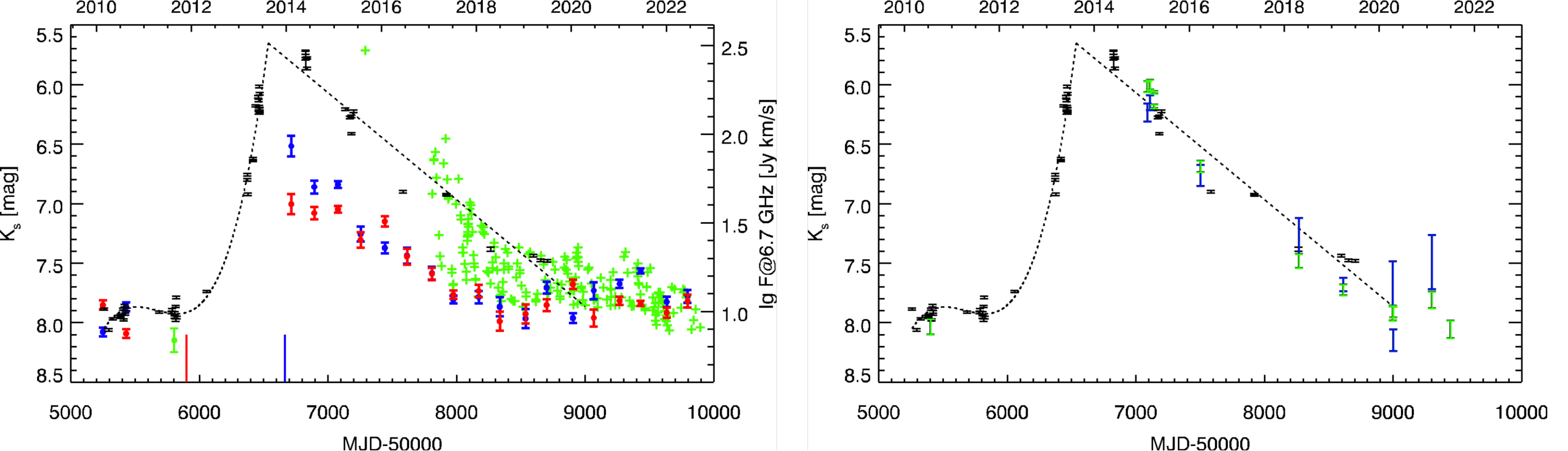}
	\caption{Left:
	Light curves based on VVV(X) (black) and (NEO)WISE photometry (W1 - blue, and W2 - red) as well as 6.7{\ts}GHz total maser flux (green, \citealp{greenExcitedstateHydroxylMaser2015, MacLeod:2021}). Vertical
 red and blue lines mark the dates of the burst onset and first flare evidence 
	from the 
	6.035{\ts}GHz exOH maser
 \citep{MacLeod:2021}.
    The $K_{\rm s}$ rise was approximated by 
    a polynomial, 
    while its decay is roughly linear on a log scale (dashed line). 
    The (NEO)WISE magnitudes are shifted 
    to match those of $K_{\rm s}$. The 
    integrated maser flux is shown on a log scale (right ordinate). 
    Its scatter is due to the short-term periodicity. 
    Right: $K_{\rm s}$ (black), $i$ (blue), and $z$ (green) light curves, with $i$ and $z$ magnitudes shifted 
    to match those of $K_{\rm s}$. 
    The $z$ pre- and post-magnitudes agree within the errors.
}
 \label{fig:lc}
\end{figure*}

The IR light curves based on VVV(X) $K_{\rm s}$-band as well as the (NEO)WISE photometry are shown in the left panel of Fig.{\ts}\ref{fig:lc}. For comparison, the (NEO)WISE $W1$ and $W2$ light curves were shifted to match the pre-burst $K_{\rm s}$-band magnitude.
Since the (NEO)WISE fluxes in both bands exceed the limit for which a meaningful bias correction can be applied, the burst light curves are not suitable for drawing quantitative conclusions. However, they represent an independent confirmation of the event, heralded by a maser-flare. The integrated maser flux is plotted in green. The scatter during and after the burst is due to the short-term periodicity of $93.5\, \rm d$ \citep{proven-adzriDiscoveryPeriodicMethanol2019}. The correlation between the maser flare and the burst is discussed in more detail in Sect. \ref{sec: K-IR}. The vertical 
red, and blue lines mark the dates of 
the burst onset, and the first flare evidence from the 6.035 GHz exOH
maser \citep{MacLeod:2021}. 

The temporal behavior of the $K_{\rm s}$-band brightness can be subdivided into two phases. The burst appears to have started in early 2012, and we designate 5 June 2012 (MJD 56083) 
as its onset date. At that date, the polynomial rise approximation started to exceed the error margin of the mean pre-burst $K_{\rm s}$-band magnitude of 7.91{\ts}mag.
A particular rapid flux increase occurred in June 2013, probably shortly before the burst peak. 
The intersection of the polynomial rise approximation and the 
exponential decay of the NIR flux variability suggests that the peak date of the burst was in late summer 2013, probably around 31 August 2013 (MJD 56535). Then it fainted with a linear trend of 0.75{\ts}${\rm mag{\ts}yr^{-1}}$ which corresponds to a flux decline with an e-folding time of 3.3 years. 
The pre-burst $K_{\rm s}$
magnitude was again reached around 27 September 2020 (MJD 59119). We consider this date to mark the end of the accretion burst, which then lasted $\sim$8.4 years. 
The right panel of Fig. \ref{fig:lc} shows the comparison of the $i$, and $z$-band light curve from the Skymapper survey with the $K_{\rm s}$ light curve. 
Both the $i$, and $z$ bands confirm the end of the burst in 2020.
The $z$ post-burst magnitude is the same as the pre-burst magnitude within the given errors. \\


The color indices $(H-K_s)$, and $(J-H)$ were estimated from images in the $J$-, and $H$-band taken about 1.8 years after the burst peak. 
They indicate that compared with its color during the pre-burst stage, the source was bluer by 0.7{\ts}mag in $(H-K_s)$, and 0.62{\ts}mag in $(J-H)$ at that time. Comparison of the $W1$, and $W2$ magnitudes at the burst peak appears to indicate that for G323 as well (see also Fig. \ref{fig:LE_lc}). 
 Young eruptive stars may become bluer or redder when brighter during the outburst (e.g., \citealp[]{Lucas:2024}) which cannot be explained by variable extinction alone. For example, \cite{pena:2023} observed bluer $(W1-W2)$ burst colors in the embedded FUor SPICY~97~855.

\subsection{The scattered light echo }
In recent years, an increasing number of light echoes (LEs) from low-/intermediate-mass young stellar objects (YSOs) are identified \citep{ortiz:2010,hodapp:2015,dahm:2017}. The first observed LE associated with a MYSO burst is that of S255IR-NIRS3 \citep{carattiDiskmediatedAccretionBurst2017,stecklum:2017}.
%
To assess 
the presence of a scattering LE of the G323 burst, optimal image subtraction (e.g., \citealp{alard:lupton:1998}) on the VVV(X) images was performed. For this purpose, the implementation using Interactive Data Language (IDL)\footnote{IDL® is a registered trademark of L3Harris Technologies, Inc.} was used \citep{miller:pennypacker:2008}. For each band, the pre-burst image with the smallest FWHM served as a reference. Spatial extinction variations across massive star-forming regions hamper LE detection in difference images. For this reason, ratio images were created by division with the reference image, which cancels these variations (assumed to be constant in time). This proved useful in revealing the LE of the S255IR-NIRS3 accretion burst \citep{carattiDiskmediatedAccretionBurst2017}.

\begin{figure*}
    \sidecaption
    \includegraphics[width=12.5cm]{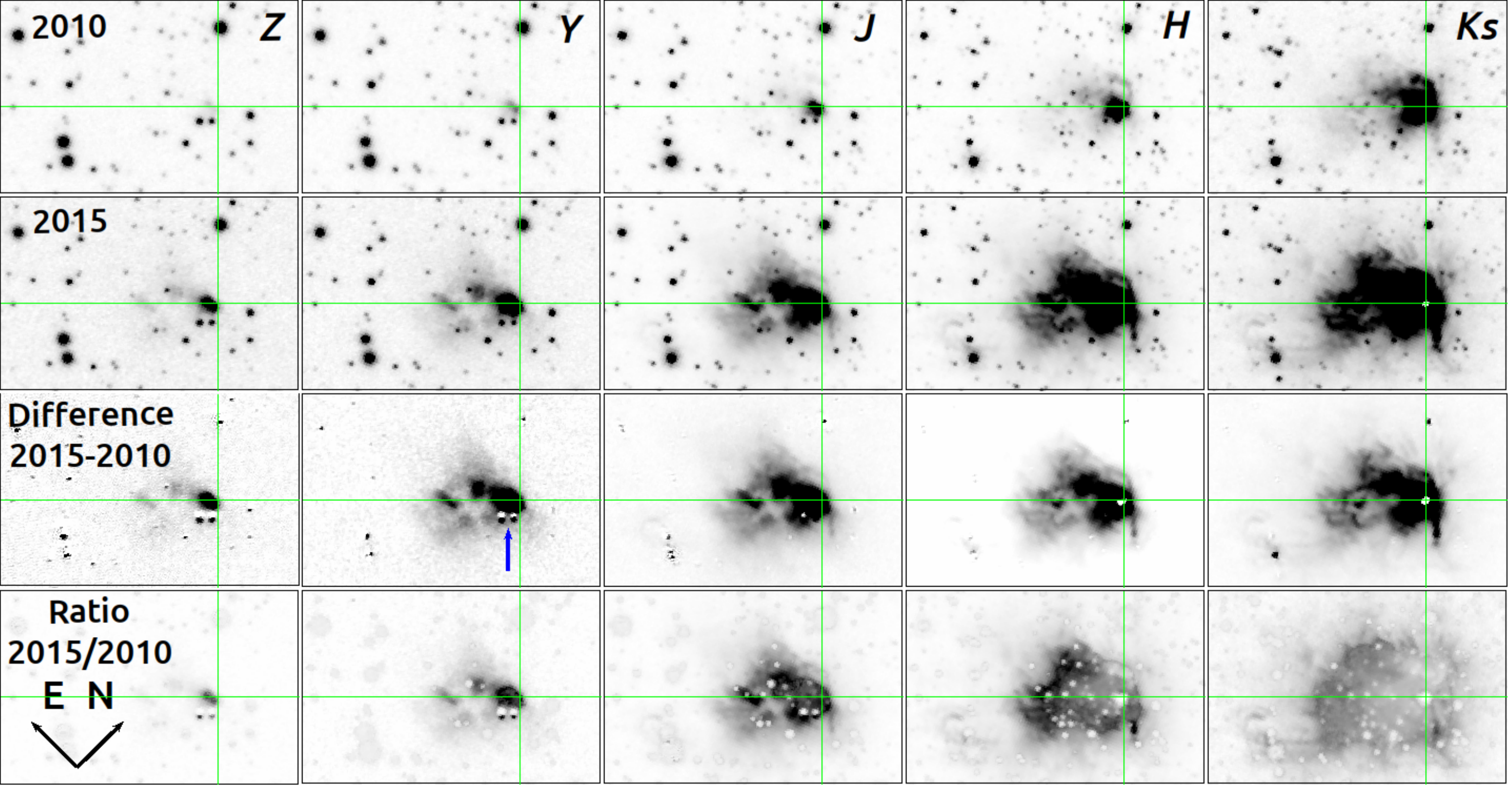}
\caption{Columns from left to right show $Z, Y, J, H$, and $K_{\rm s}$ images and rows from top to bottom show those of 2010, 2015, difference, and ratio (range 0$\dots$17.5). The crosshair marks the MYSO position. For the upper two rows, the values comprise 98 percentiles, displayed using a linear stretch.
At red optical wavelengths (two leftmost columns), the object appears bipolar. Both blobs are probably due to scattering, with the eastern one brighter in $Z$.
Both are offset from the nominal MYSO position and indicated by the red $Z$ contours in Fig.{\ts}\ref{fig:naco}. The prompt LE is visible in all five bands. Its large size in the $K_{\rm s}$ ratio image is due to a smaller number of scatterings compared to the other bands.
A common foreground proper motion binary (blue arrow) appears in $Z$ and $J$ next to G323. The FoV amounts to 70\arcsec$\times$45\arcsec. 
}
 \label{fig:10-15}
\end{figure*}

\begin{figure*}[t]
    \centering
    \includegraphics[width=\textwidth]{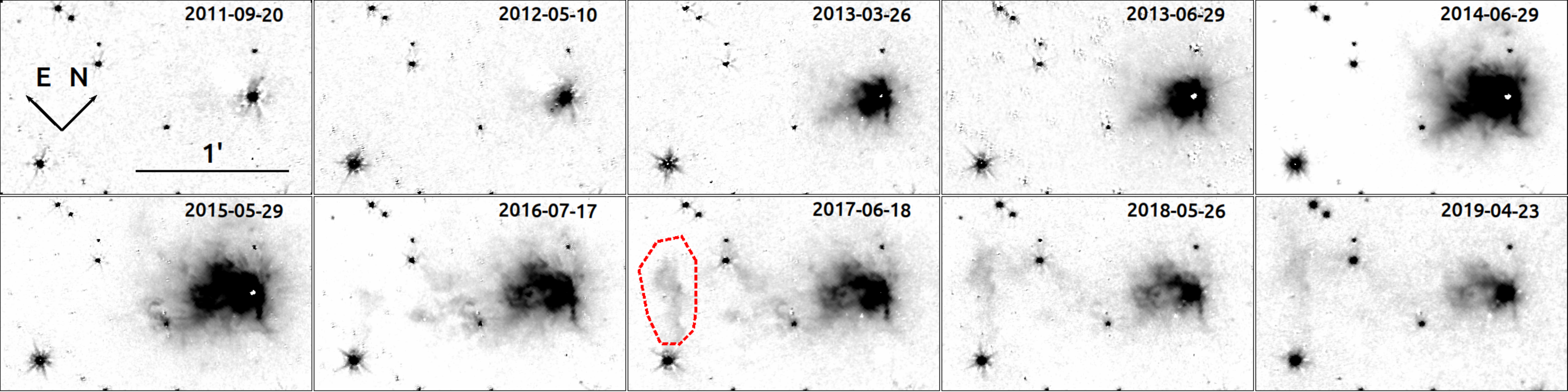}
\caption{The $K_{\rm s}$-band difference images show the temporal evolution of the burst and the associated LEs, displayed using a log scale. The prompt echo, which originates from the cloud core, is monopolar and spreads to the southeast at an PA of $\sim$ 135\degr. A remote LE that appeared later traces denser structures of the ISM. The red polygon encloses the area from which its light curves were established. The FoV is $135\arcsec\times80\arcsec.$ 
} 
\label{fig:seq}
\end{figure*}

An overview of the burst-induced change in the appearance of G323
is given by Fig.{\ts}\ref{fig:10-15}, showing the pre-burst, difference, and ratio images for the various VVV(X) filters. 
The difference and ratio images
provide clear evidence  of a light echo associated with G323. Another one 
1\farcm35 southeast of G323, which appeared later, can be seen in the sequence of $K_{\rm s}$ images (Fig.{\ts}\ref{fig:seq}). We denote those as `prompt' and `remote' LEs. In the $J$, $H$, and $K_{\rm s}$ ratio images of Fig.{\ts}\ref{fig:10-15}, the highest values are not located close to the source, but at a distance that increases with wavelength. 
This illustrates the echo propagation and the dependence of its velocity on the optical depth, which is lower for longer wavelengths (e.g., \citealp{draine:2003}).

 Light echoes (LEs) are also detected in the NEOWISE $W1$ and $W2$ images, which were retrieved using the ICORE tool \citep{masci:2013}.
This is not self-evident, as the scattering cross-section of interstellar dust grains at longer wavelengths is smaller (e.g., \citealp{draine:2003}). However, the lower scattering efficiency is more than compensated by higher flux densities
in the NEOWISE bands compared to $K_{\rm s}$ (cf. Fig. \ref{fig: G323 pool}).
The features of the LEs and their evolution seen by NEOWISE correspond to what is found in the $K_{\rm s}$-band. 


Near the source, the spatial brightness distribution of the LE echo represents a record of the burst history. Thus, the maximum of the ratio values provides an estimate of the maximum burst strength. This holds for the $J$- and $H$-bands, where the maximum values at various locations are around 20. 
Further away, the echo strength decreases as a result of spatial propagation and superposition along the line of sight. This is already the case for the $K_{\rm s}$-band.
Clearly, the LE is not circular-symmetric
but extends southeast and is missing in the opposite direction. This indicates a nonuniform dust distribution in the environment of G323. The absence of an LE in the northwest
could be related to the curved rim-shaped structure in this area (cf. Fig. \ref{fig:burst}) that could block or shadow it.

The temporal evolution of the source appearance during the burst can be seen in the sequence of $K_{\rm s}$-band frames (Fig.{\ts}\ref{fig:seq}), taken from 2010 to 2019. For illustrative purposes, images were selected with a time difference of about one year, except for a half year during the burst rise. Although this is only a subset of all $K_{\rm s}$-band images, it nonetheless provides the most important information. To emphasize structural changes, the difference images are shown in Fig.{\ts}\ref{fig:seq}. They can be traced on the ratio images as well, albeit at a lower dynamic range. Early signs of the burst are visible in 2012 which reached its full swing in the following year.
The larger number of saturated pixels at the MYSO position indicates an increase in brightness. One year later (2014), the propagation of the prompt LE 
becomes obvious. An arc-like feature becomes visible that winds toward the east, as well as
a straight one that is oriented toward the south. Until the next epoch (2015), the eastern arc-like feature vanished, and the straight one became arcuate with a strong turn. Over time, the burst light propagated into the interstellar medium, leading to the appearance of a remote LE in 2016, featuring variable illumination. The last frames of Fig.{\ts}\ref{fig:seq} also show the general fading of the source and its environment. Finally, an elliptical structure with the MYSO at its western apex is worth mentioning, as it is visible from 2015 to 2019. Due to its pertinent morphology, it has to be stationary. Perhaps this could be light scattered from the wall of an outflow cavity.

\begin{figure}[ht]
\centering
 \includegraphics[width=\hsize]{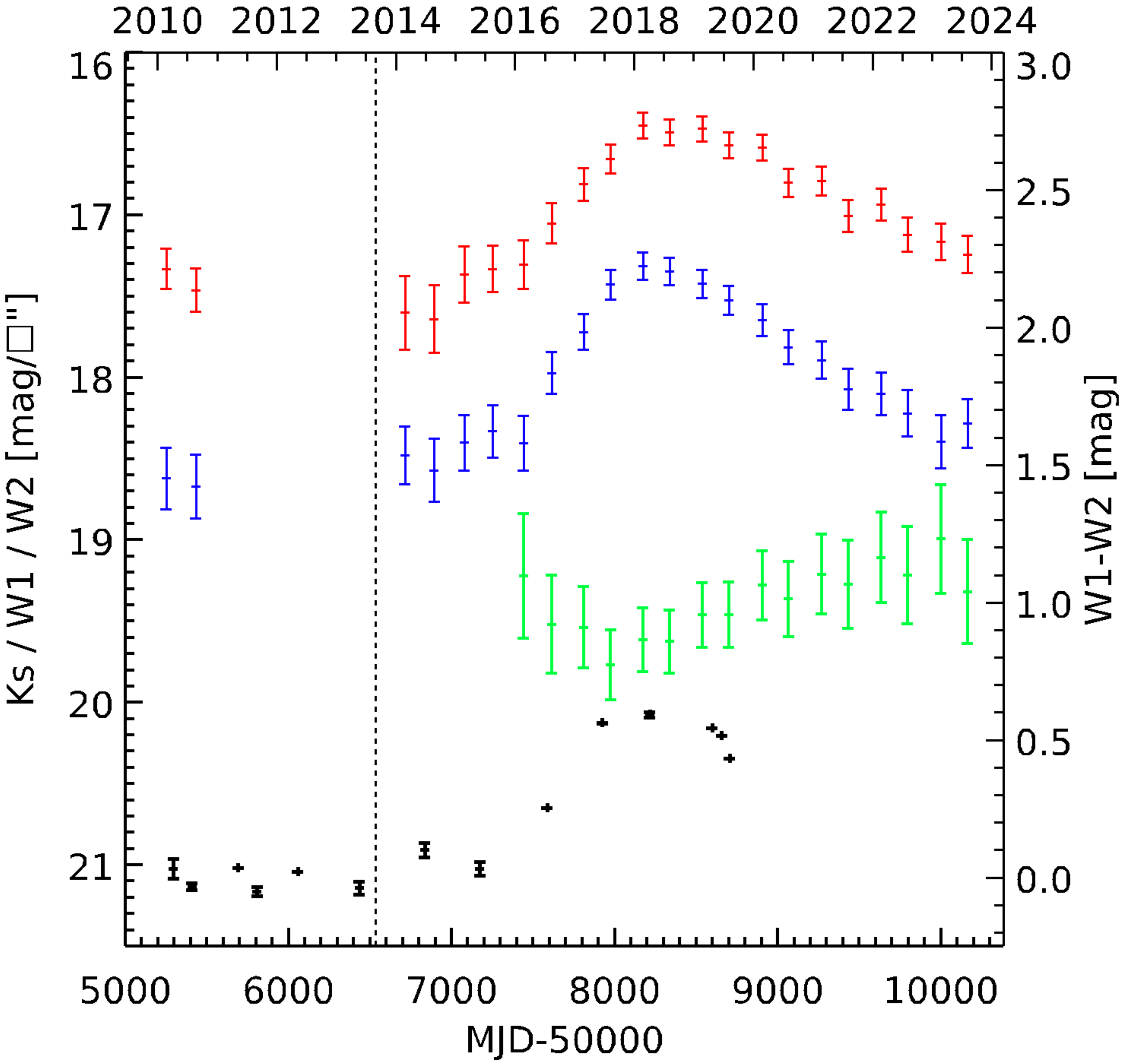} 
 \caption{  Light curves of the remote LE where colored error bars denote the following: $K_{\rm s}$ (black), $W1$ (blue), $W2$ (red), and $W1-W2$ color (green, right ordinate). The vertical dashed line marks the date of the burst peak. The LE peak occurs slightly more than four years after the burst peak.} 
 \label{fig:LE_lc}
\end{figure}

The remote LE is still faintly visible in NEOWISE images taken in 
2023
(cf. Fig. \ref{fig:W1_LE}).
Its light curves (Fig. \ref{fig:LE_lc}) were established by integrating the fluxes in the $K_{\rm s}$, $W1$, and $W2$ bands over this region, marked in Fig. \ref{fig:seq}. The subtracted background was derived from an area northwest of the MYSO, devoid of stars. Notably, these light curves cover the burst rise which is missing in the on-source data (cf. Fig.\,\ref{fig:lc}) because of the time gap between the $WISE$ and NEOWISE missions.
The primary characteristics that can be drawn from them
are the following. The LE peak occurs slightly more than four years after the burst peak. Its amplitude is shallower than that of the burst, which is due to echo propagation. The $W1-W2$ color change confirms the `bluening' during the burst. Due to the wavelength dependence of the scattering cross section, the pre-burst colors of the LE are bluer than those of the source, as indicated by $K_s-W1=2.44\pm0.20$ vs. $3.20\pm$0.12 and $W1-W2=1.25\pm0.23$ vs. $1.56\pm0.16$.

For common cases of LEs, as for example those associated with novae, supernovae, and
variable stars (\citealp{sugerman:2007,jang:dou:2016,kervella:2008}), a superluminal motion of the echo is present. 
It is a pure geometrical effect that follows from first principles in the case of single scattering \citep{nemiroff:2015}.
Whether a prompt YSO LE is superluminal is questionable, since its environment is quite dusty.
Thus, multiple scattering may overwhelm the geometric effect on which the superluminal motion rests.
For G323, the apparent propagation velocity can be derived from the projected separation between the echo boundary and the MYSO, as well as the time difference between the date of the burst peak and the epoch of observation. For the $K_{\rm s}$-band, the respective quantities are $\sim${\ts}22\arcsec{} and one year. Together with a distance of 4.08{\ts}kpc, this yields 1.4{\ts}$c$. Although this confirms the superluminal motion in the $K_{\rm s}$-band, it implies that the apparent velocity 
is smaller for $H$ and $J$. Given the small extent of the $J$ echo, its apparent velocity 
is probably close to subluminal.

In the optically thick regions, the photons follow random walk paths, where the effective travel distance is given by the diffusion path length.
For random walk processes, the diffusion path length $l_{diff}$ is $\propto l_{free} \cdot \sqrt{n_{scat}}$ (assuming isotropic scattering), where $l_{free}$ is the mean free path and $n_{scat}$ is the number of scattering events. 
Within a fixed time $t$ the photons can travel a path with a "real" length of $l_{real} = n_{scat} \cdot l_{free} \cdot c$, with c being the speed of light.
The "real" path is the same for all wavelengths (it depends only on the travel time and c), but the interaction free path length (and hence the number of scattering events) depends on the optical depth (and hence the wavelength).
When the equations are combined, the diffusion length at a given time and wavelength is $l_{diff}(\lambda, t)\propto l_{real}(t) \cdot (n_{scat, \lambda} )^{-1/2}$.
This implies a difference in the diffusion path length (and hence the apparent LE extent) for different wavelengths. 
With $n_{scat}\propto \tau^2$ (see e.g., \citealp{2021A&A...645A.143K}), 
the expected extent of the LE in $K_{\rm s}$ compared to that seen in Z can be written as the inverse ratio of the optical depths $l_{diff, Ks}/ l_{diff, Z} \sim \tau_{Z}/ \tau_{Ks}$.
For MRN dust \citep{Mathis1977}, the optical depth ratio between the Z and $K_{\rm s}$ filters is $\tau_Z/\tau_{Ks}\sim 3.8$. 
This would imply that LE propagation is almost a factor four slower in Z than in $K_{\rm s}$. 
However, 
the density within the cloud core is highly variable, and a measurable slowdown is expected only in the densest regions (close to the protostar) and therefore the effect will be smaller. 

\subsection{IR-maser correlation}\label{sec: K-IR}
Class II methanol masers are pumped by MIR radiation (e.g., \citealp{1997A&A...324..211S}, \citealp{Cragg:2005}). 
Therefore, variations in the pumping rate will change the maser flux.
The recent accretion bursts from MYSOs were all accompanied by flares of those masers (e.g., \citealp{sugiyama:2015}, \citealp{hunterExtraordinaryOutburstMassive2017}, \citealp{2018A&A...617A..80S}, \citealp{2019ATel12446....1S}), which confirmed their radiative excitation.
The correlation between mid-IR and maser radiation exists also 
for sources, that show periodic simultaneous maser and IR flares, such as G107.298 + 5.639 \citep{stecklum:2018}, G36.705\,+0.096 \citep{stecklum:2022}, as well as G45.804\,-0.356, and G49.043\,-1.079 \citep{olech:2022}.

For the G323 burst, there is also a clear correlation between the maser flux and the $K_{\rm s}$-band flux (cf. Fig. 
\ref{fig:lc}
). We note that although NIR $K_{\rm s}$-band photons do not pump the masers, the long-term variation of the integrated maser flux closely follows the $K_{\rm s}$ light curve, simply because the NIR and MIR maser-exciting radiation vary in the same fashion. A log-log regression yields a power-law exponent of $1.4\pm0.12$.
Thus, the G323 burst strengthens the evidence that Class II methanol maser flares are a reliable tracer of episodic accretion variability of young massive stars.

An attempt to detect the maser periodicity in the VVV(X) photometry failed because the cadence of the NIR survey was insufficient. 
While (NEO)WISE is able to trace intraday variability during a visit for sources that are
not too bright, the large photometric errors for saturated targets such as G323 precludes concluding whether its brightness varies in sync with the maser.





\subsection{HAWC+ FIR photometry}\label{sec: HAWC}
As stated above, the main objective of the SOFIA/HAWC+ observations is to constrain the burst energy by assessing the strength of a possible FIR afterglow due to the burst.
This was done by comparing the pre-burst fluxes from the HPPSC with those measured with HAWC+. 
Although the field sizes of the 53, 62, and 89{\ts}$\mu$m bands do not cover additional HPPSC sources, the larger fields of the 154 and 214{\ts}$\mu$m bands comprise one or more. The detection of the source HPPSC160A\_J152920.2-562522 in the 154{\ts}$\mu$m image provides the opportunity to check the flux calibration. PSF photometry of this object revealed that its HAWC+ flux exceeds the PACS flux by (15$\pm$7)\%. Consequently, this was taken into account in the analysis. Furthermore, the HAWC+ 62 and 154{\ts}$\mu$m fluxes were interpolated to the PACS wavelengths using a polynomial approximation of the post-burst FIR SED established from all HAWC+ bands.

\begin{figure}
    \includegraphics[width=\hsize]{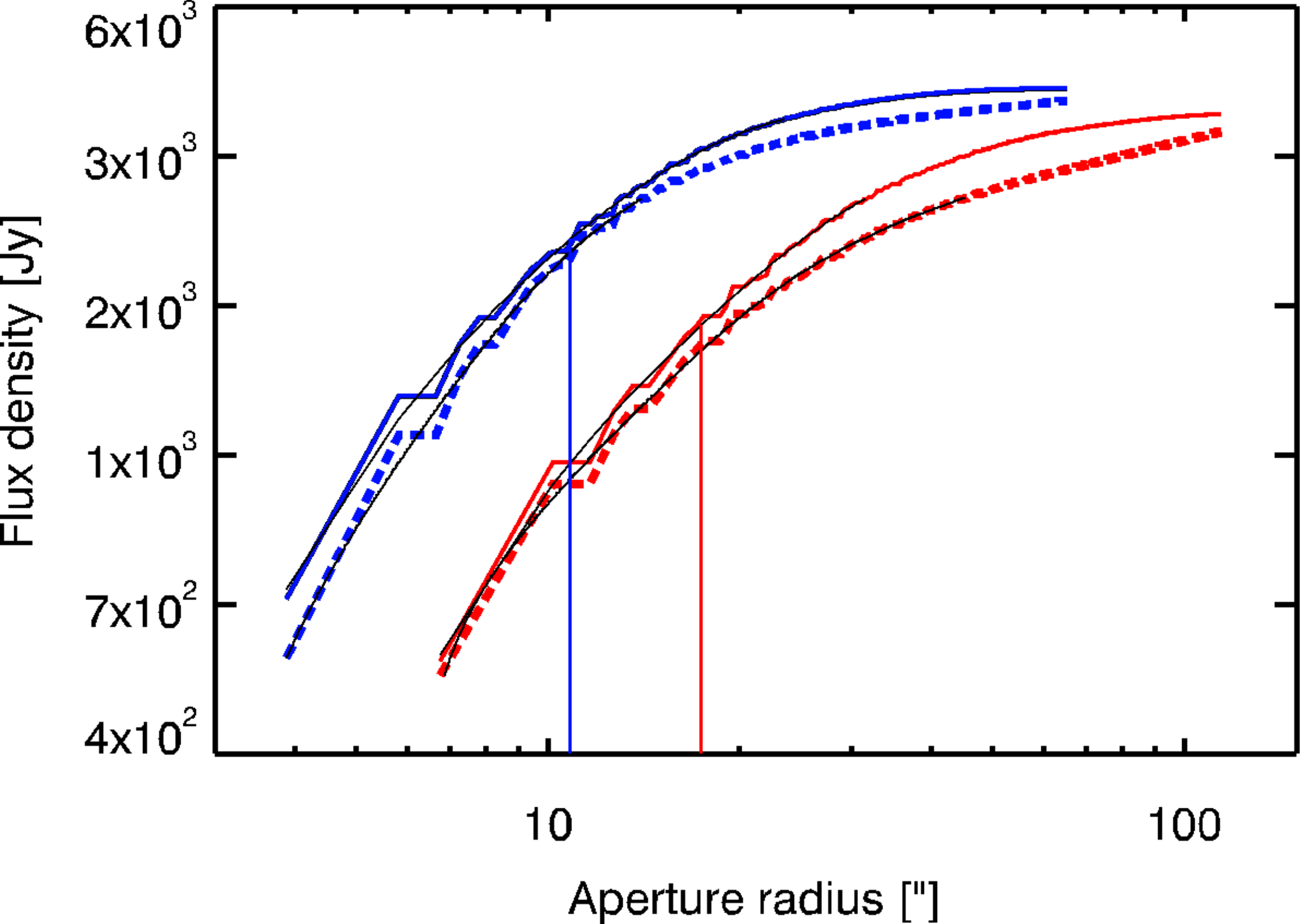}
\caption{Flux growth curves of PACS (dashed) and HAWC+ (solid). The wavelength bands are indicated by color, where blue is $62/70$ and red $154/160\, \rm \mu m$ for HAWC+/PACS, respectively. The vertical lines mark the aperture radii, which reproduce the HPPSC fluxes. Black lines are polynomial approximations, reducing the influence of finite pixel size. The mean of the growth curve ratios gives the increase at the HAWC+ epoch.
}
 \label{fig:ha_pa}
\end{figure}

The flux growth curves with increasing aperture size are shown in Fig. \ref{fig:ha_pa} for the PACS 70 and 160{\ts}$\mu$m as well as for the HAWC+ counterparts.
Red and blue mark the wavelengths, while dashed and solid lines indicate PACS and HAWC+ fluxes. The vertical lines point to aperture radii, which reproduce the HPPSC fluxes.
The flux ratios HAWC+ vs. PACS for the blue and red bands
were derived by dividing the values of the respective growth curves and taking their average. They amount to $1.142\pm0.046$ and $1.085\pm0.061$.  
The errors take into account the scatter of the growth-curve ratios and the relative uncertainties of the PACS fluxes as well as those of the HAWC+ calibration. 
The ratio values agree within their errors, which a posteriori confirms the 154{\ts}$\mu$m flux correction. 
They 
suggest the presence of a weak
FIR flux excess, which stems from the burst afterglow.


The HAWC+ fluxes were derived for all bands using aperture sizes tied to those obtained as outlined above using a linear relation for the wavelength dependence.  The formal uncertainty obtained from the error images provided by the processing pipeline (DRP version 3.0.0) is in the range of a few Jansky only.
The uncertainty of the flux calibration can be assessed from the residuals of a low-order polynomial fit to the observed fluxes, since the dust-continuum SED of the YSOs is continuous and featureless in the FIR (cf. \citealp{Stecklum:2021}). This yields a relative error of 10\%.

\begin{figure}
    \includegraphics[width=\hsize]{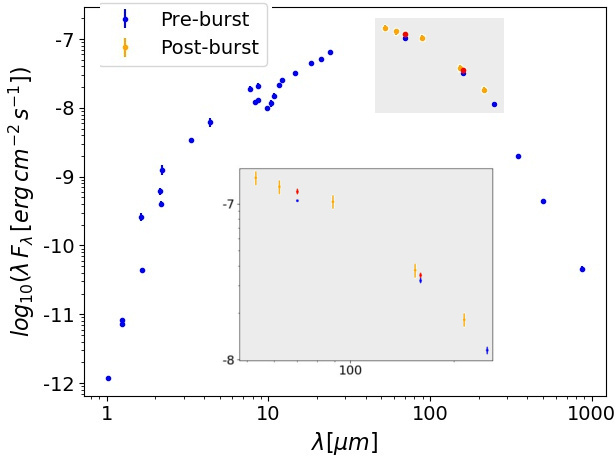}
\caption{Pre-burst SED (blue), together with the HAWC+ post-burst observations (orange). The HAWC+ observations were interpolated to match the wavelengths of the pre-burst observation. The resulting data points are colored red. The inset shows a zoom-in on the region of interest. The flux excess in the post-burst epoch is small (only $\sim 10\%$ at $70, 160 \, \rm \mu m$). 
}
 \label{fig: all obs SEDs}
\end{figure}

The HAWC+ photometry is given in Table{\ts}\ref{tab: Hflux} along with beam sizes, as well as image and deconvolved FWHMs. The FWHMs of the image profile were derived by fitting an ellipse to the cut at 50\% peak level. The uncertainty of the length of the axis ranges from 0\farcs5 to 1\farcs5, while that of the position angles is $\sim$13\degr. The deconvolved sizes are based on the geometric mean of the major and minor axes, from which the beam size was subtracted in quadrature.
The post-burst SED is plotted in Fig. \ref{fig: all obs SEDs} in orange (HAWC+ filters)
, with the interpolation to the PACS wavelengths in red. The pre-burst SED is shown in blue for comparison. The inset shows a zoom-in on the region of interest. 

\subsection{AO and thermal IR imagery}
\begin{figure}
    \includegraphics[width=\hsize]{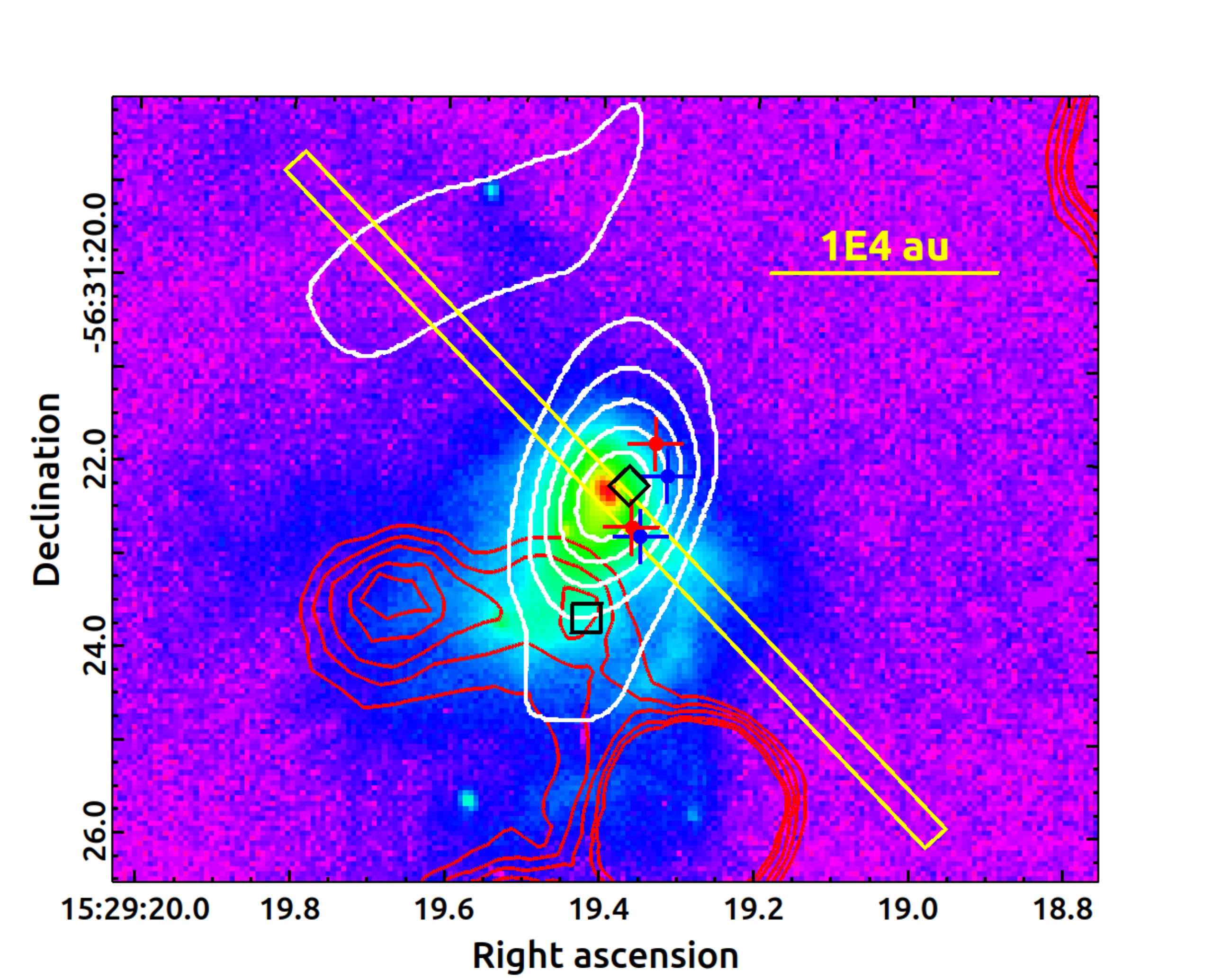}
    \caption{NACO AO $K_{\rm s}$-band image (epoch 2009) using a logarithmic stretch with contours of the 19{\ts}GHz radio continuum (white, \citealp{murphy:2010}) and from the $Z$-band pre-burst image (red, epoch 2010). 
    Maser spots are marked by crosses in blue (\citealp{Caswell:2001}, epoch 1994) and red (\citealp{greenExcitedstateHydroxylMaser2015}, epoch 2011), with sizes indicating the position error. The black diamond is at the peak of the 1.4{\ts}mm emission while the black square marks the {\it GAIA} source. The yellow rectangle shows orientation and width of the ISAAC slit.
}
 \label{fig:naco}
\end{figure}

The pre-burst NACO $K_{\rm s}$-band image is shown in Fig.{\ts}\ref{fig:naco}. A logarithmic scale is used to display the lower surface brightness. 
The black diamond marks the peak position of 1.4{\ts}mm emission, based on the ALMAGAL data obtained with the ALMA 12-m array. This probably corresponds to the location of the MYSO. 
As mentioned in the caption of Fig.{\ts}\ref{fig:10-15}, the contours of the emission of the $Z$ band are superimposed in red. The two scattering blobs are located southeast of the $K_{\rm s}$-band peak. The western one coincides with the position of the GAIA source. The white contours delineate the 19{\ts}GHz radio continuum \citep{murphy:2010}, which is centered on the MYSO. The remaining symbols, filled circles with error bars, mark the positions of 6.7{\ts}GHz masers (blue - \citealp{caswellCoincidenceMaserEmission1997a}, red - \citealp{greenExcitedstateHydroxylMaser2015}), as well as the width and orientation of the ISAAC slit (yellow).
The tightly winded circular $Z$-band contours at the bottom are due to a foreground star that is barely visible at their center.

The field stars in the NACO AO image have FWHMs of the order of 0\farcs11-0\farcs16. Therefore, the image can be well compared with the available radio data of similar resolution and position accuracy. It provides information on linear spatial scales of about 500{\ts}au.
The elliptical image core is marginally resolved. The FWHMs obtained by a bivariate Gaussian fit are 0\farcs254 and 0\farcs177, and the major axis is at a position angle (PA) of 28\fdg5. This corresponds to the beam-deconvolved linear sizes of 890{\ts}au and 490{\ts}au, respectively. The ISAAC slit (PA=43\fdg8) applied for K band spectroscopy is almost aligned with the major axis.
Both PAs are not too far from those of the HAWC+ images listed in Tab{\ts}\ref{tab: Hflux}.

\subsection{K-band spectroscopy}\label{sec:Kspec}
\begin{figure}
    \centering
    \includegraphics[width=\hsize]{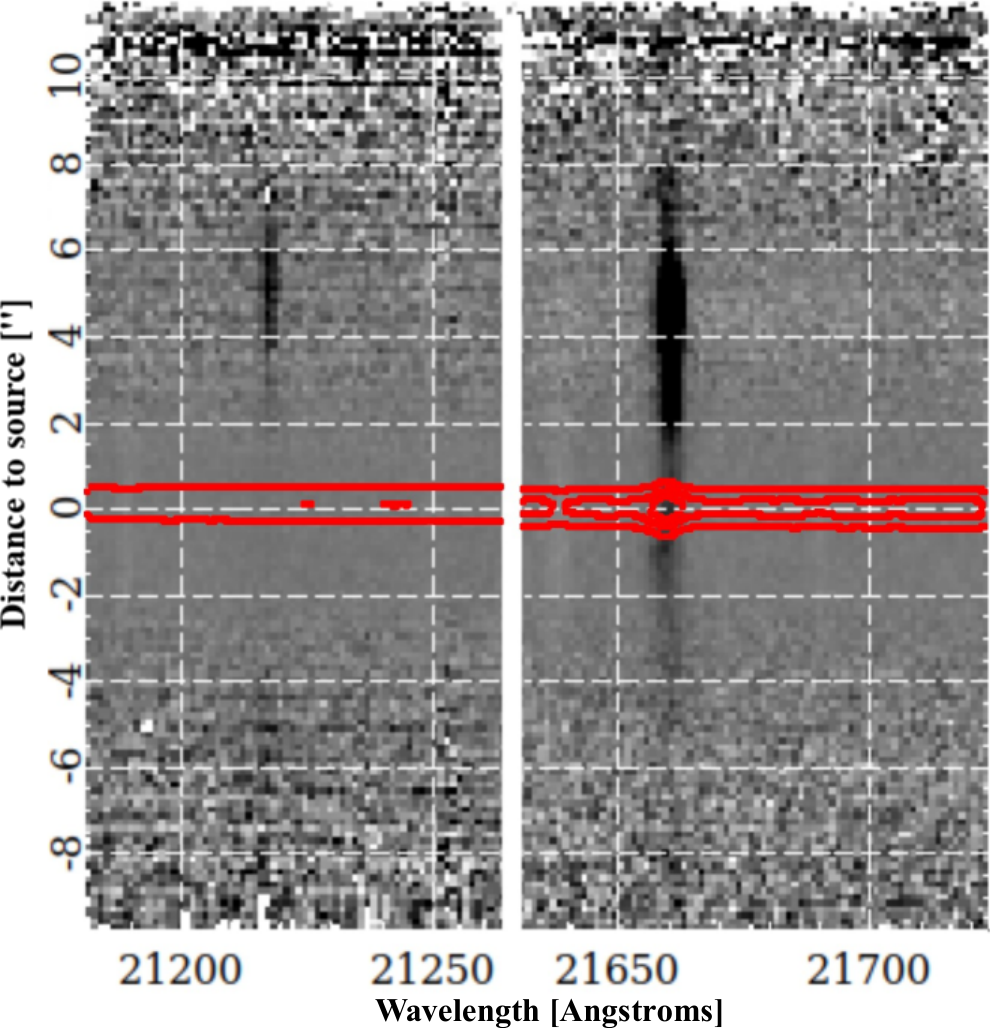}
	\caption{Zoom-in  sections of the continuum-subtracted spectral image
	showing 2.122{\ts}$\mu$m H$_2$ (left) and 2.166{\ts}$\mu$m Br$\gamma$ (right), both featuring extended emission. Red contours show the source continuum emission 
	\mbox{$(1 - 3)\times$10$^{-14}${\ts}erg{\ts}s$^{-1}${\ts}cm$^{-2}${\ts}\r{A}$^{-1}$}. NE is up.
	}
 \label{fig:specima}
\end{figure}

The spectral image shows a rising continuum and two bright emission lines, namely the \ion{H}{i} Br$\gamma$ line at 2.166{\ts}$\mu$m and the H$_2$ 1--0{\ts}S(1) line at 2.122{\ts}$\mu$m. A zoom-in on the continuum-subtracted spectral image is given in Fig. \ref{fig:specima}. There may be a third line, well below the 3{\ts}$\sigma$ detection, centered around $\sim$2.206{\ts}$\mu$m. If real, this line would be one of the two \ion{Na}{i} doublet lines often detected in the $K$-band~\citep{carattiDiskmediatedAccretionBurst2017}, the second one (at 2.208{\ts}$\mu$m) falls beyond our spectral coverage. In fact, the spectral image covers a small portion of the $K$ band. Therefore, other important features, such as the CO band heads (between 2.28 and 2.5{\ts}$\mu$m) that trace disk-mediated accretion are not included. 

Both the continuum and detected lines extend a few arc seconds farther away from the point source position and are probably scattered along the outflow cavities. While the Br$\gamma$ line extends toward NE and SW along the slit, the H$_2$ emission is detected toward NE and is not seen at the source or toward SW (see Figure~\ref{fig:specima}). There may be a faint H$_2$ emission toward the SW, but it is well below the 3{\ts}$\sigma$ detection limit.

Table~\ref{tab: lines} lists line fluxes, full width at half maximum ($FWHM$), peak radial velocities ($v_{\rm{r}}$) and full width at zero intensity ($FWZI$) of Br$\gamma$ and H$_2$ along the three different regions extracted from the spectral image, namely at the source, toward the NE and SW of the source. The H$_2$ emission, only detected in the NE region, is spectrally unresolved ($FWHM\sim$2.4{\ts}$\text{\r{A}}$ or 34{\ts}km{\ts}s$^{-1}$) and its maximum radial velocity is around 0{\ts}km{\ts}s$^{-1}$ or slightly redshifted (4$\pm$4{\ts}km{\ts}s$^{-1}$). The H$_2$ kinematics points to fluorescent emission from the HCH{\sc ii} region around the source. On the other hand, the kinematics of the Br$\gamma$ line is more complex. The line is spectrally resolved in the three extracted regions. It has a radial velocity close to 0{\ts}km{\ts}s$^{-1}$ on source, whereas the SW region is slightly blueshifted (-14$\pm$5{\ts}km{\ts}s$^{-1}$) whereas the NE region has three different velocity components at -28, 2, and 32{\ts}km{\ts}s$^{-1}$. Furthermore, the $FWZI$ values of the line in the three regions range from 150 to 170{\ts}km{\ts}s$^{-1}$. 
Spectroscopy was performed during the light-curve rise, close to the NIR peak of the outburst. At that time, no P-Cygni profile was present in the Br$\gamma$ line. This feature is commonly attributed to an emerging ionized wind.
Its absence in the G323 spectrum may indicate that the launching of the wind is not a prompt process, but delayed with respect to the temporal evolution of the burst. This was observed for the S255IR-NIRS3 outburst \citep{2018A&A...612A.103C}. 


\begin{table}
\caption{Fluxes and kinematics of the Br$\gamma$ and H$_2$ (2.12{\ts}$\mu$m) lines.} 
\begin{tabular}{|c|c|c|c|c|}
\hline
Quantity & On source & \multicolumn{2}{c|} {NE region} & SW region \\
\hline
 & Br$\gamma$ & Br$\gamma$  & H$_2$& Br$\gamma$ \\ \hline
$F\pm\Delta F$ & 61$\pm$1 & 20$\pm$1 & 0.43$\pm$0.04 & 6.5$\pm$0.2\\
$[\rm FU]$&&&&\\ \hline
 $v_{\rm{r}}\pm\Delta v_{\rm{r}}$  & -3$\pm$4 & -28$\pm$5 & 4$\pm$4 & -14$\pm$5\\ \
$[\rm km{\ts}s^{-1}]$ &&2$\pm$5&&\\
&&32$\pm$5&&\\ \hline
$FWHM$ &58$\pm$5&73$\pm$5&34$\pm$5&74$\pm$5\\
$[\rm km{\ts} s^{-1}]$&&&&\\ \hline
$FWZI$ &150$\pm$5&170$\pm$5&99$\pm$10&170$\pm$5 \\
$[\rm km{\ts} s^{-1}]$&&&&\\ \hline
\end{tabular}
\begin{tablenotes}\footnotesize
\item[] FU denotes the flux unit, which is in $10^{-14}{\ts}\rm erg{\ts}cm^{-2}{\ts}s^{-1}$
\end{tablenotes}
\label{tab: lines}
\end{table}


\subsection{The ALMA view of G323}

\begin{figure}[ht]
    \centering
    \includegraphics[width=\hsize]{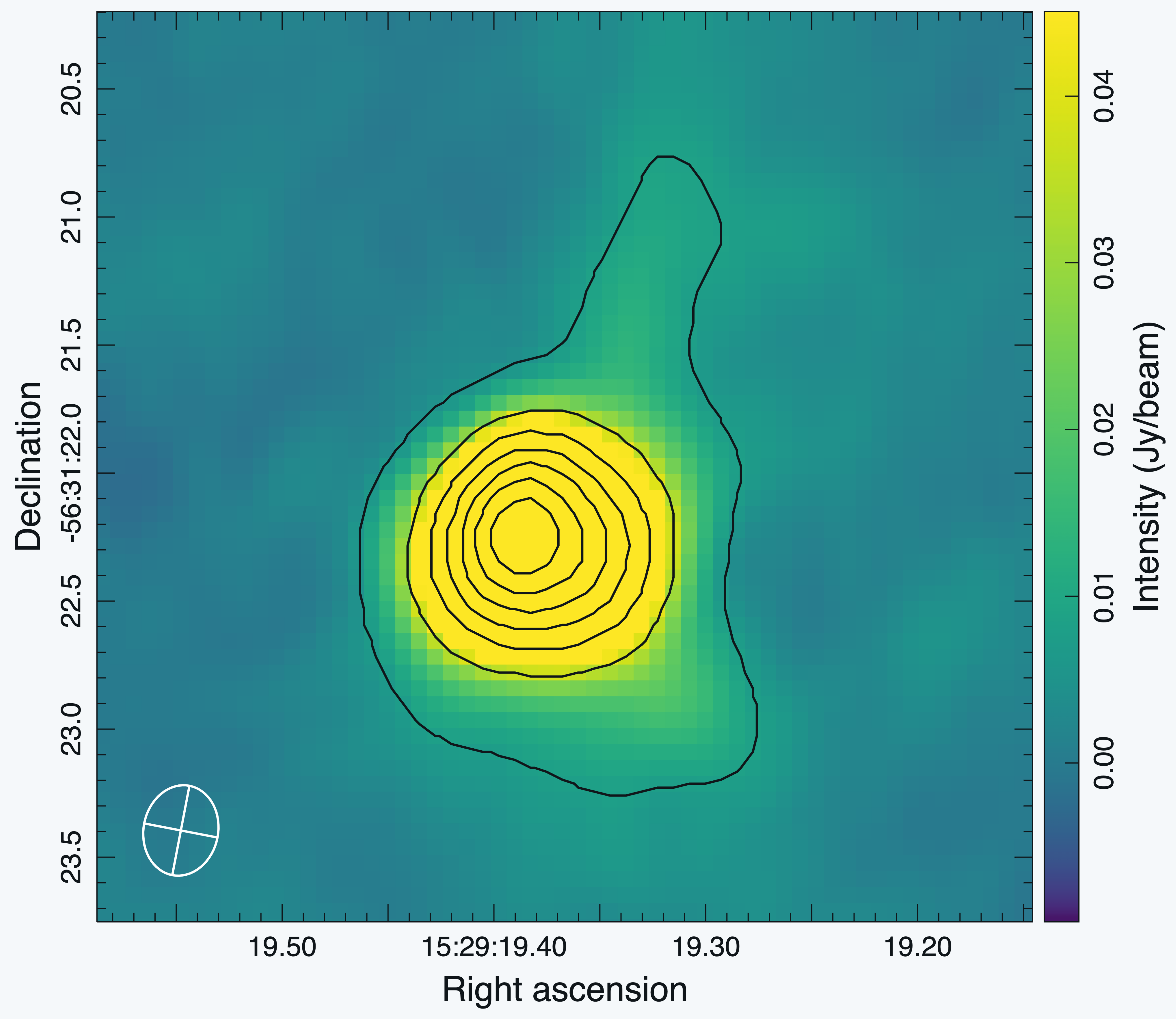} 
	\caption{
    1.4{\ts}mm dust continuum map with superimposed contours. The ellipse on the lower left indicates the beam size. The object is resolved and shows faint extended emission.
	}
 \label{fig:ALMA0}
 \end{figure}

The ALMA continuum and line data are a crucial supplement for the IR observations to complete the characterization of the MYSO. We emphasize that the data were taken at the end of 2019, less than one year before the burst end. %
Fig.{\ts}\ref{fig:ALMA0} shows the 1.4{\ts}mm dust continuum map taken with an 80th percent baseline length (L80BL) of 511\,m which indicates that the source is resolved by a few beams. The contours start at 5{\ts}$\sigma$ ($\sigma = 1.6${\ts}mJy/beam) and increase in steps of 20{\ts}$\sigma$. At the lowest level, it is not 
circular but shows extended emission. 


\begin{figure}[ht]
    \centering
    \includegraphics[width=\hsize]{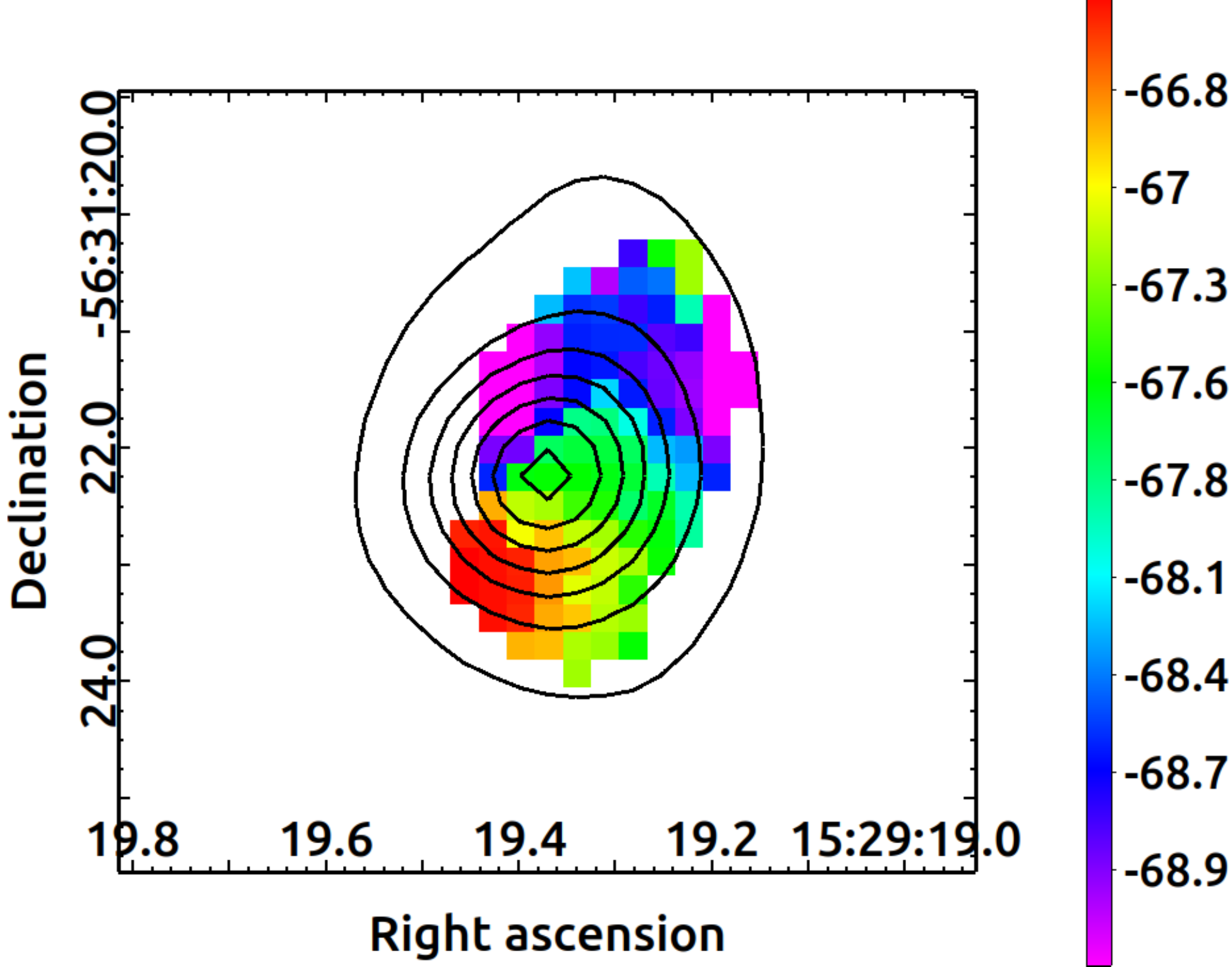}
	\caption{
    First moment of the SiO emission with superimposed dust continuum contours from the ALMA 12-m array. The color bar indicates the velocity interval, in units of $\rm km\,s^{-1}$. There is a  slight gradient from the SE to the NW.
	}
\label{fig:ALMA1}
\end{figure}

Fig.{\ts}\ref{fig:ALMA1} displays the first moment of SiO line emission along with the contours of the dust continuum at a coarser resolution, based on observations with an L80BL value of 126\,m. The line emission is centered on the millimeter source and shows a velocity gradient from the SE to the NW. This indicates that the gas to the SE is redshifted. The velocity interval is rather small for the source being seen face-on, which argues against an origin of the emission in an outflow.
There is no sign of Keplerian rotation for which the largest velocity differences would be at the center. The coarse resolution of the ALMA 12-m array is not sufficient to resolve a possible disk.

\begin{figure}[ht]
    \centering
    \includegraphics[width=\hsize]{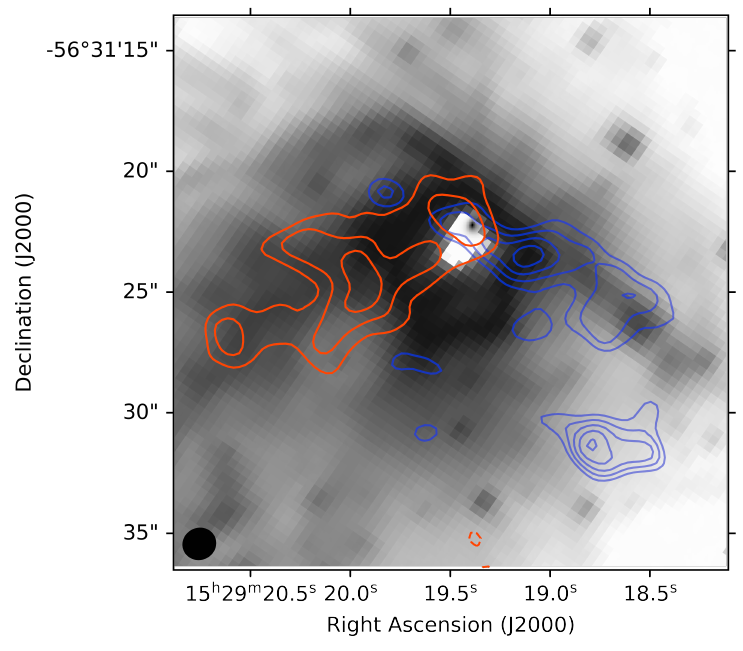}
	\caption{Distribution of the redshifted and blueshifted $^{13}$CO{\ts}(2-1) line emission as mapped with the ALMA 12-m Array superimposed on a $K_{\rm s}$  image taken in  mid-2015. The synthesized ALMA beam size for the line observations is shown in the lower left. The compact 1.4{\ts}mm emission, mapped at much higher resolution, appears as a small blob within the white saturated pixels.
	}
 \label{fig:ALMA2}
\end{figure}

The contours of the intensity map of the blueshifted and redshifted $^{13}$CO(2–1) emission 
are shown in Fig.{\ts}\ref{fig:ALMA2}, superimposed on a $K_{\rm s}$ image taken in mid-2015 along with the 1.4{\ts}mm continuum emission. The latter appears as a small blob in the white area of the saturated central region. The synthesized beam size ($1\farcs3\times1\farcs2$) of the  $^{13}$CO(2–1) observations is shown on the bottom left. The line emission was integrated in the LSR velocity ranges from 7 to 18{\ts}km{\ts}$\rm s^{-1}$ for the redshifted gas and from -7 to -15{\ts}km{\ts}$\rm s^{-1}$ for the blueshifted gas. 
The contours begin at 5$\sigma$ and increase in steps of 2$\sigma$ where $\sigma$ is 0.18{\ts}$\rm Jy/beam\, km\, s^{-1}$ and 0.13 $\rm Jy/beam\, km\, s^{-1}$ for the redshifted and blueshifted lobes, respectively.
Remarkably, the blueshifted gas appears to coincide with the dusty ridge west of the MYSO while the redshifted emission toward the east is co-spatial with the NIR nebulosity. This is not typical for outflow cavities, which are usually devoid of dust and scatter from their walls. If the CO emission traces an outflow of a source seen almost pole-on, its bipolar lobes should overlap more strongly than is the case for G323. Moreover, considerable bending seems to be present. Although this challenges the outflow interpretation and possibly argues for bulk-gas motion, a deflection of the blueshifted lobe at the western dust rim cannot be ruled out. Unfortunately, the ALMA 1.4{\ts}mm map is not sufficiently sensitive to shed light on the dust distribution at larger scales.
Inspection of the SPITZER/IRAC images does not provide evidence for a bipolar outflow as well. Unlike other MYSOs, for which outflows were found (e.g., \citealp{Caratti:2010}), the IRAC2 image, which covers a wealth of shock-excited H$_2$ lines (e.g., \citealp{Noriega-Crespo:2004}), does not show typical bow-like emission features on either side of the source.



\section{Radiative transfer modeling}
\label{RTM}
\subsection{The static pre-burst model}\label{sec: premod}


\begin{table}
\caption{Adapted parameter spaces and sampling for the TORUS pre-burst models. }
\resizebox{0.48\textwidth}{!}{
\setlength\extrarowheight{2.5pt}
\begin{tabular}{|c|c|c|c|}
\hline
Parameter & Samp. & Adapted range & Mean \\
\hline
$L_* [\rm L_\odot]$&log&$8 \times 10^3\ts -\ts 1.6 \times 10^5$& $(6.1\mypm^{4.2}_{2.5}) \times 10^{4}$\\  \hline
$M_{disk} \rm [M_\odot]$& log&$4 \times 10^{-8} \ts -\ts  2.5$&$(1.70\mypm_{1.67}^{1.07}) \times 10^{-3}$\\
$r_{disk} [\rm au]$&log&$80 \ts -\ts  7\,000$&$680 \mypm ^{1\,600}_{470}$\\
$\beta_{disk}$ &lin& $1 \ts -\ts  1.3$& $1.12 \mypm0.08$ \\
$\alpha_{disk}$ &lin& $1 \ts -\ts 3.3$& $2.2 \mypm 0.6$ \\
$h_{disk} [\rm au]$ &log&$0.5 \ts -\ts 33$ &$3.1 \mypm ^{5.9}_{2.0}$ \\ \hline 
$\dot{M}_{env} \rm [M_\odot\, \rm yr^{-1}]$&log&$10^{-5}\ts -\ts  1$& $(3.2 \mypm^{3.8}_{1.7}) \times 10^{-2}$\\ 
$r_{env} [\rm au]$&log&$(1\ts -\ts  4) \times 10^{4}$& $(2.4 \mypm^{1.2}_{0.8})\times 10^4$  \\ \hline 
$\rho_{cav} [\rm g\,cm^{-3}]$&log&$10^{-22} \ts -\ts  10^{-17}$& $(4.5\mypm _{1.5}^{2.3})\times 10^{-20}$\\
$\Theta_{cav} [\degr]$ &lin&$10 \ts -\ts  60$ & $42 \mypm 11 $\\ \hline  
$i [\degr]$&lin&$0 \ts -\ts  60$&$26 \mypm 17$\\
\hline
\end{tabular}
}
\tablefoot{
All densities/masses are total values (dust+gas), where we assume a dust-to-gas ratio of 100. The right column gives the fit results as obtained with the static RT model grid.}

\label{tab: param-ranges}
\end{table}


In order to obtain a first guess 
pre-burst model, we chose a similar approach as introduced in \cite{2017A&A...600A..11R}, where a pool of SEDs is generated for a sample of different YSO configurations, that consist of the same set of components.
For this purpose, we applied the TORUS code \citep{Harries2019} its radiative equilibrium mode to generate a database of pre-burst SEDs. 
This is the same code used for time-dependent modeling as well.

We assumed an axisymmetric geometry, and the state variables (temperature, density etc.) were stored as a 2-D cylindrical adaptive mesh. The code employs Monte Carlo (MC) radiative transfer (MCRT), splitting the radiation field into a large number of indivisible photon packets that propagate through the mesh. An iterative method is used to determine the dust temperature based on the radiative-equilibrium method of \cite{lucy:1999} and described in detail in \cite{Harries2019}. Once the temperature distribution has been determined a second MCRT calculation is used to produce the SED at the given inclination.

All models are composed of a protostar, a passive disk,
a curved bipolar cavity, and an Ulrich-type envelope \citep{1976ApJ...210..377U}. 
The disk can be described as a flared disk in hydrostatic equilibrium \citep{Chiang;1997}, but, similar to \cite{2017A&A...600A..11R}, the parameter ranges are chosen such that flat disks are included as well.
The passive disk approach is justified for FIR/(sub)mm wavelength regions. Active disks (e.g., \citealp{Shakura1973}) differ from passive disks mainly in the innermost region, where most of the energy is released by viscous dissipation \citep{1981ARA&A..19..137P}. Thus, the difference manifests primarily in the MIR, 
which might rise earlier in the case of active disks and maybe 
even before the source and/or NIR peak.
To model the thermal FIR afterglow, the passive disk approach can be applied. 

We used MRN dust \citep{Mathis1977}, which consists of compact, homogeneous, and spherical
grains with a composition of 62.5\% silicate and 37.5\% graphite, where we took the optical properties from \cite{2001ApJ...548..296W}. The size distribution can be described by a power law with $n \propto a^{-3.5}$ with grain sizes ranging
from $a_{min} = 5\, \rm nm$ to $a_{max} = 250\, \rm  nm$. 
All parameters were sampled from the ranges given in Table\,\ref{tab: param-ranges}. 
The protostellar luminosity $L_*$ includes all possible mechanisms of energy release (fusion, contraction, accretion). Throughout the paper the term `luminosity' represents the bolometric one, obtained by integrating the SED over the whole wavelength range observed.
The inner radius of the circumstellar disk is governed by the dust sublimation radius (assuming $T_{\rm sub}{\ts}{=}{\ts}1\,600{\ts}{\rm K}$). 
For our models, we cut the envelope at an outer radius $r_{env}$, which was sampled between $(1\,-\,4)\times 10^4\, \rm au$ ($0.05\,-\,0.2\,\rm pc$).
This is in the range of the size of a typical cloud core.  
The extent of the FWHM in the ALMA images is smaller by a factor of two to ten. Due to the high resolution, the interferometer may not be sensitive to the extended component of the emission. However, most of the emission should be covered by ALMA. 
The extent of the prompt LE is greater than the outer radius of our models by a factor of $\sim 5$. However, the density in the outermost parts of the parent cloud core is probably pretty low.

\begin{figure}[hbt]
	\begin{center}
 \includegraphics[width=\hsize]{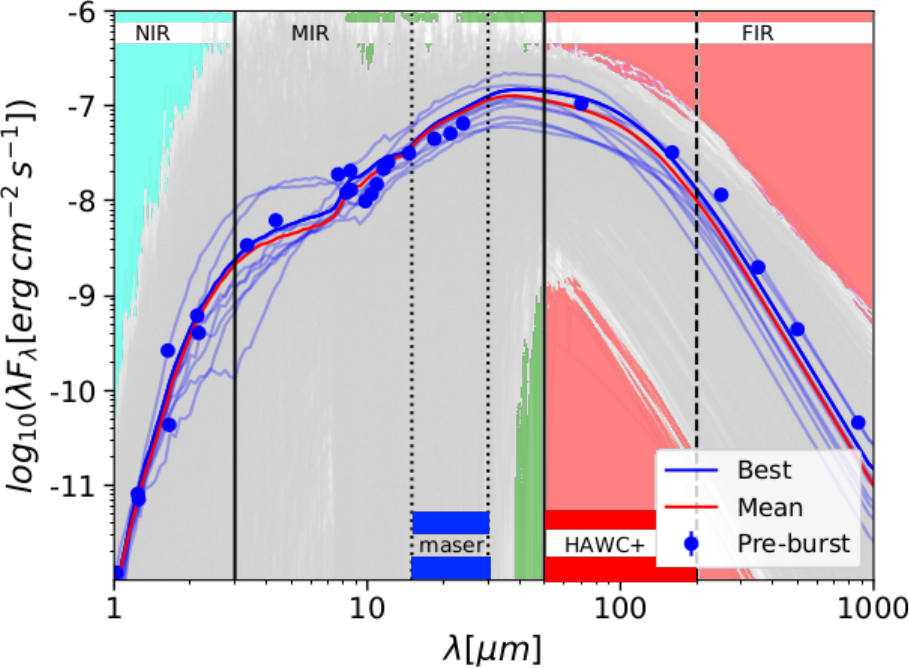}
 \end{center}
	\caption[G323 pre-burst SEDs (fit result)]{All TORUS models (gray), together with the pre-burst SED (blue dots), the ten best fits (blue, the best one is darkest), and the mean model (red). The model SEDs are reddened according to the foreground extinction of 18\,mag \citep{murphy:2010}. Background colors indicate different wavelength ranges. The SED peaks in the MIR/FIR.
   The wavelength range of HAWC+ is indicated at the bottom. That for radiative maser excitation \citep{2002IAUS..206..183O} is highlighted in blue.
   Flux densities are listed in Table \ref{tab: G323 preSED}. The mean model fits the pre-burst SED quite well.}
    
	\label{fig: G323 pool}
\end{figure}

In total, we include 2500 SEDs. Although this seems small compared to the YSO grid of \cite{2017A&A...600A..11R}, we applied a much smaller range of protostellar luminosities. Therefore, the density of models in the parameter space is actually similar.
Fig. \ref{fig: G323 pool} shows the pre-burst SED (blue) together with the model pool (gray). The data is reproduced pretty well. 
The spikes visible in the MIR are due to low photon counts. 
To save computing time, we initially used only a relatively small number of photon packets for the models. Models that provide good fits were recalculated with a sufficient number.
The pre-models were fitted assuming a distance of $4.08\mypm^{0.40}_{0.38}\, \rm kpc$ and a foreground extinction of $A_V=18\mypm1$\,mag as given in Sect. \ref{intro}. 

The G323 pre-burst SED is well covered from NIR to (sub)mm. 
The number of free parameters is as large as 11 
and there are degeneracies among the models. 
Therefore, we do not use just the best model, but 
rely on the ten best models instead.
This approach was applied already in 
\cite{Stecklum:2021}. 
All parameters of the ten best pre-burst models (blue) are averaged 
using weights according to the respective $\goodchi^2$ value. 
For linear sampled parameters, we use the arithmetic mean, whereas for logarithmically sampled parameters, it is the geometric mean. 
The resulting mean model, that is, the model where all parameters are mean values, is colored red. 
The mean values can be found in the right column of Table\,\ref{tab: param-ranges}. 
The table with the ten best models is given in the Appendix (Table\,\ref{tab: param-ranges-all}). 
Note that the best models point to an almost face-on view. 
The inclination is smaller than the opening angle of the cavity, which implies that the optical depth toward the object is minimized.

In addition to the mean model, two additional settings are included (Tmin and Tmax). 
These settings minimize and maximize the afterglow timescales and compromise parameters, which are in agreement with the pre-burst fit within the given confidence intervals.
The Tmin setting has a lower dust content, and a wider opening angle of the cavity compared to the mean setting. The Tmax setting has a higher dust content, and a smaller cavity opening angle. 
The parameters of all three models are summarized in Table\,\ref{tab: settings}, where only those that differ are given. 
We changed only those parameters that are considered to strongly affect the duration of the afterglow.
All other parameters 
are the same as for the mean model (right column of Table\,\ref{tab: param-ranges}). Constant parameters are the protostellar luminosity, inclination, and most of the disk parameters, since their impact is small or unclear. 
The Tmin, and Tmax settings are meant to give limits on the burst energy that take into account that the afterglow depends not only on the burst energy but also on the local dust distribution. 
Both the mean, and the Tmax models agree well with the pre-burst SED (their $\chi^2$ is comparable or better than that of the best-fit model). 
The Tmin model has too little cold dust and cannot reproduce the FIR fluxes. However, TDRT models were also performed for this model.

\begin{table}[hbt]
\caption{
 Overview over the simulation settings. }
\resizebox{0.48\textwidth}{!}{
	\begin{tabular}{|c|c|c|cc|cc|}
		\hline
		&Fit&Disk&Envelope&&Cavity&\\ \hline
		&$\goodchi^2$&$m_d$& $r_{max}$ &$\dot{M}_{env}$&$\rho_0^{cav}$&$\Theta_c$\\ %
		&& $\rm M_\odot$ & $10^4\, \rm au$&$\rm M_\odot\,\rm yr^{-1}$&$10^{-20}\, \rm g\,cm^{-3}$&\degr\\ \hline 
		mean&90&$2\times 10^{-3}$&$ 2.4$ & $0.032$& $ 4.5$&$42$\\ 
		Tmin&610&$3\times10^{-5}$ &$1.6 $ & $0.015$& $3.0$&$53$\\
		Tmax&180&0.1 &$ 3.6$ & $0.070$& $ 6.8$&$31$\\\hline
	\end{tabular}
 }
 \tablefoot{The models Tmin and Tmax represent settings with a minimal and a maximal afterglow timescale respectively. The mean model represents the result of the pre-burst fit, with an intermediate afterglow timescale. 
 Only the parameters that are different are displayed.
 All other parameters (i.e., those of the disk, and protostar, as well as the inclination) are fixed to the values  given in the right column of Table\,\ref{tab: param-ranges}. 
 Densities, and masses are total values (adapting a gas/dust ratio of 100). 
 The $\goodchi^2$- values refer to a distance of $3.9\, \rm kpc$ and a foreground extinction of $19\, \rm mag$
 , which is the best fit for the mean model within the error margins of these quantities. 
 The mean model fits the pre-burst SED best. The Tmin model lacks cold dust and therefore cannot reproduce the fluxes.
 }
 \label{tab: settings} 
\end{table}

 \subsection{TDRT modeling}


\subsubsection{TORUS-Code}

To model the temporal evolution of the SED, the TORUS radiative transfer code was used in its time-dependent mode. The time-dependent algorithm follows the random walk of photon packets as they propagate over a time step, with the energy absorption rates for each grid cell calculated using a MC estimator and the energy emission rate calculated from the local dust temperature. The TDRT method is described in detail in \cite{Harries2011} and has been benchmarked against several standard thermodynamic test problems. Unlike traditional methods, such as flux-limited diffusion, the code is able to simultaneously and accurately treat both the free-streaming and diffusive regimes. The role of the time step is crucial to the accuracy of the calculation. It needs to be short enough for the luminosity variation to remain tractable, but long enough to avoid creating too much computational overhead \citep{Harries2019}. We used a time step of $3.65 {\ts} \rm d$, which was verified with an interval shorter by a factor {$\sim$}7. With this choice, the envelope heating, which is responsible for most of the FIR radiation, is traced very well. But for the disk, a shorter time step is better. 
TORUS is a very versatile code with several microphysics modules from which we only apply those to calculate the dust radiative equilibrium (static RT) and nonequilibrium (TDRT). The former was used, for example by \cite{Esau;2014} while the latter was validated against the Pascucci disk benchmark \citep{Pascucci2004}.



\subsubsection{Assumptions and constraints}
\label{sec:assumpt}

To derive the burst energy using TDRT, we 
make several assumptions, which are explained below.
TORUS computes dust continuum emission as a function of time and wavelength for a given source-luminosity variation and local dust distribution.

First, we examine whether the energy radiated away by dust is a good proxy for the total energy released by the burst. The two main coolants for YSOs are molecular lines, in particular from CO,
and dust radiation. Dust grains are solids, which implies that they are thermal emitters.
Their opacities cover more than four orders of magnitude in wavelength of the electromagnetic spectrum (e.g., \citealp{Min;2015}) and are thus much more capable of affecting radiative heat transfer than gas opacities. In fact, the luminosity of the FIR molecular cooling lines of MYSOs is only a tiny fraction ($\sim10^{-5}$, \citealp{Karska;2014}) of the dust luminosity.  
The same holds for the energy release by free-free emission from a compact H{\sc ii} region (R. Cesaroni, priv. comm.).
Therefore, the luminosity estimate derived from the SED can be considered as the total value.

Although the gas cooling is not relevant for the energy considerations, we note in passing that the heating/cooling timescales for dust and gas may differ.
At volumetric densities $\gtrsim 1.2\times10^4\rm\, cm^{-3}$ (e.g., \citealp{Merello;2019}) gas and dust are thermally coupled through collisions.
In the (dense) innermost regions of MYSOs 
this assumption is certainly justified. 
In the extended outer envelope 
the coupling may be weaker. 
Then, the gas will heat/cool much slower than the dust (see, e.g., \citealp{johnstone:2013}). Therefore, the response of the gas cooling lines to an accretion burst may be strongly delayed in comparison to that predicted by our TDRT model.

The dust distribution is given by our three reference models: Tmin, mean, and Tmax (as introduced in Sect. \ref{sec: premod}).
We assume that the $K_{\rm s}$ light curve may serve as a proxy for the variation of the accretion rate, which is hereafter referred to as the `burst profile'. It is essentially defined by the dates of the burst rise, its peak, and the return to the pre-burst $K_{s}$ level. According to Wien's law, the effective wavelength of this band is closest to the dust sublimation temperature. 
Even with the recent Skymapper release, the VVV(X) $K_{s}$ photometry still provides the best coverage of the burst evolution. Moreover, 
this choice is justified by the almost face-on view of the best-fitting pre-burst models (see Sect. \ref{sec: premod}), allowing for a `direct look' at the innermost regions, where the VIS/NIR-photons originate. 



For TDRT modeling, we use the most simple approach possible, namely, we include only heating and cooling but omit any kind of density variation as well as any changes in the disk chemistry. The latter is generally not met. Instead, the burst will most likely sublimate the innermost disk, change the chemistry of the disk, and shift the ice line outward (e.g., \citealp{lee:2007}, \citealp{2012ApJ...754L..18V}, \citealp{2013A&A...557A..35V}, \citealp{2017A&A...604A..15R}). 
In principle, dust sublimation is implemented in TORUS 
but it requires a time step of 
less than 1{\ts}d to remove the disk on typical disk dispersal timescales. 
However, the impact on the evolution of the (F)IR luminosity will be minor. 
Therefore, we use a longer time step, where we move the inner radius outward to ensure that the dust does not become hotter than $T_{\rm sub}$. 
We shift it to 
60\,au, which is $\sim 3\, R_{sub}$ according to \cite[Eq. 1, with $T_{sub}=1\,600${\ts}K]{2004ApJ...617.1177W}, 
where we kept the disk mass constant.

\begin{figure}[!hbt]
    \includegraphics[width=\hsize]{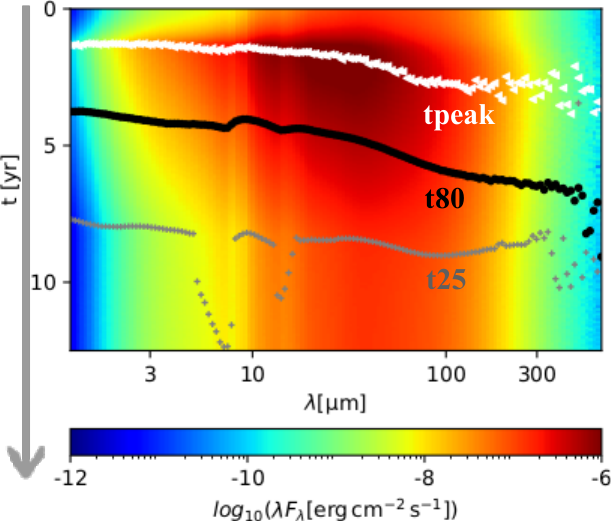} 
	\caption{Dynamic SED shows the  flux density over wavelength and time (increasing from top to bottom) of the mean-setting and a burst with an energy of $2.3\times 10^{47} \, \rm erg$. 
    Time 0 marks the onset of the burst. Horizontal lines indicate the peak time (white triangles), the time when 80\% of the energy is released in the respective band (black dots), and 
    when
    the flux increase is back at 
    1.25 times 
    the pre-burst level (gray crosses). The increase in scatter in the (sub)mm range is due to lower synthetic photon counts. }
 \label{fig:dynSED}
\end{figure}

\subsection{The dynamic SED of the mean model}
\label{sec:tsed}

This is the first astrophysical application of dust-continuum TDRT. Therefore, we briefly discuss one particular model in more detail before fitting the burst energy. 
We used the mean model with an outburst energy of $E_{\rm acc}=2.3 \times 10^{47}{\ts} \rm erg$. 
This first guess for the burst energy
is based on the assumption that the increase in $K_{\rm s}$ is the same as the increase in 
luminosity 
at all times. 
With the model setup and the burst profile at hand, TDRT simulations were performed. The result is shown in Fig. \ref{fig:dynSED}. The figure displays the dynamic SED, which is the flux density over wavelength and time. This probably provides the most concise overview of the entire afterglow.
Time increases from top to bottom, where zero corresponds to the burst onset. Three particular times are indicated: Flux peak time (tpeak, white triangles), the time when $80 \%$ of the energy is released
(t80, black dots), and the time when the flux density dropped to 1.25 times of its pre-burst value (t25, gray crosses) for each wavelength.

Both $\rm tpeak$ and $\rm t80$
increase with wavelength. This is expected since the radiation at longer wavelengths can be attributed to regions more distant from the star (i.e., colder regions). Interestingly, the delay between NIR and FIR is almost two (and three) years for $\rm tpeak$ (and $\rm t80$) at $100\, \rm \mu m$. 
For comparison, the light travel time from the source to the outer edge of the grid is only $160\, \rm d$. 
The delay between the peak of the burst profile (1.2 yr after onset) and that in the FIR is by far more than what could be explained by geometrical/projection effects and distinct spatial origins alone. It clearly indicates a measurable slowdown of the energy transfer toward the FIR-emitting regions by numerous absorption and re-emission processes owing to the high optical depths in between the protostar and FIR-emitting regions. Toward longer wavelengths ($\lambda \geq 300\, \rm \mu m$) both curves flatten. This is expected because at some point even the coldest and most distant regions are `processed’ and we are reaching the Rayleigh-Jeans tail of the emission from the coldest dust with 
$\sim 20 \, \rm K$. Furthermore, these regions are generally optically thin at these wavelengths.

The t25-curve looks somewhat different. The time increases only slightly; it is almost constant between $30$ and $80 \, \rm \mu m$ and decays for $\lambda$  exceeding $100 \, \rm \mu m$. Although the timescales increase in principle with wavelength, the peak level is much lower at higher wavelengths (at some point hindering a further increase of t25). At $4$ to $8{\ts}\rm\mu m$ there is a `prominent' 
feature and a weaker one at $\sim 15{\ts}\rm\mu m$.
These MIR features can most likely be attributed to the densest regions (disk midplane), which cannot efficiently cool because of the high optical depths in the silicate absorption bands.
For weaker bursts or less dense environments, this will probably not occur since the disk will not completely heat or cooling is faster. However, this requires further investigation. Toward long wavelengths, the synthetic noise increases dramatically because of the low number of photon packets. We emphasize that t80 is the timescale least sensitive to numerical scatter. 

\subsection{Results of the time-dependent fitting} \label{sec: TDRT res}
\subsubsection{Burst energy estimate for the mean model}\label{Sec:burst}

In general, the reprocessed FIR radiation provides an indirect but reliable measure of the burst energy, since it covers most of the emission (SED peak) and the role of extinction is rather small (compared to the NIR/MIR), while the maximum increase due to the burst is still strong (contrary to the (sub)mm). 
FIR measurements were used successfully to estimate the energy of MYSO bursts in the past
\citep{carattiDiskmediatedAccretionBurst2017, Stecklum:2021}.
For a more general discussion of the power of FIR data, based on static RT for a sample of low-mass YSOs, see, for example, \cite{fischer:2024}.

In the following, we use the SOFIA post-burst FIR measurements together with a set of time-dependent models to constrain the burst energy.
Our HAWC+ measurements indicate that the post-burst flux densities are slightly higher than the pre-burst ones.
They are elevated by $(14.2\pm4.6)\%$ at $70{\ts} \rm \mu$m and $(8.5\pm6.1)\%$ at $160{\ts}\rm\mu$m 
(see Sect. \ref{sec: HAWC}). 
As the excess is quite small, we fit the flux density ratios rather than the absolute values. 
The ratios trace the increase in the protostellar luminosity and hence the released energy. 



\begin{figure*}[hbt]
\begin{center}
    \resizebox{\textwidth}{!}{
    \includegraphics[height=5cm]{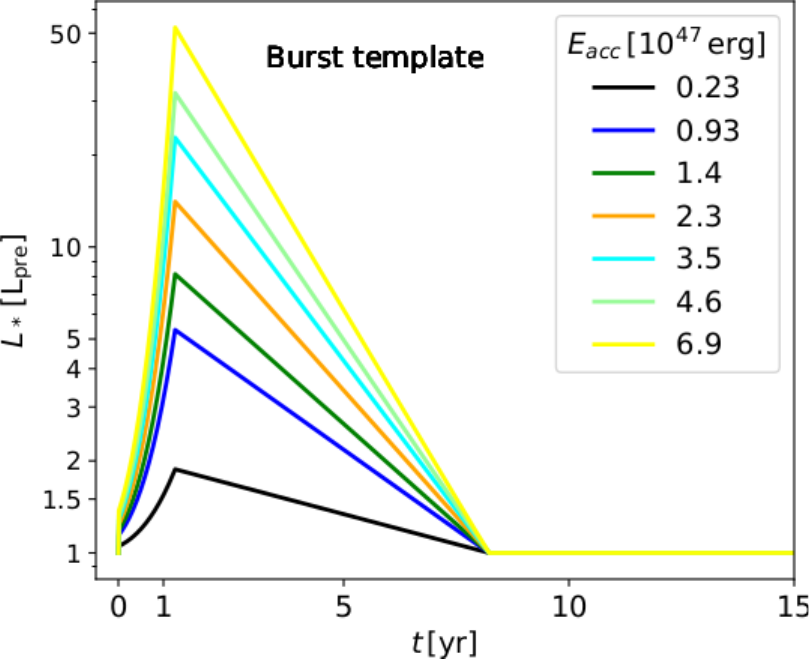}
    \includegraphics[height=5cm]{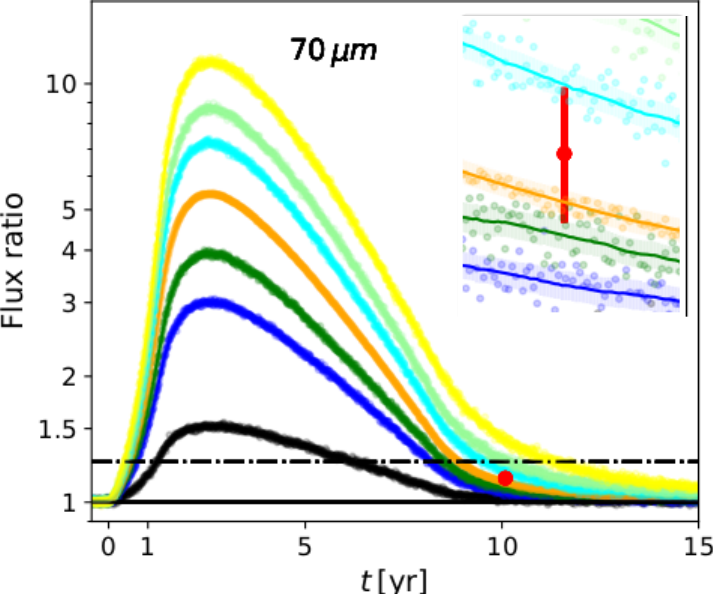}
    \includegraphics[height=5cm]{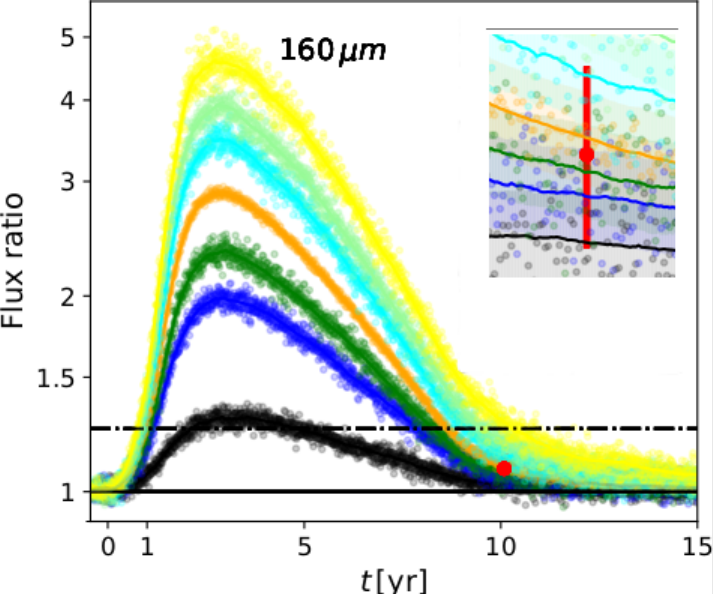}
    }
	\caption{Input burst profiles (left) and output flux ratios as a function of time for $70$ (middle) and $160\, \rm \mu m$ (right) for the mean model. The burst energy is varied between (0.23 and $6.9)\times 10^{47}\,\rm erg$ (color-coded). Obviously, the afterglows are longer for more energetic bursts. The geometry is the same for all simulations. The HAWC+ measurements are colored red. The horizontal solid and dashed-dotted lines mark the pre-burst level and 1.25 times that level. The insets show a zoom-in on the region of interest. The simulations (colored dots) were interpolated in time to reduce numerical scatter (corresponding lines and shaded regions). The observations are given in red.}
 \label{fig:Eacc-mod}
 \end{center}
\end{figure*}

\begin{figure}[hbt]
	\begin{center}
		\includegraphics[width=\hsize]{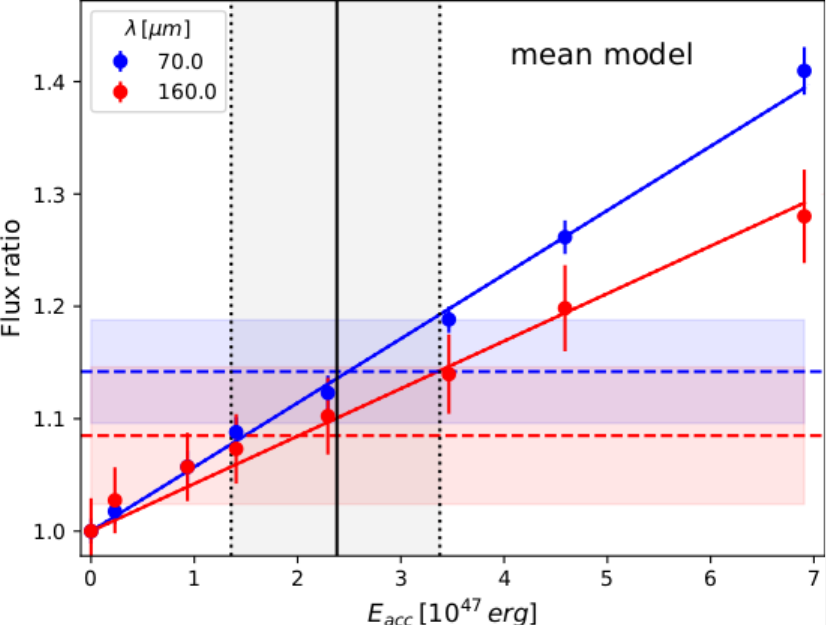}
		\caption{Modeled flux ratio at 70 (blue) and $160\,\rm \mu m$ (red dots) for different burst energies, all featuring the mean setting. Solid lines are linear fits. The dashed horizontal lines are the observations with confidence intervals (overlapping colored areas). The vertical black line indicates the best burst energy $E_{acc}$ for both wavelengths. The 1$\sigma$- confidence interval is indicated. We use a $\goodchi^2$-minimization to determine $E_{acc}$, where we use linear fits (colored solid lines) as model values for both wavelengths. We take into account the observational errors and uncertainties of the modeled ratios, where the contribution of the latter is minor. }
        \label{fig: joint mean}
	\end{center}
\end{figure}

To constrain the burst energy, we performed seven simulations with burst energies between (0.2 and $7)\times 10^{47}\, \rm erg$. The corresponding 
burst templates (simulation input) are displayed in the left panel of Fig. \ref{fig:Eacc-mod} with the different burst profiles color-coded.
The middle and right panels show the corresponding flux density ratios over time for $70$, and $160\ts\mu$m respectively. 
As the dust distribution is the same in all cases, the changes are due to the different energy inputs of the respective bursts.
Obviously, the more energetic the burst, the longer the afterglow. Zoom-in to the HAWC+ data (red points in the inset) shows that the observed excess can be explained by a burst within the adapted energy range. The simulations (colored dots) show a lot of scatter. Therefore, we apply a stepwise interpolation in time (solid lines with confidence intervals in transparent). \\
The predicted flux ratios at the date of the HAWC+ observation are shown in Fig. \ref{fig: joint mean} as functions of the burst energy for 70 (blue), and $160\,\rm \mu m$ (red). The errors correspond to the confidence interval of the interpolation in time (at the HAWC+ date) for each model and wavelength (cf. the inset in Fig. \ref{fig:Eacc-mod}).
The HAWC+/PACS ratios are indicated by the dashed horizontal lines with the corresponding $1\sigma$ confidence intervals indicated by the colored areas.
The ratios can be fitted with linear functions in the given range (indicated by the colored solid lines), whereby the y-axis intercept equals one, as no energy input means no flux increase. 
The fit leads to $r70 = 0.06\cdot E_{acc}+ 1$ and $r160 = 0.04 \cdot E_{acc} +1 $ 
with r70 and r160 as ratios at 70 and $160\, \rm \mu m$ respectively, and $E_{acc}$ as the burst energy.
A $\goodchi^2$ minimization yields a burst energy of $E_{acc}=(2.4\pm1.0) \, \times 10^{47}\,\rm erg$ (gray-shaded area with vertical black lines).
The error ranges extend until the $\goodchi^2$ value becomes worse than $1+\goodchi^2_{min}$ (the $1\sigma$ confidence interval according to {\em kafe}\footnote{\url{https://etpwww.etp.kit.edu/~quast/kafe/htmldoc/index.html}}). 
 
\subsubsection{Limits on the burst energy}\label{sec: Eacc} 

In the previous section we estimate the burst energy by using the mean model.
Although the result agrees very well with our initial estimate, to assess its credibility, it is necessary to investigate the influence of the local dust distribution. 

\begin{figure*}[!hbt]
\begin{center}
    \resizebox{\textwidth}{!}{
    \includegraphics[height=7.5cm]{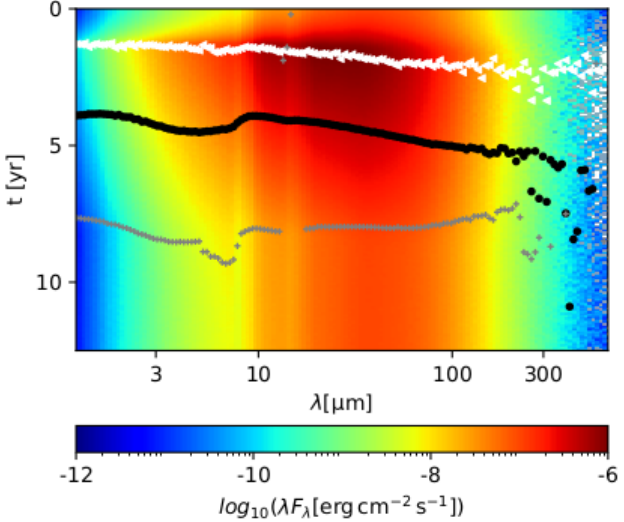}
   \includegraphics[height=7.5cm]{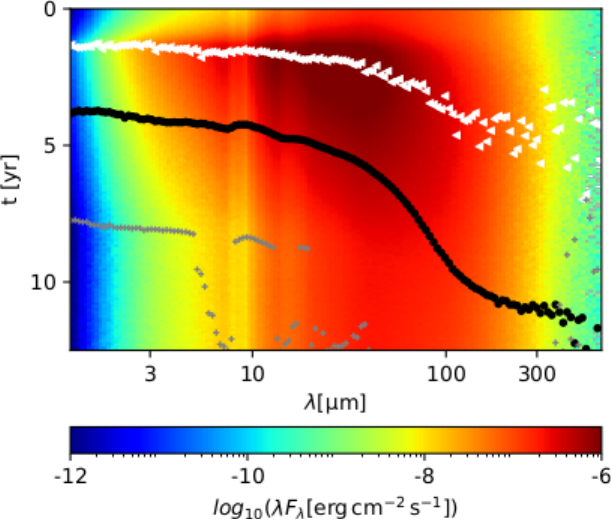}
   }
	\caption{Same as Fig. \ref{fig:dynSED} but for the Tmin (left) and Tmax (right) setting. The burst is the same for both. The plot shows that the dust distribution strongly imprints the afterglow. For denser and more extended envelopes, the MIR/FIR afterglow can be significantly longer.
    Horizontal lines indicate the times when the peak is reached (white triangles), when 80\% of the energy is released (black dots), and the time when the flux density is back at $1.25\times L_{pre}(\lambda)$ (gray). }
 \label{fig:dynSEDmax}
 \end{center}
\end{figure*}

\begin{figure}[hbt]
	\begin{center}
		\includegraphics[width=\hsize]{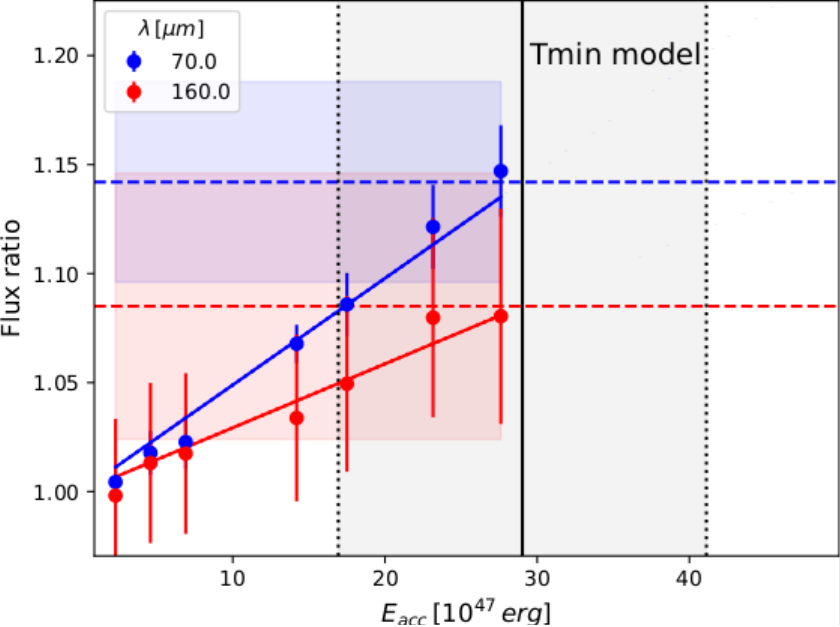}\\
  \hspace{0.1cm}\\
    \includegraphics[width=\hsize]{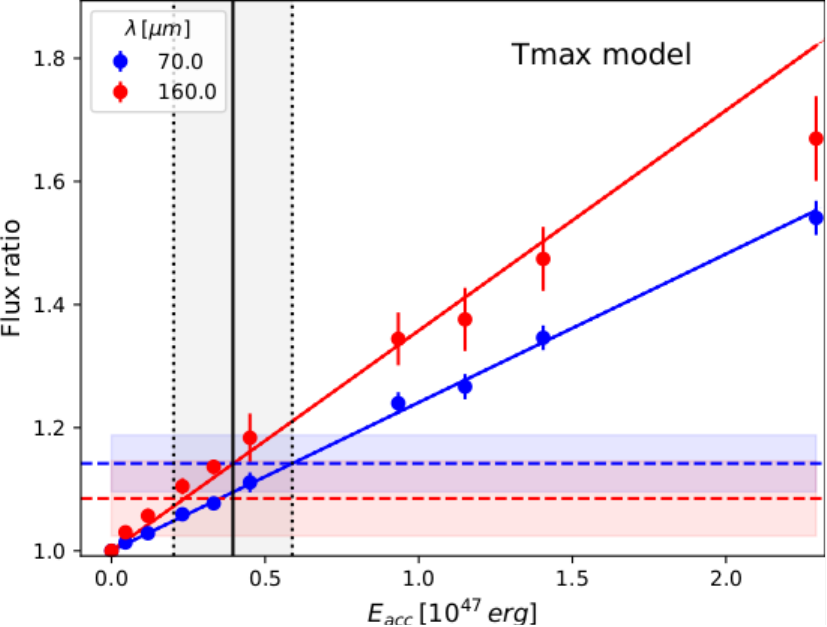}
    \caption{Same as Fig. \ref{fig: joint mean}, but for the Tmin and Tmax model (top and bottom, respectively). For these settings, the burst energy (indicated by the vertical black line) needed to explain the HAWC+ data (horizontal red and blue lines) is maximized or minimized respectively. Interestingly, for the Tmax setting, which is the most extended and densest, the flux ratio at $160\mu$m (red) exceeds that at $70{\ts}\mu$m (blue). The dependency is almost linear. The best fits are indicated by the solid colored lines. Note the different scales.} 
  \label{fig: joint minmax}
	\end{center}
 \end{figure}

To show that the afterglow durations can be quite different 
we plot the dynamic SEDs for two different YSO configurations in Fig.\ts{}\ref{fig:dynSEDmax}.
The left panel shows the dynamic SEDs for the Tmin and the right panel shows the Tmax setting. Both settings feature the same burst 
and the huge differences can be explained solely by the different YSO configurations. 
We emphasize that the Tmin and Tmax settings are limiting cases with minimal and maximal afterglow durations, respectively. 
This implies that the energy needed to explain the HAWC+ measurements can be considerably larger or smaller than that of our previous mean model estimate. \\
To determine the limits for the burst energy, we repeat the analysis from Sect. \ref{Sec:burst}. 
We ran seven simulations for the Tmin setting with burst energies between (2.3 and $28) \times 10^{47}\, \rm erg$ and nine simulations for the Tmax setting with burst energies between (0.05 and $2.3) \times 10^{47}\, \rm erg$.
The ratios on the HAWC+ dates are shown in Fig. \ref{fig: joint minmax} for Tmin (top) and Tmax (bottom), respectively. A linear fit leads to 
$r70 = 0.24\cdot E_{acc}+ 1$ and $r160 = 0.35 \cdot E_{acc} +1 $ for Tmax and 
$r70 = 0.005\cdot E_{acc}+ 1$ and $r160 = 0.003 \cdot E_{acc} +1 $ for Tmin with r70 and r160 as ratios at 70 and $160\, \rm \mu m$
respectively and $E_{acc}$ as burst energy. 
For the Tmax setting, the ratio at $160{\ts}\mu$m exceeds the one at $70{\ts}\mu$m. This is in agreement with the dynamic SED (Fig. \ref{fig:dynSEDmax}), which shows a strong increase of the indicated timescales 
in between $\sim (50$ and $300){\ts}\mu$m. 
Again, the best estimate for the burst energy is indicated by vertical lines. 
It amounts to $30\pm 12$ for Tmin and $0.40\mypm^{0.20}_{0.19}$ for Tmax in units of $10^{47}\,\rm erg$. 
For the Tmin and Tmax settings, the values of $E_{acc}$ are about a factor 13 above and a factor six below the mean model respectively. 
We emphasize that the higher value is not likely, as the pre-burst fit of this model is much worse than for the other two settings. 
However, it may still serve as a conservative upper limit. A summary of the derived burst parameters is provided in Table\,\ref{tab: G323 burst param}.

\renewcommand{\arraystretch}{1.05}
\begin{table}[hbt]
\caption[G323's burst parameters]{Summary of the burst parameters for all three settings. }
	\begin{center}
	\setlength\extrarowheight{4pt}
 \resizebox{0.5\textwidth}{!}{
 \begin{tabular}{|c|c|c|c|c|}
		\hline 
		Model & $E_{acc}$& $M_{acc}$ & $<\dot{M}_{acc}>$&$L_{peak}$\\
	     & [$10^{47}\, \rm erg$] &[$\rm M_{Jup}$] & [$10^{-3} \rm M_\odot\,  yr^{-1}$] &[$\rm L_{pre}$]\\[1.5pt] \hline
	     Tmin &$30\pm12$&$230\pm110$&$27\pm12$ &$310\mypm_{140}^{130}$\\[1.5pt] \hline
      mean &$2.4 \pm1.0$&$19\pm9$&$2.1\pm1.0 $& $14\mypm_6^8$\\[1.5pt] \hline
      Tmax&$0.4 \pm 0.2$&$3.1\pm{1.7} $&$0.4\pm0.2$&$2.6\mypm_{0.8}^{0.9}$ \\[1.5pt] \hline\rowcolor{lightgray}
	    $K_{\rm s}$-based& $0.9\mypm_{0.7}^{2.5}$ &$7.3\mypm_{5.9}^{20}$&$0.8 \mypm_{0.6}^{2.2}$& $5.4\mypm_{3.6}^{16.6}$\\[1.5pt] \hline
	\end{tabular}}
	\end{center}
\label{tab: G323 burst param}
\tablefoot{The derived values span almost two orders of magnitude. The highest values (reached for Tmin) are probably too large. We consider the $K_{\rm s}$-based range (highlighted) the most reliable (see text). The mass estimate is based on the assumption that the protostar is close to the ZAMS. If it is bloated, the mass (and the mass accretion rate) will be higher (cf. Sect. \ref{sec: bloat}). }
\end{table}
\renewcommand{\arraystretch}{1}



\begin{figure*}[hbt]
	\begin{center}
 \resizebox{\textwidth}{!}{
		\includegraphics[height=5cm]{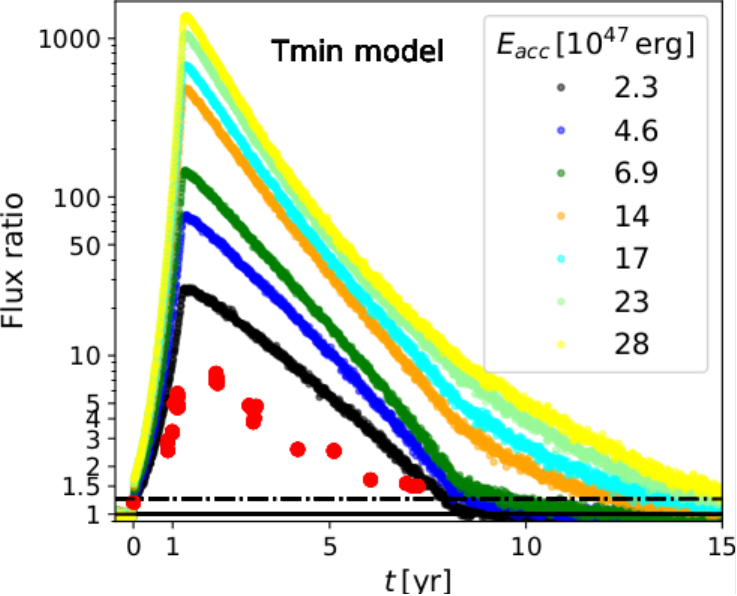}
		\includegraphics[height=5cm]{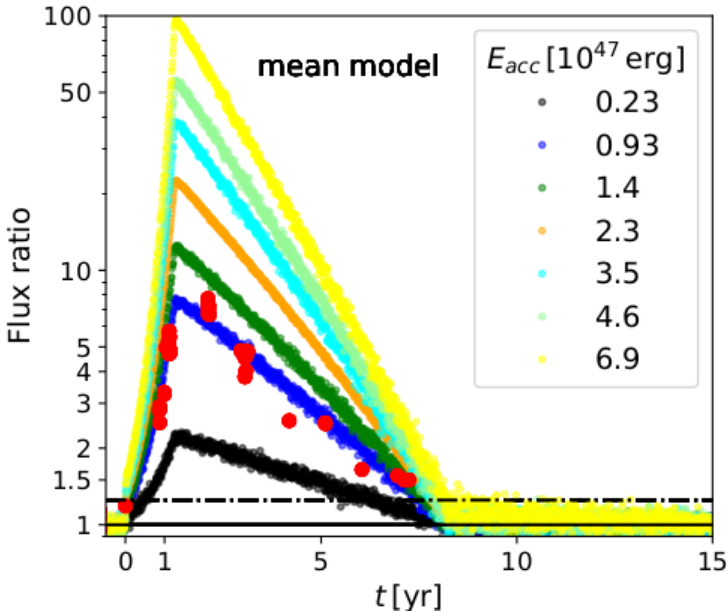}
		\includegraphics[height=5cm]{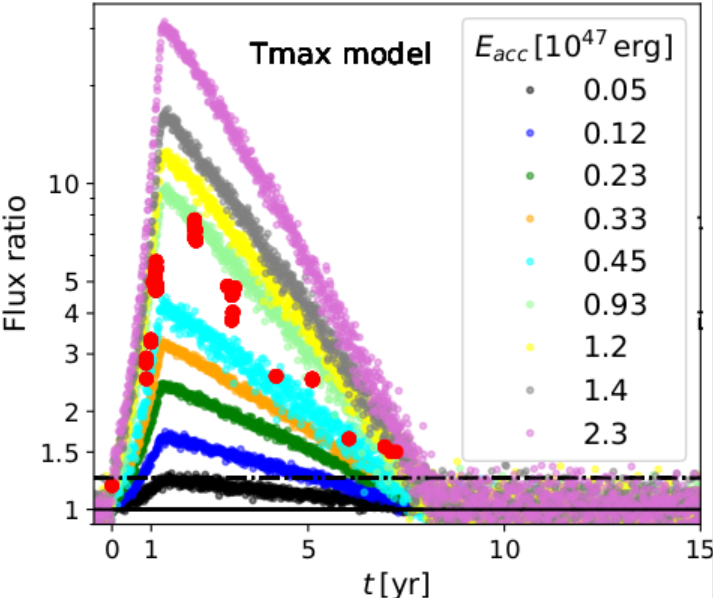}
  }
	\end{center}
	\caption[Ks light curves (G323 models)]{Modeled $K_{\rm s}$ ratio curve for the Tmin (left), mean (middle), and Tmax (right) model setups featuring different bursts (in $10^{47}\, \rm erg$, color-coded). The observed $K_{\rm s}$ increase is shown in red for comparison. The ratio depends more on the energy input than on the setting. The best agreement is reached for $E_{acc}=0.93 \times 10^{47}\,\rm erg$.
    For the Tmin setting and the highest burst energies, the $K_{\rm s}$ curves decay delayed, this might be an artifact due to the innermost dust becoming unrealistically hot ($r_{min}$ is fixed to $60 \, \rm au$, that is, $3 \times R_{sub}$). We note that the ordinate scales are different. 
    }
	\label{fig: minmax Eacc lcK}
\end{figure*}

In addition to the HAWC+ flux density ratios, the models can be compared with the $K_{\rm s}$ measurements. Similar to the HAWC+ fit, we used the $K_{\rm s}$ flux density ratios.
Fig. \ref{fig: minmax Eacc lcK} shows the $K_{\rm s}$ ratio curves for all three configurations (from left to right) and all burst energies (color-coded). The observation is shown in red. For the Tmin configuration and the highest burst energies, the $K_{\rm s}$ flux does not return to its pre-burst level at the end of the burst. This is probably caused by dust that becomes hotter than $T_{\rm sub}$. 
We shifted the innermost radius to 60 au ($\sim3 R_{sub}$) to avoid too high dust temperatures for all models. This is not sufficient for the most energetic bursts.
All bursts overestimate the observed $K_{\rm s}$ ratio (red) for the Tmin setting. This is another indication that the burst energy estimate obtained with this setting is too large.
We emphasize that the $K_{\rm s}$ ratio depends only slightly on the setting. 
For nearly the same burst energy for Tmin and the mean setting, the best match is reached with the $K_{\rm s}$ curve, despite the vastly different afterglows. 
Therefore, the $K_{\rm s}$ ratio provides a good measure for the burst energy.
For both the mean and the Tmax setting, a good agreement is reached for a burst with an energy of $0.93\times 10^{47}\,\rm erg$.
This value lies well within the determined range and would imply a dust configuration in the range between mean and Tmax. 
Probably the best estimate of the burst energy is $E_{acc}=(0.9\mypm_{0.7}^{2.5})\times 10^{47}\, \rm erg$, which is based on the $K_{\rm s}$ value and covers the entire range spanned by the mean and Tmax settings.
This would correspond to a peak value of $L_{max}=(3.2\mypm_{2.1}^{10.3})\times 10^5L_\odot$, and hence an increase of the luminosity of the protostar by a factor $5.4\mypm_{3.6}^{16.6}$ (or $\Delta L\sim 2.6 \times 10^5L_\odot$).


\subsubsection{Accreted mass and mass accretion rate}\label{sec: mass}

From the burst energy, the accreted mass 
can be inferred with $M_{acc}=\frac{ E_{acc}\cdot R_*}{G \cdot M_*}$. 
where G is the gravitational constant and $R_*$ and $M_*$ are the protostellar radius and mass. 
This approach is used for a passive disk in \cite{Stecklum:2021}. The underlying assumption is that the infalling material falls from a large distance onto the protostellar surface, where it releases its entire gravitational energy within the burst. 

Although the SED shape, the NIR brightness, and the presence of an HCH{\sc ii} suggest that G323 is a more evolved MYSO, it must be kept in mind that its spectral appearance is dominated by the face-on view. Nevertheless, assuming that it is on the ZAMS will provide a lower limit of the accreted mass as protostars contract toward this locus of stellar evolution. 
For a star with a luminosity of $(6.1\mypm^{4.2}_{2.5}) \times 10^4 L_\odot$, 
the ZAMS mass amounts to $(23\mypm^5_4)\,M_\odot$ and the corresponding radius $R_*=(6.5\mypm_{0.7}^{0.9})\,R_\odot$ \cite[Eq. 1 and 2]{1996MNRAS.281..257T}. 
The resulting parameters are summarized in Table\,\ref{tab: G323 burst param}.  
Adopting these parameter values, an accreted mass between 1.4 and 30 Jupiter masses (or 450 and 9\,000 Earth masses) is derived. 
The errors
are dominated by the uncertainty on $E_{acc}$.

With the accreted mass at hand, the average mass accretion rate ${<\dot{M}_{acc}>}\, = M_{acc}/\Delta t$ follows as
$(0.8\mypm_{0.6}^{2.2})\times 10^{-3}M_\odot\, \rm yr^{-1}$, where $\Delta t$ is the burst duration of $8.4\,\rm yrs$.
For comparison, an upper limit of the quiescence accretion rate is given by $\dot{M}_{acc}\leq \frac{L_{pre} R_*}{G M_*}$, assuming that the entire pre-burst luminosity is due to accretion. 
Inserting the values yields $5.4 \times 10^{-4} M_\odot \rm \, yr^{-1}$. 


\section{Discussion}\label{disk}

\subsection{Accretion burst evidence}

The $K_{\rm s}$ and (NEO)WISE light curves, together with the maser measurements, provided clear evidence of the previous accretion burst. 
Although discovered later, it actually started about three years before the discovery of the S255IR-NIRS3 \citep{carattiDiskmediatedAccretionBurst2017} and NGC 6334I MM1 \citep{hunterExtraordinaryOutburstMassive2017} events in 2015. 
This is another example of a disk-mediated accretion outburst.
We emphasize the fact that, thanks to the almost face-on view, the $K_{\rm s}$ light curve of G323 is the most direct trace of accretion variability during a MYSO burst obtained so far. Although the burst of S255IR-NIRS3 was also photometrically monitored in the $K_{\rm s}$-band \citep{Uchiyama:2020}, its emergent light curve suffered from multiple scattering due to the close to edge-on view. Thus, the case of G323 will be crucial for comparing accretion burst models with real events. Progress in this direction is ongoing (see, e.g., \citealp{Elbakyan:2023}).

\subsection{Reliability of the derived burst energy}

Since the burst energy is crucial for understanding the burst physics, restrictions of our approach and how they can be overcome will be addressed in the following.
The estimated range for the accretion energy of the G323 burst is fairly large (Sect. \ref{sec: Eacc}). 
We use only three settings, which are meant to provide limits. 
In principle, the best method to quantitatively analyze the afterglow requires a preferably huge set of different settings, which are all capable of reproducing the pre-burst SED (see Sect. \ref{sec: TRRT pros}). 
This is planned for the near future and is expected to lead to refined values.

In our analysis, we did not consider different burst shapes, but only varied the burst energy. This 
is justified by the availability of the $K_{\rm s}$ light curve which is a good proxy for the temporal variation of $\dot{M}_{acc}$ as confirmed by our modeling, cf. \ref{sec: Eacc}. Further support of this choice comes from the $i$ and $z$-band measurements that show the same qualitative behavior. 

\begin{figure}[hbt]
	\begin{center}
		\includegraphics[width=\hsize]{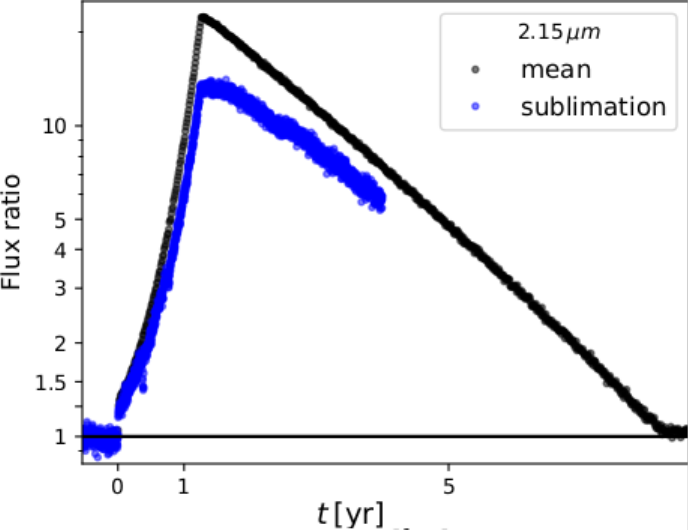}
	\caption[Ks ratio with sublimation (G323 mean model)]{Predicted $K_{\rm s}$ ratio for the mean model with (blue) and without (black) dust sublimation. Without dust sublimation, the increase is overpredicted by a factor of ${\sim} 1.5$.} 
	\label{fig: mean maser lcK}
    \end{center}
\end{figure}

The burst energy estimate was obtained using two approaches: we compared the HAWC+ post-burst flux density ratios and the Ks ratio curve individually (for all three settings). The results agree pretty well, which strengthens our approach.
Our final estimate of the burst energy is based on the fit to the $K_{\rm s}$ ratio curve
, while the confidence intervals are based on the HAWC+ fits. 
In the following, we discuss how well the $K_{\rm s}$ flux increase is reproduced by our models. We neglect all the changes that may occur during the burst. But this will not be the case in reality. 
The most obvious change is probably the shift of the sublimation radius due to the burst. To investigate the effect, we performed a simulation that includes dust sublimation. We set the innermost radius at $R_{sub}$ (20\,au) and chose a time step of 0.7 days. Dust with a temperature greater than 1\,600\,K was removed at each time step. The $K_{\rm s}$ radiation is produced by the hottest dust. Therefore, we cut the grid at the outer edge of the disk. With this choice, the resolution in the innermost regions is higher, while the entire $K_{\rm s}$ flux is captured.
The result is plotted in Fig. \ref{fig: mean maser lcK}.  The maximum $K_{\rm s}$ ratio is about a factor of 1.5 smaller for the more realistic sublimation model (blue) compared to the mean model (black). The burst energy (and shape) are the same in both cases. 
This result might be surprising, as the emitting surface area should increase as the inner rim moves outward. 
Therefore, the ratio should be higher when dust sublimation is included. 
However, the inner radius of the mean model was set at $3 R_{sub}$ to avoid unrealistic high temperatures. Thus, the area where the $K_{\rm s}$-band photons are emitted is larger. 
This shows that the most probable value (for $E_{acc}$) is slightly higher than our estimate, 
which does neglect dust sublimation. 
One could argue that a small fraction of the burst energy is needed to sublimate the disk, which might compensate for the above effect. However, this fraction is probably negligible. 
Typical values for the sublimation enthalpy range from 1 (hydrogen) to almost $1\,000 \rm \,kJ\,mol^{-1}$ (graphite). With a molar mass of $2 \rm \,g\,mol^{-1}$ ($\rm H_2$) and a gas-to-dust ratio of 100, it would take less than $10^{44}\, \rm erg$ to sublimate $1\, M_\odot$ (on the order of a total disk mass or more). This is less than $1\%$ of the burst energy.

Another issue that could impact the visible NIR flux is line-of-sight extinction variations. As the extinction is highly wavelength dependent, small changes may be enough to alter the $K_{\rm s}$-magnitude.
Tidal disruption or evaporation of the accreted object may increase the extinction, while on the opposite 
FUor/EXor outbursts are often accompanied by winds, which 
will blow out dust grains and thus reduce it 
(see, e.g., \citealp{2004ApJ...606L.119R, 2013A&A...552A..62M}).
However, the $K_{\rm s}$ light curve does not show hints of sudden extinction changes. 
This is confirmed by the $i$ and $z$-band flux densities, which are the same within the errors after and before the burst.

The cooling efficiency depends not only on the local temperature but also on the grain size and material. We use a mixture of amorphous silicate (astronomical silicate, astrosil) and carbon with small grains \citep[MRN dust]{Mathis1977}, which are efficient in cooling. However, especially within the disk and also within the envelope, grain growth processes occur. In addition, photoevaporation can reduce the size of the largest particles through vaporization of the surface, whereas the smallest grains may be evaporated entirely. While grain growth is most important in the densest regions, photoevaporation merely affects the disk surface.
Together, these processes may significantly alter the grain size distribution.  
If the grains are larger on average, the average cooling efficiency will be lower. This means that the afterglow duration is longer than predicted by our models. In that case, the limits obtained with the HAWC+ fit overestimate the burst energy. 
It is beyond the scope of this work to use different dust compositions or grain size distributions.

\subsection{The accreted object} \label{sec: bloat}
To understand the nature of the burst it is important to figure out what kind of object was accreted. 
The accreted mass is consistent with a disk fragment, a heavy planet, or a brown dwarf. 
The burst lasted $8.4\, \rm yrs$, which is still short enough to be explained by the accretion of a compact object that can be disrupted by tidal forces only within the accretion event. This is consistent with the rapid increase, which lasted about $1.4\,\rm yrs$.

Another indication of the accretion of a compact body comes from the observed color change. 
During protostellar growth, the circumstellar disk also evolves. In very early stages, the disk lacks self-gravitating objects. The initial high accretion rates in connection with viscosity 
drive it to be active. A larger accretion rate 
due to enhanced infall from the envelope or streamers will lead to an increase in its energy release, which becomes more intense as the additional matter approaches the protostar. This will result in stronger emission from the warm outer areas before the extra dust and gas reach the inner disk and raise its temperature. In this case, a reddening of the IR colors can be expected at the beginning of the accretion burst, since the increase in emission from warm regions will precede that from hotter ones. 
This is observed for the rise of an ongoing outburst of a low-mass YSO \citep{Guo_2023}.
Once the disk becomes less active and self-gravitational eddies and planetesimals build up on the route to planet formation, another behavior of the spectral change may emerge. These bodies traverse the disk almost undisturbed, but will be disrupted by tidal forces and/or evaporation close to the protostar. The central energy release starts to heat the disk at its inner radius, possibly leading to dust sublimation, which moves the inner disk rim outward. Then, the ongoing temperature rise 
leads to a `bluening' of the IR color. Therefore, the change in color during burst onset might represent an indicator of which mechanism is at work. 
Both types of color change are found for eruptive protostars in the VVV survey \citep{Lucas:2024}.
If taken at face value
, the observed `bluening' suggests that G323 probably swallowed a compact body. The pre-burst observations described above did not have sufficient sensitivity or spatial resolution to detect it before the event.
The compact nature of the swallowed object supports the idea that it was a big Jupiter (or a brown dwarf) or a compact disk fragment, such as a clump. 


The accreted mass was derived by assuming the ZAMS values for $R_*$ (and $M_*$). 
However, massive protostars could be extremely bloated (e.g., \citealp{2010ApJ...721..478H}). 
Their lower surface gravity would then require a much higher accreted mass to release the observed burst energy. 
Such protostars may be unstable to pulsation when heavily accreting (at
$\dot{M}_{acc}\ge 10^{-3}\,M_\odot \rm \, yr^{-1}$).
Then, regular changes in the pumping IR radiation due to pulsation could lead to mid-term maser periodicity (on the order of several $10$ to a several $100\, \rm d$, \citealp{2013ApJ...769L..20I}). 
Cyclic variability was observed in this range for the G323 maser \citep{proven-adzriDiscoveryPeriodicMethanol2019}.
Its period of 93.5 days
would correspond to that of a pulsating protostar with a  luminosity of $\sim 4\times 10^4 L_\odot$ \cite[see their Eq. 1, for a spherical accretion model]{2013ApJ...769L..20I} or slightly more for a thin-disk model \cite[see their Fig. 2, blue stars]{2013ApJ...769L..20I}. 
This agrees well with our models with $L_*=(6.1\pm_{2.5}^{4.2})\times 10^4\,L_\odot$. 
Therefore, we suggest that G323's mid-term maser variability might not be caused by the variable background seed radiation, as proposed by \cite{proven-adzriDiscoveryPeriodicMethanol2019}. Instead, it could be explained by the pulsation of a bloated protostar.

Pulsation occurs only in bloated protostars, which are cooler than ZAMS stars.
When these protostars contract toward the ZAMS, the He$^+$-ionization layer at their surface is destroyed, and the pulsation instability is no longer possible (see \citealp{2013ApJ...769L..20I}). 
For the spherical accretion model, the oscillation period and the protostellar mass and radius can be related according to \cite{2013ApJ...769L..20I} [Eqs. 2 and 3]. In that case, the radius would be as large as $R_*=336\,R_\odot$ and the mass would be $17\,M_\odot$ (slightly below the ZAMS value). With our estimate of the burst energy, this would give an accreted mass of roughly half the solar mass (instead of $7\,M_{Jup}$). In reality, the spherical case is certainly not fulfilled, but it can be considered as a limit. 
The expected bloating depends on the accretion rate for the thin-disk model \cite[see their Fig. 12]{2010ApJ...721..478H}. 
If G323 is heavily accreting ($\dot{M}_{acc}\sim4\times 10^{-3}M_\odot\rm yr^{-1}$) it could be bloated to a few 100 solar radii ($\sim 0.3$ times the value of the spherical accretion model)
, but if the protostellar accretion rate is $\sim 10^{-4}M_\odot\rm yr^{-1}$ its radius could indeed be close to the ZAMS value. 
This implies that the accreted mass (and hence the mass accretion rate) can be significantly higher, probably even on the order of a smaller companion. 
Unfortunately, the K-band spectrum (cf. Sect. \ref{sec:Kspec}) does not show photospheric lines, which could be used to derive the surface gravity of the central source.
If the maser periodicity is caused by protostellar pulsations, this would be a remarkable finding. Maser observations before 2017 are too scarce to detect a periodicity \citep{proven-adzriDiscoveryPeriodicMethanol2019}. 
Although periodic background variations of the maser seed radiation cannot be excluded, the fact that the 6.7\,GHz integrated maser flux density during and after the burst resembles a damped oscillation \citep{MacLeod:2021} with an amplitude that is tightly correlated with the IR flux suggests
that the accretion burst may have triggered protostellar pulsation. 
We emphasize that in principle a heavily bloated protostar is consistent with the pre-burst SED, although some modifications to the setup are required to achieve a $\goodchi^2$-value comparable to our mean model. 


The properties of the HCH{\sc ii} region can be used to assess the evolutionary status of G323. The free-free emission of most hyper/ultra-compact H{\sc ii} regions associated with MYSOs points to an excess of Lyman photons over pure photospheric emission, which is attributed to accretion \citep{Cesaroni:2015, Cesaroni:2016}. For G323, the Lyman continuum photon flux $N_{\rm Lyc}$ can be derived using Eq. 10 in \cite{Martin-Hernandez:2005} by 
plugging in the 3\,mm flux density of $(1.25\pm 0.02)$\,Jy measured by us, the electron temperature of $(7140\pm 680)$\,K obtained by \cite{Zhang;2023}, and the distance given above. 
The resulting value of $(9.3\pm 1.8)\times 10^{47}\,\rm s^{-1}$ is lower than the prediction of $(2.0\pm 0.2)\times 10^{48}\,\rm s^{-1}$ from the $L_{\rm bol}-N_{\rm Lyc}$ relation for ZAMS stars \citep{Cesaroni:2016} for the luminosity range of our models. The discrepancy becomes even greater when a possible contribution of an accretion shock is taken into account. 
This suggests that G323 is still in the pre-ZAMS state at a lower effective temperature, making the bloating scenario seem feasible.
Together, we conclude that the swallowed object was very likely much heavier than our estimate given in Sect. \ref{sec: mass}. 

\subsection{The G323 burst in the context of known MYSO bursts -- possible triggering mechanisms}\label{sec: G323_context}

\begin{table*}[hbt]
\caption[Observed MYSO bursts]{Accretion outbursts observed so far from MYSOs. 
 }
\begin{center}
 \resizebox{\textwidth}{!}{
	\begin{tabular}{|cccccccccc|}
		\hline
		Object & $M_*$ &$L_{pre}$& $L_{peak}$ & $\Delta L$   & $t_{rise}$ & $\Delta t$ & $\dot{M}_{acc}$ & $E_{acc}$ & $M_{acc}$\\ 
		& [$\rm M_\odot$] &[$\rm 10^3 \, L_\odot$]& [$\rm L_{pre}$]& [$\rm 10^3 \, L_\odot$]&[yr]  & [yr] & [$10^{-3}\,M_\odot\,yr^{-1}$] & [$10^{45}$ \, erg] & [$\rm M_{Jup}$]  \\ \hline
  \rowcolor{lightgray} 	G323.46-0.08 (G323)* & 23 &60 &5.4&260 & 1.4 & 8.4 & 0.8 & 90 & 7\\
		S255IR NIRS3* & 20 &30 &5.5 &130 & 0.4 & 2.5 & $5$ & 12 & 2 \\
         G358.93-0.03-MM1* & 12 &5.0 & 4.8 &19 & 0.14 & 0.5 & 1.8 & 2.8 & 0.5\\
		NGC 6334I MM1* & 6.7&3 & 16 & 44 & 0.6 & >8 & $2.3$ & >40 &>0.4 \\
		V723 Car & 10? & $\sim 4$ & & & 4 & $\sim 15$ &  &   & \\
        M17 MIR & 5.4 &1.4 & 6.4 & 7.6& & 9-20 & $\sim 2$ &  & \\
		\hline
	\end{tabular}
 }
 \end{center}
 \tablefoot{The asterisk indicates an accompanying  Class {\sc II} methanol maser flare. 
 Since the NGC 6334I MM1 event is ongoing, the given energy and mass represent lower limits. 
\tablebib{NIRS3 (S255IR NIRS3): \cite{carattiDiskmediatedAccretionBurst2017, 2018A&A...617A..80S, 2020ApJ...904..181L}, G358 (G358.93-0.03-MM1): \cite{Stecklum:2021, brogan:2019, 2020NatAs.tmp..144C, ross:2020, 2022A&A...664A..44B, 2023NatAs.tmp...42B}, G323 (G323.46-0.08): \cite{proven-adzriDiscoveryPeriodicMethanol2019}, NGC (NGC 6334I MM1): \cite{hunterExtraordinaryOutburstMassive2017, 2021ApJ...912L..17H, 2018ApJ...866...87B, 2019A&A...624A..82B}, V723 Car: \cite{2013prpl.conf1H030T, 2015MNRAS.446.4088T, 2015MNRAS.448.1402T}, M17 MIR: \cite{Chen_Sun_Chini_Haas_Jiang_Chen_2021}.
}
   }
	\label{tab: overview bursts}
\end{table*} 

Although the number of MYSO bursts discovered so far is quite small, they span a considerable range of burst characteristics. 
This raises the question whether different trigger mechanisms are responsible. 
An overview of all known bursts is provided in the Table\,\ref{tab: overview bursts}. 
The G323 outburst is, with an energy of $\sim 10^{47}\, \rm erg$, the most energetic observed so far.
The accreted mass is probably the largest, even without taking into account protostellar bloating (see above). 

It might be surprising that all events imply objects heavier than $0.4$ times the mass of Jupiter. 
In principle, one would expect the accretion of lighter objects to occur more frequently. However, a flare caused by an earth-mass planet is about 300 times less energetic than a flare caused by a Jupiter. 
Consequently, the corresponding increase in the luminosity of the protostar is much smaller (less than about $\sim 1\%$ for G323). Furthermore, the increase in the exciting IR radiation might be too small to cause strong methanol maser flares, which served as a burst alert for most of the known outbursts.
Therefore, such events are much more unlikely to be found. 


Interestingly, the accretion rates during the burst are quite similar for all known objects (on the order of a few $10^{-3}M_\odot \rm yr^{-1}$). 
The timescales of the G323, the G358.93-0.03-MM1, and the S255IR NIRS3 events are short enough to be explained by the accretion of a compact object. 
Longer burst durations (similar to V723 Car and possibly M17 MIR) point to a more diffuse object. 
None of the bursts agree with magneto-rotational instability (MRI) or thermal instability (TI) of the disk, where the (peak) accretion rate is much lower ($\sim 10^{-4}M_\odot \rm yr^{-1}$) and the rise time ($\sim 50$ years) and duration ($\sim 100$ for TI and $\sim 1000$ years for MRI) are much longer \citep{2021A&A...651L...3E}.

Gravitational instabilities (GIs) may provide an important burst triggering mechanism as they can form compact objects, such as clumps, planets, or companions.
It is reasonable to assume that massive stars are accompanied by massive disks and that these disks are prone to fragmentation. 
High-resolution hydrodynamical simulations of collapsing massive cloud cores by \cite{2020A&A...644A..41O} show that during formation the primary forms a massive disk. The disk develops spirals and clumps, where some of the clumps reach the innermost grid cell, which contains the protostar. 
A quite famous observational example for a MYSO disk with disk fragmentation is G358.93-0.03 MM1, which also showed an outburst. 
During the burst, an expanding maser ring was visible, which 
revealed the spiral structure of the disk \citep{ross:2020, 2023NatAs.tmp...42B}. 
More evidence of disk fragmentation was sought by \cite{Ahmadi:2023} using CH$_3$CN lines, who found 13 disks in dense cores, of which 11 are massive enough to fragment. 


The accreted masses for the known MYSO bursts
range from more than 0.4 times the 
Jupiter mass (NGC 6334I MM1) to $7 \, M_{Jup}$ (G323), 
all in the range of a heavy planet. Therefore, planet accretion cannot be excluded as the dominant triggering mechanism for MYSO outbursts. This idea was already considered by \cite{1977ApJ...217..693H}, who pointed out that the infall of a Jupiter onto a solar mass protostar would release enough energy to power the outburst of FU~Orionis.
Planets can form directly through fragmentation or via core-accretion. 
In the latter, no GI is required to explain the outbursts. 
For MYSOs the timescales of planet formation via core-accretion are problematic, given their much shorter life times. 
Nevertheless, the possibility of -- heavy -- planet accretion is very interesting, independent of the possible formation pathway,
also from the perspective of planetary population studies. 
Close-in hot Jupiters (with a period of a few days) are observed, but their formation is challenging (see, e.g., \citealp{2018ARA&A..56..175D} and references therein).
The accretion of compact Jupiter mass bodies shows that migration processes are taking place during the formation of massive stars. 
Some migration processes might not end in accretion bursts but form systems with hot Jupiters. However, some care is required, as migration processes around low- and high-mass YSOs might show significant differences. MYSOs have more ionizing radiation and stronger winds. Furthermore, their disks can be more massive and hotter. This might lead to a more efficient transfer of angular momentum and hence to faster migration processes. 

If the massive protostars are bloated, the accreted masses are significantly higher, as discussed for G323 in Sect. \ref{sec: bloat}. Therefore, accretion of a (proto-)stellar companion may also be considered an important burst-triggering mechanism, not only for G323. 
Most stars form as binaries or multiple systems, and the presence of companions is generally expected, particularly for high-mass stars (e.g., \citealp{Bordier:2024, Li:2024}). 
According to recent simulations of \cite{Elbakyan:2023}, accretion from or merger with a large clump or a small companion is possible. Earlier simulations (e.g., \citealp{2020A&A...644A..41O}) did support this as well but were unable to resolve the protostar. 
\cite{Elbakyan:2023} applied a 3D+1D approach to achieve a higher spatial/temporal resolution.
Interestingly, while their timescale is still larger than the observed one, the shape of the luminosity change during the tidal disruption and accretion of an object resembles the $K_{\rm s}$ light curve of the G323 event. 
A similar scenario was suggested by \cite{Nayakshin:2012} which is based on Roche-overflow from a gaseous protoplanet that is tidally disrupted when orbiting close to the protostar. While it was developed with the focus on FU~Ori-like outbursts, it may also be applicable to MYSO bursts, in particular because the model is sensitive to the initial ionized hydrogen fraction $X_i$ of the planet. Their simulations suggest that the larger $X_i$, the shorter the rise time of the burst, and the larger the maximum value of $\dot{M}_{acc}$. In fact, their model with the highest $X_i$ features a time scale and a peak accretion rate which are similar to those of the G323 event.

\subsection{The inner region and the protostellar disk}\label{sec: r_inner} 



\begin{figure}[hbt]
	\begin{center}
		\includegraphics[width=\hsize]{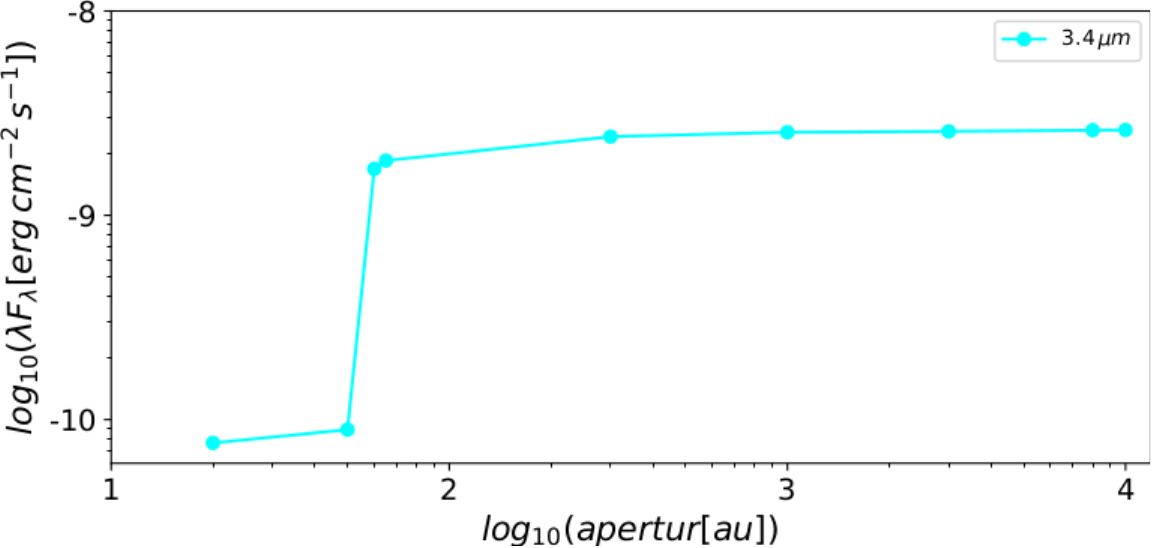}
	\caption{Integrated $L$-band emission for the mean model (cyan) within different radial apertures (dots). The solid line 
             (between the dots) is for guidance. 
            There is a jump at the inner edge of the disk (60\,au), where most of the $L$-band flux is produced.} 
	\label{fig: Lband ap}
    \end{center}
\end{figure} 

Disk-mediated accretion is one pathway for massive star formation, as evidenced by theory and observations. Thus, at the high accretion rates involved in this process, a disk is present, which likely extends inward to the dust sublimation radius (e.g., \citealp{carattiDiskmediatedAccretionBurst2017, beuther_2018, ross:2020, Ahmadi:2023}). 
During later phases of the MYSO evolution, disks can be destroyed by photoevaporation.
In fact, G323 is associated with an HCH{\sc ii} region of $0.06\, \rm pc$ in size \citep{Zhang;2023} that exceeds the diameter of the disk of the mean model. This might challenge our assumption of the presence of a disk. The photoevaporation rate of ${\sim}8 \times 10^{-5}M_\odot \, \rm yr^{-1}$, derived using \cite{1994ApJ...428..654H}, Equation 7.3, together with the extent of the HCH{\sc ii} region and the Lyman continuum flux (cf. Sect. \ref{sec: bloat}), amounts to ${\sim}15 \%$ of the upper limit of $\dot{M}_{acc}$ in quiescence for the ZAMS case (cf. Sect. \ref{sec: mass}). While this might be seen as a hint for substantial disk erosion, it does not apply since, as outlined above, the MYSO likely did not reach the ZAMS yet, which implies higher accretion rates.


Evidence for hot dust 
comes from the $L$-band VLTI/MATISSE observations. Fringes would be seen if the emission were primarily photospheric. Their absence points to a dominant extended component, namely scattered and thermal emission from dust. 
Thermal $L$-band emission originates from dust with $T{\gtrsim}900\, \rm K$, which implies that it must be located within ${\sim} 100\, \rm au$.
The inner radius of the mean model, governed by dust sublimation, amounts to ${\sim}20$\,au according to \cite{2004ApJ...617.1177W}, Eq. 1, with $T_{sub}{=}1\,600\,\rm K$.
The lack of interferometric fringes implies a minimum angular size of 14\,mas (or ${\sim57}\,\rm au$ at the distance of G323, see Sect. \ref{sec: matisse}), which is consistent with that value.
The extent of the $L$-band emission can be compared with the mean model. To do so, synthetic SEDs on different photometric apertures were established.
Fig. \ref{fig: Lband ap} shows a strong increase in the flux density of the $L$-band, which is mainly thermal in origin, exactly at the position of the inner disk rim, followed by an outward shallow rise. 
Direct protostellar emission is more than one order smaller. 


We note that during the burst, the sublimation radius is expected to shift outward by a few 10\,au. 
The $L$-band observation was done nearly a decade after the peak of the burst.
This is too short for the material to recondense. Thus, the inner rim at the date of the $L$-band observations may be further away than ${\sim20}\, \rm au$.
Indeed, for low-mass YSO bursts, dust sublimation fronts are an indirect tracer of past outbursts (see, e.g., \citealp{2020ARA&A..58..727J}). 
However, for massive protostars, the matter will be replenished quickly, given the high accretion rates. With an accretion rate of $10^{-3}\,M_\odot \, yr^{-1}$, it follows that in a decade $10^{-2}\,M_\odot$ can be transported to the inner region, enough to refill the inner disk. 
If the sublimation radius were still further out, higher fluxes would result from the increased surface area. However, the $i$-band post- and pre-burst fluxes are the same within the errors. 
Together, this indicates that there are no significant changes present in the dust distribution in 2023 and that the $L$-band emission originates at the inner rim of the disk, which seems to be as close as the pre-burst dust sublimation radius. 

\subsection{Prospects for TDRT modeling of MYSO bursts} \label{sec: TRRT pros}


The present paper describes the first application of TDRT based on the TORUS code in a real-world case. In that context, we want to outline its potential for the future. 
The time-dependent approach we use is only the first step in the direction of a fully self-consistent treatment of outbursts. 
Improved and more systematic modeling is planned for existing FIR bursts (including G323).
The next step is to use larger sets of time-dependent models, which all agree with the pre-burst. With this in mind, the previous parameter ranges (for both burst and MYSO geometry) shall be refined in the near future.
Due to the SOFIA shutdown, no further FIR observations will be possible in the mid-term. However, future MYSO bursts can be studied with ground-based facilities in the (sub)mm and in the NIR/MIR provided the burst hosts are no longer deeply embedded. JWST is capable of targeting younger and more enshrouded objects. 

With TORUS the dust continuum can be modeled in those spectral ranges as well, while line emission is currently not included in the time-dependent version.  
The temperature grids are delivered at each time step. They might be used as input for chemical burst models, similar to the approach used in \cite{2017A&A...604A..15R, Guadarrama:2024}. 
TORUS does not include chemical networks, which are needed to directly obtain chemical burst models.
Another application of the temperature output is to constrain the location of possible methanol maser sites. This was done in a simplified manner by \cite{Stecklum:2021} for static model grids.

For all our models, we used very simplified assumptions, namely no protostellar bloating, axis-symmetry, same
dust in all regions and no changes during the burst besides heating and cooling 
(i.e., no changes in the dust distribution and chemistry due to sublimation, etc.). Some of these limitations might be explored in the future.
A simple form of dust sublimation is included in which a fraction of the dust exceeding the sublimation temperature is removed each time step. Currently, resublimation is immediate once the temperature drops below the sublimation temperature. This might be improved in the future.
Eventually, we emphasize that TDRT can be applied to other classes of variable objects and transient phenomena where dust plays a major role.

\section{Conclusions}\label{conc}

G323 featured a powerful accretion burst that peaked in 2013. 
It extends the small sample of known MYSO bursts. The objective of this work is to draw conclusions about the nature of the burst from the derived limits for the released energy and the accreted mass. 
G323's burst is a multiwavelength phenomenon that was accompanied by flares of different maser species.
It was observed in the NIR, MIR, FIR, and radio at different timescales. 
There is a clear correlation between the Class II methanol maser flare and the IR radiation.
A light echo is present in $J, H, K_{\rm s}$, and $Z$ as well as in the $WISE$ images. 
Interestingly, its expansion appears to be faster at longer wavelengths, possibly 
because at these wavelengths the scattering optical depth is reduced.

Our SOFIA/HAWC+ observations, performed in 2022, two years after the end of the 
burst, constrained the strength of the thermal afterglow and were crucial to derive limits on the burst energy. 
These measurements and the $K_{\rm s}$ light curve from the VVV(X) survey, indicate that the
burst is probably the most energetic of all known MYSO bursts. The energy released is equal to $E_{acc}=(0.9\mypm_{0.7}^{2.5})\times 10^{47}\, \rm erg$. Its duration, i.e, $8.4\, \rm yrs$, is in line with the accretion of a compact object.
This is in accordance with the observed bluening during the burst, which is not expected if the material flow arises in a more diffuse stream from an active disk. 
For a protostar close to the ZAMS, the minimum mass needed to release this amount of energy is $7.3\mypm_{5.9}^{20}$ times the Jupiter mass, which is in the range of a big Jupiter, a brown dwarf, or a disk fragment. 
The pre-burst luminosity of G323 is consistent with that of a bloated protostar, which might feature pulsations with the same period as the maser. If the bloating scenario holds, the accreted mass may be as high as half a solar mass.
The post-burst emission in the $L$-band can be attributed to the inner disk region. It indicates a minimum extent of the emitting region of 14\,mas or 57\,au,
which is consistent with a disk inner rim situated at the dust sublimation radius during quiescence.


For the first time, we used time-dependent radiative transfer (TDRT) to study a real science case.
With TDRT, we can model the timescales self-consistently. 
The thermal afterglow is
much longer than the grid crossing time. 
G323 is most likely seen face-on. Therefore, the NIR timescales are equal to those of the accretion rate variation. 
The $K_{\rm s}$ light curve reflects the protostellar luminosity variation throughout 
the entire burst duration. At longer wavelengths, the timescales are much longer. 
We find that the visibility of the (MIR/FIR) afterglow can vary by up to years, depending on the dust distribution and the burst characteristics.
The example of G323 shows that TDRT simulations are a powerful tool for investigating transient phenomena of dusty objects.
For episodic accretion of MYSO, it opens up the possibility to obtain reliable burst parameters,
which are needed to understand the unsteady protostellar growth, especially in the 
high-mass regime.
The shutdown of SOFIA is a serious drawback, since it was the only facility that could observe in the FIR which is crucial to derive the burst energy. Future studies must focus on MIR and (sub)mm. 
The rare face-on view of G323 and its NIR record of the burst make this source a key object for studying the consequences of episodic accretion and a benchmark for the corresponding simulations.

\begin{acknowledgements}
We thank the anonymous referee for accurate review and highly appreciate comments and suggestions which helped to improve the
quality of this paper. The authors thank Dr. Gordon MacLeod for support concerning the maser observations and Dr. Riccardo Cesaroni for input concerning the free-free emission luminosity.
VW was supported by the German Aerospace Center (DLR) under grant number 50OR1718. A.C.G. acknowledges from PRIN-MUR 2022 20228JPA3A "The path to star and planet formation in the JWST era (PATH)" funded by NextGeneration EU and by INAF-GoG 2022 "NIR-dark Accretion Outbursts in Massive Young Stellar Objects (NAOMY)" and Large Grant INAF 2022 "YSOs, Outflows, Disks, and Accretion: toward a global framework for the evolution of planet-forming systems (YODA)". 
This publication uses data products from the Near-Earth Object Wide-field Infrared Survey Explorer ((NEO)WISE), which is a joint project of the Jet Propulsion Laboratory/California Institute of Technology and the University of Arizona. (NEO)WISE is funded by the National Aeronautics and Space Administration. Based on data products from observations made with ESO Telescopes at the La Silla or Paranal Observatories under ESO programmes ID083.C-0582, ID 179.B-2002 and 198.B-2004. Based on data obtained from the ESO Science Archive Facility under request numbers 557557 and 573746. The ALMA data are from the programme ALMAGAL (2019.1.00195.L) and were retrieved from the ALMA Science Archive \citep[e.g.,][]{StoehrALMAScienceArchive2014}. This paper used information from the Red MSX Source survey database at \url{http://rms.leeds.ac.uk/cgi-bin/public/RMS_DATABASE.cgi}, which was constructed with the support of the Science and Technology Facilities Council of the UK. This work used data from the European Space Agency (ESA) mission
{\it Gaia} (\url{https://www.cosmos.esa.int/gaia}), processed by the {\it Gaia}
Data Processing and Analysis Consortium (DPAC,
\url{https://www.cosmos.esa.int/web/gaia/dpac/consortium}). Funding for the DPAC
has been provided by national institutions, in particular, the institutions
participating in the {\it Gaia} Multilateral Agreement. 
The national facility capability for SkyMapper has been funded through ARC LIEF grant LE130100104 from the Australian Research Council, awarded to the University of Sydney, the Australian National University, Swinburne University of Technology, the University of Queensland, the University of Western Australia, the University of Melbourne, Curtin University of Technology, Monash University and the Australian Astronomical Observatory. SkyMapper is owned and operated by The Australian National University's Research School of Astronomy and Astrophysics. Survey data were processed and provided by the SkyMapper Team at ANU. 
This research has made use of NASA's Astrophysics Data System. This research has used adstex (\url{https://github.com/yymao/adstex}).\\

\end{acknowledgements}

\bibliographystyle{aa_url} 

\bibliography{G323.46-0.08} 

\begin{appendix}
\begin{table*}[hbt]
\section{Additional information}\label{app}

\caption{HAWC+ photometry and beam sizes.} 
\begin{center}
\begin{tabular}{| c c c c r c|}
\hline
$\lambda$ & F$_\nu $ & Aperture & Beam size  & Image size & Deconvolved\\
\hline
[$\mu$m] & [Jy] & radius [\arcsec] & FWHM [\arcsec] & FWHM [\arcsec]/PA [\degr] & FWHM [\arcsec]\\
\hline
53 & $2\,602\pm260$ & 7.8 & 4.85 & 6.0$\times4.8@58$ & 2.4$\pm$0.7\\
62 & $2\,669\pm270$ & 12.2 & 5.59 & 9.9$\times7.4@46$ & 6.4$\pm$0.7\\
89 & $3\,067\pm310$ & 12.2 & 7.8 &11.0$\times7.2@46$ & 4.4$\pm$1.0\\
154 & $1\,928\pm190$ & 17.0 & 13.6 &18.2$\times14.0@52$ & 8.4$\pm$1.6\\
214 & $1\,295\pm130$ & 21.0 & 18.2 &24.8$\times20.5@56$ & 13.3$\pm$1.9\\
\hline
\end{tabular}
\end{center}
\label{tab: Hflux}
\tablefoot{HAWC+ photometry along with beam sizes of the instrument and extent of the object. The 62{\ts}$\mu$m beam size was interpolated since it could not be measured \citep{harper:2018}.}

\caption{(NEO)WISE photometry.}
\begin{center}
\begin{tabular}{| c c c c c |}
\hline
MJD & $W1$ & $\sigma$ & $W2$ & $\sigma$ \\
\hline
[d] & [mag] & [mag] \ & [mag] & [mag] \\
\hline
55250.93 & 4.88 & 0.06 & 3.09 & 0.04 \\
55432.44 & 4.67 & 0.03 & 3.33 & 0.04 \\
56714.99 & 3.32 & 0.14 & 2.24 & 0.09 \\
56894.50 & 3.66 & 0.04 & 2.32 & 0.05 \\
57077.54 & 3.64 & 0.10 & 2.28 & 0.03 \\
57253.79 & 4.06 & 0.06 & 2.54 & 0.06 \\
57442.45 & 4.17 & 0.05 & 2.39 & 0.04 \\
57614.82 & 4.24 & 0.09 & 2.68 & 0.07 \\
57808.87 & 4.38 & 0.07 & 2.83 & 0.05 \\
57975.66 & 4.61 & 0.17 & 3.00 & 0.03 \\
58173.18 & 4.59 & 0.06 & 2.97 & 0.05 \\
58335.97 & 4.67 & 0.14 & 3.22 & 0.08 \\
58537.30 & 4.77 & 0.12 & 3.16 & 0.08 \\
58700.21 & 4.51 & 0.10 & 3.09 & 0.05 \\
58904.44 & 4.76 & 0.08 & 2.91 & 0.04 \\
59067.13 & 4.53 & 0.11 & 3.19 & 0.07 \\
59268.60 & 4.47 & 0.08 & 3.05 & 0.03 \\
59431.64 & 4.37 & 0.05 & 3.07 & 0.02 \\
59633.90 & 4.63 & 0.04 & 3.15 & 0.04 \\
59796.55 & 4.58 & 0.09 & 3.05 & 0.05 \\
\hline
\end{tabular}
\end{center}
\label{tab:Wmag}

\end{table*}

\begin{figure*}
    \sidecaption
    \includegraphics[width=12.5cm]{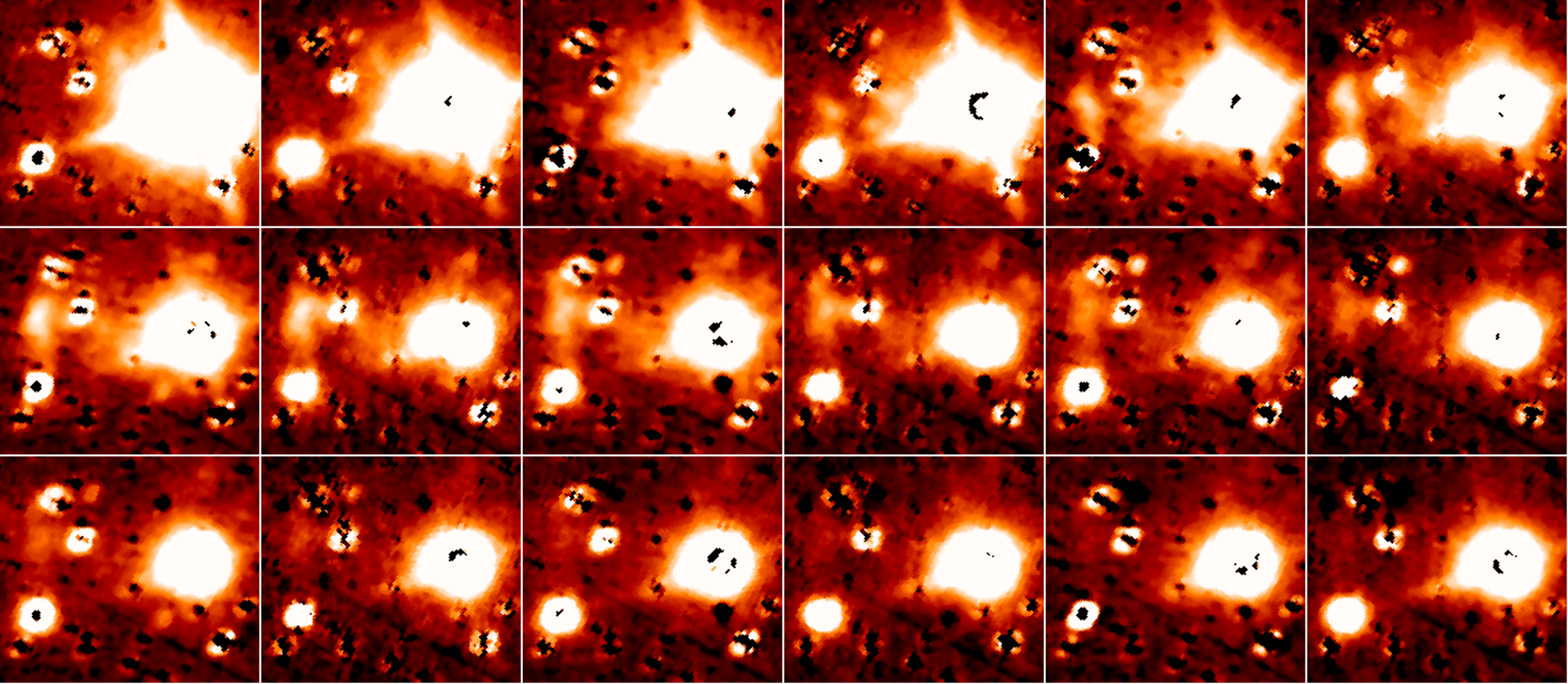}
\caption{Evolution of the remote LE (left border) in the W1 band, illustrated by bi-annual difference images that were created by subtracting those from 2014. FoV and orientation correspond to Fig.\,\ref{fig:seq}. Each row covers three years, starting on top left in early 2015. The W2 images look similar. A few variable stars are present as well.}
 \label{fig:W1_LE}
\end{figure*}


\begin{table}[t]
\begin{center}
\caption[G323 pre-burst SED]{G323 pre-burst flux densities.}
\label{tab: G323 preSED}
\begin{tabular}{|cccc|}
\hline
$\lambda$ & $F_\nu $&$\Delta F_\nu $ &Ref.\\
$[\mu$m] & [Jy]& [Jy]&\\
\hline
1.02 &   0.000407  &      0.000005& 1\\
1.24  &  0.00336 &0.0002  & 2\\
1.25   & 0.00297 &0.00001 & 1\\
1.63  &  0.143   &0.02    & 1\\
1.65  &  0.0239  &0.0018  & 2\\
2.13  &  0.437   &0.05    & 1\\
2.16  &  0.291   &0.03    & 2\\
3.35  &  3.79    &0.21    & 3\\
4.35  &  9       &1.48    & 4\\
7.67  &  48.5    &5       & 9 \\
8.28  &  33.6    &1.4     & 4\\
8.58  &  59.1    &6       & 9 \\
8.61  &  37.3    &0.3     & 5\\
9.83  &  32.4    &3       & 9 \\
10.4  &  40.7    &5       & 9 \\
10.9  &  54.2    &6       & 9 \\
11.6  &  84.1    &5       & 6\\
12.1  &  103     &5       &4\\
14.6  &  154     &9       &4\\
18.4  &  276     &4       &5\\
21.3  &  364     &22      & 4\\
23.9  &  522     &26      & 6\\
\textbf{70}    &  \textbf{2\,459}    &\textbf{23}      &7\\ 
\textbf{160}   &  \textbf{1\,721}    &\textbf{70}      & 7\\ 
250   &  962     &54      & 8\\
350   &  232     &8       & 8\\
500   &  73.6    &1.8     &  8\\
870   &  13.3    &1.33    &8\\
\hline
\end{tabular}\\
\end{center}
\tablefoot{ Facilities and instruments are given behind corresponding references. 
Observations with complementing post-burst measurements are boldface.}
\tablebib{(1) \cite{2017yCat.2348....0M}  (VISTA/VIRCAM); (2) \cite{Jarrett_2000} 2MASS; 
(3) \cite{2014yCat.2328....0C} WISE; (4) \cite{lumsden:2013} (5) \cite{2010cosp...38.2496Y} AKARI; (5) \cite{1994yCat.2125....0J} IRAS; (5) \cite{2013A&A...553A.132M} (Herschel/PACS); (5) \cite{ATLASGALSchuller2009} (ATLASGAL/APEX) The flux was scaled, as mentioned in \cite[Section 3.1]{2017MNRAS.471..100E}.; (9) \cite{IRAS_LRS}, extracted from the IRAS-LRS Spectrum.} 
\label{Tab: preSED}
\end{table} 

\begin{table}
\begin{center}
\caption{Skymapper photometry.}
\begin{tabular}{| c c c c c |}
\hline
MJD & $i$ & $\sigma$ & $z$ & $\sigma$ \\
\hline
[d] & [mag] & [mag] & [mag] & [mag] \\
\hline
55404.2 & 17.72 & 0.07 & & \\
57090.8 & 15.71 & 0.04 & 18.62 & 0.08\\
57109.6 & 15.69 & 0.04 & 18.54 & 0.06\\
57142.2 & 15.76 & 0.01 & & \\
57144.4 & 15.87 & 0.02 & & \\
57504.7 & 16.38 & 0.05 & 19.15 & 0.09\\
58269.5 & 17.17 & 0.06 & 19.66 & 0.15\\
58615.5 & 17.41 & 0.04 & 20.06 & 0.06\\
58997.5 & 17.60 & 0.06 & 20.06 & 0.19\\
59000.5 & 17.60 & 0.05 & 20.53 & 0.09\\
59303.7 & 17.50 & 0.07 & 19.88 & 0.23\\
59445.4 & 17.74 & 0.07 & & \\
\hline
\end{tabular}
\label{tab:Skymp}
\end{center}
\end{table}

\begin{table}
\caption{$K_{\rm s}$ VVV(X) photometry.}
\begin{center}
\begin{tabular}{| c c c |}
\hline
MJD & $K_{\rm s}$ & $\sigma$ \\
\hline
[d] & [mag] & [mag] \\
\hline
55260.3805 & 7.88 & 0.01 \\
55298.2012 & 8.06 & 0.01 \\
55327.2239 & 7.97 & 0.01 \\
55387.1757 & 7.94 & 0.01 \\
55389.1413 & 7.95 & 0.01 \\
55407.1306 & 7.89 & 0.01 \\
55411.0769 & 7.98 & 0.01 \\
55422.0257 & 7.86 & 0.01 \\
55423.0842 & 7.91 & 0.01 \\
55425.0818 & 7.93 & 0.01 \\
55690.3558 & 7.91 & 0.01 \\
55787.1077 & 7.92 & 0.01 \\
55788.0946 & 7.89 & 0.01 \\
55809.0276 & 7.92 & 0.01 \\
55810.0342 & 7.95 & 0.01 \\
55818.0067 & 7.98 & 0.01 \\
55819.0259 & 7.86 & 0.01 \\
55821.0143 & 7.79 & 0.01 \\
55824.0080 & 7.96 & 0.01 \\
56057.3551 & 7.74 & 0.01 \\
56373.4077 & 6.80 & 0.01 \\
56376.1931 & 6.76 & 0.01 \\
56377.2970 & 6.92 & 0.01 \\
56418.1619 & 6.62 & 0.01 \\
56420.2644 & 6.64 & 0.01 \\
56438.2499 & 6.18 & 0.01 \\
56459.1327 & 6.11 & 0.01 \\
56461.0305 & 6.24 & 0.01 \\
56462.2432 & 6.13 & 0.01 \\
56464.2453 & 6.20 & 0.01 \\
56466.2352 & 6.06 & 0.33 \\
56467.2296 & 6.02 & 0.01 \\
56468.2418 & 6.19 & 0.01 \\
56469.1801 & 6.08 & 0.01 \\
56470.1437 & 6.23 & 0.01 \\
56471.0450 & 6.20 & 0.01 \\
56472.1275 & 6.24 & 0.01 \\
56826.9659 & 5.84 & 0.04 \\
56827.0582 & 5.75 & 0.03 \\
56827.0712 & 5.70 & 0.09 \\
56827.1085 & 5.73 & 0.02 \\
56828.9879 & 5.78 & 0.01 \\
56831.0109 & 5.83 & 0.10 \\
56837.0041 & 5.78 & 0.01 \\
56839.9728 & 5.86 & 0.01 \\
57134.1239 & 6.21 & 0.01 \\
57171.0419 & 6.28 & 0.01 \\
57172.0697 & 6.47 & 0.19 \\
57181.9897 & 6.27 & 0.01 \\
57182.2159 & 6.41 & 0.01 \\
57198.9787 & 6.22 & 0.01 \\
57586.0967 & 6.90 & 0.01 \\
57922.1865 & 6.92 & 0.01 \\
57924.1343 & 6.93 & 0.01 \\
58264.0137 & 7.38 & 0.02 \\
58596.3896 & 7.44 & 0.01 \\
58655.2577 & 7.48 & 0.01 \\
58704.0489 & 7.48 & 0.01 \\
\hline
\end{tabular}
\end{center}
\label{tab:Ksmag}
\end{table}

\begin{sidewaystable}[hbt]
\caption{G323 pre-burst fit result.}
	\begin{tabular}{|c|c|ccc|ccccccccccc|}
		\hline
Nr & name & $\goodchi^2$ & d & Av & $m_{disk}$ & $r_{disk}$ & $\alpha_{disk}$ & $\beta_{disk}$ & $h_{100}$ & $\rm r_{env}$ & $\rm \dot{M}_{env}$ & $\Theta_{cav}$ & $\rho_{cav}$ & $i$ & $L_*$ \\
&  &  & kpc & mag & $\rm M_\odot$ & au &  &  & au & au & $1.33 \, \rm M_\odot\,yr^{-1}$ & $\degr$ & $ 10^{-20} \, \rm g\,cm^{-3}$ & $\degr$ & $\rm L_\odot$  \\ \hline
\rowcolor{lightgray}
&&& log & lin & log & log & lin & lin & log & log & log & lin & log & lin & log \\ \hline
\rowcolor{lightgray}
min &&&&& $4\times 10^{-8}$ & 80 & 1.0 & 1.0 & 0.5 & 10\,000 & $2\times 10^{-5}$ & 14 & 0.01 & 0 & 8\,000 \\
\rowcolor{lightgray}
max &&&&& 2.5 & 7\,000 & 3.3 & 1.3 & 33 & 40\,000 & 1 & 60 & 1\,000 & 60 & 160\,000 \\ \hline
\hline
1 & MCX48 & 140 & 4.48 & 17 & 0.04 & 498 & 2.542 & 1.058 & 2.244 & 34\,877 & 0.03084 & 52 & $5.5$ & 25 & 88\,059 \\
2 & MCX24 & 263& 3.9 & 19 & 0.0201 & 3\,381 & 1.603 & 1.014 & 1.69 & 23\,928 & 0.03654 & 54 & 5.2 & 37 & 64\,540\\ 
3 & MCY437 & 306 & 3.7 & 17 & 0.00062 & 3\,605 & 2.22 & 1.221 & 26.51 & 11\,982 & 0.05759 & 38 & 6.5 & 40 & 44\,781\\ 
4 & MCY916 & 346 & 3.7 & 17 & 0.00079 & 2\,083 & 2.485 & 1.138 & 1.836 & 16\,950 & 0.01633 & 50 & 3.7 & 56 & 101\,215\\ 
5 & MCX831 & 395 & 4.48 & 19 & 0.00473 & 106 & 1.159 & 1.222 & 0.6861 & 38\,660 & 0.0111 & 39 & 2.4 & 6 & 103\,690\\ 
6 & MCX152 & 396 & 3.7 & 19 & $3.4\times 10^{-7}$ & 555 & 1.859 & 1.101 & 3.885 & 20\,568 & 0.01352 & 22 & 2.0 & 3 & 19793\\ 
7 & MCX391 & 475 & 4.48 & 18 & 0.02483 & 429 & 3.16 & 1.259 & 1.231 & 30\,800 & 0.03746 & 28 & 6.83 & 4 & 40\,115\\ 
8 & MCZ27 & 489 & 3.7 & 18 & 0.09312 & 108 & 2.397 & 1.18 & 2.72 & 17\,546 & 0.08764 & 31 & 6.2 & 28 & 24\,158\\ 
9 & MCY891 & 529 & 4.0 & 17 & 0.00049 & 542 & 2.613 & 1.147 & 25.37 & 15\,747 & 0.01049 & 26 & 4.6 & 30 & 97\,363\\ 
10 & MCX92 & 556 & 3.9 & 18 & $6.8 \times 10^{-8}$ & 220 & 1.757 & 1.067 & 4.295 & 32\,383 & 0.003268 & 41 & 3.6 & 10 & 58087\\  \hline
\rowcolor{orange} 
mean &&90&3.9&19& 0.0017 & 684 & 2.2 & 1.12 & 3.1 & 24\,000 & 0.024 & 42 & 4.5 & 26 & 6\,0587 \\
sigma &&&&& 64 & 3.25 & 0.6 & 0.08& 2.9 & 1.5 & 2.2 & 11& 1.5 & 17 & 1.7 \\ \hline
min &&&&& $2.7 \times 10^{-5}$ & 200 & 1.7 & 1.03 & 1.1 & 16\,000 & 0.011 & 31 & 3.0 & 9 & 36\,000 \\
max &&&&& 0.11 & 2\,100 & 2.8 & 1.20 & 9.0 & 36\,000 & 0.053 & 53 & 6.8 & 43 & $10^5$ \\ \hline
\hline
\rowcolor{yellow} 
Tmin &&610&3.9&19& $2.7 \times 10^{-5}$ & 684 & 2.2 & 1.12 & 3.1 & 16\,000& 0.011 & 53 & 3.0 & 26 & 60\,587 \\
\rowcolor{yellow} 
Tmax &&180&3.9&19& 0.11& 684 & 2.2 & 1.12 & 3.1 & 36\,000 & 0.053 & 31 & 6.8 & 26 & 60\,587 \\
		\hline
	\end{tabular}
	\label{tab: G323 mean}\label{tab: param-ranges-all}
 \tablefoot{Parameters of the mean model (orange), the ten best fits, and the models with a minimal (Tmin), and a maximal (Tmax) afterglow duration (both in yellow). The adapted ranges for the pre-burst models are given in gray for comparison. Log-sampled values imply geometric means (with $x_{min}=<x>\cdot\sigma$). All values are total (gas+dust). We note the factor 1.33 in the unit of $\dot{M}_{env}$.}
\end{sidewaystable}

\end{appendix}

\end{document}